\documentclass[prd,aps,eqsecnum,floatfix,nofootinbib,preprint,tightenlines]{revtex4}
\usepackage{latexsym}
\usepackage{graphicx}
\usepackage{amsmath}
\usepackage{hyperref}

\def\NON{\nonumber\\}
\def\bibi{\bibitem}

\def\a{\alpha}
\def\b{\beta}
\def\c{\chi}
\def\d{\delta}
\def\e{\epsilon}                
\def\g{\gamma}

\def\j{\psi}

\def\l{\lambda}
\def\m{\mu}
\def\n{\nu}

\def\p{\pi}                     
\def\th{\theta}                  
\def\r{\rho}                    
\def\s{\sigma}                  
\def\t{\tau}

\def\x{\xi}

\def\D{\Delta}

\def\G{\Gamma}
\def\J{\Psi}
\def\L{\Lambda}

\def\ca{{\cal A}}
\def\cb{{\cal B}}

\def\cd{{\cal D}}

\def\cf{{\cal F}}
\def\cg{{\cal G}}

\def\ck{{\cal K}}

\def\cm{{\cal M}}
\def\cn{{\cal N}}
\def\co{{\cal O}}

\def\car{{\cal R}}
\def\cs{{\cal S}}
\def\ct{{\cal T}}
\def\cu{{\cal U}}
\def\cv{{\cal V}}
\def\cw{{\cal W}}

\def\cz{{\cal Z}}

\def\cbo{{\,\raise-.15ex\Sc [\,}}                       

\def\sl#1{\rlap{\hbox{$\mskip 1 mu /$}}#1}      

\def\svev#1{\left\langle #1\right\rangle}       

\def\ddt#1{{\buildrel {\hbox{\LARGE .\kern-2pt.}} \over {#1}}}

\def\ie{\mbox{\it i.e.} }
\def\eg{\mbox{e.g.} }

\def\leqx{\,\raisebox{-1.0ex}{$\stackrel{\textstyle <}{\sim}$}\,}

\def\tr{{\rm tr}\,}

\def\half{{1\over 2}}
\def\Re{{\rm Re\,}}

\def\det{{\rm det}}

\def\bq{\overline{q}}
\def\bj{\overline\j}
\def\bc{\overline\c}

\def\bq{\overline{q}}

\def\tx{\tilde{x}}
\def\ty{\tilde{y}}
\def\tz{\tilde{z}}

\def\tJ{\widetilde{J}}

\def\tD{\widetilde{D}}

\def\Rf{\hat{\car}}

\def\bltz{\mbox{\boldmath $B$}}

\def\ts{{\rm tr}_{ts}}
\def\Seff{S_{eff}}

\def\perm{\hat{\omega}}
\def\Perm{\hat{\Omega}}
\def\rel{\rho}
\def\sumx#1{\sum{}^{(#1)}} 
\def\sumx#1{\sum_{(#1)}}
\def\ww{{\cal W}}
\def\slide#1#2{\raisebox{#1ex}{$#2$}}

\long \def \blockcomment #1\endcomment{}

\def\Dstag{D_{stag}}

\def\Zroot{Z^{root}}
\def\czr{\cz^{root}}

\def\quart{ \mbox{${1\over 4}$} }

\def\id{{\bf 1}}
\def\tcs{\widetilde{\cal S}}
\def\scpt{S$\chi$PT}
\def\ahat{\hat{A}}

\begin{document}
\hyphenation{fer-mio-nic per-tur-ba-tive}

\title{Renormalization-group analysis of the validity of\\[2mm]
  staggered-fermion QCD with the fourth-root recipe\\[4mm] }

\author{Yigal Shamir}
\email{shamir@post.tau.ac.il}
\affiliation{School of Physics and Astronomy, Raymond and Beverly
Sackler Faculty of Exact Sciences,
\mbox{Tel~Aviv University, 69978 Tel~Aviv, Israel}\\ }

\begin{abstract}

\vspace {3ex}
I develop a renormalization-group blocking framework for lattice QCD
with staggered fermions.  Under plausible, and testable, assumptions,
I then argue that the fourth-root recipe
used in numerical simulations is valid in the continuum limit.
The taste-symmetry violating terms, which give rise to non-local
effects in the fourth-root theory
when the lattice spacing is non-zero, vanish in the continuum limit.
A key role is played by reweighted theories that are local and
renormalizable on the one hand, and that approximate the fourth-root
theory better and better as the continuum limit is approached
on the other hand.
\end{abstract}

\maketitle

\section{\label{intro} Introduction}
Lattice QCD simulations with staggered fermions
\cite{stg,Suss,STW,KwSm,taste,saclay,SvdD,mgjs} have been
producing remarkably accurate predictions of various hadronic observables
\cite{milc}.
The staggered-fermion field has only one component per color per lattice site,
making the numerical computations relatively cheap,
as well as a non-anomalous $U(1)$ chiral symmetry
in the massless limit, which is
important for the phenomenology of the light-quark sector.

All staggered-fermion simulations with three flavors of light quarks
make use of the fourth-root recipe \cite{trick}.
The up, down, and strange quarks are each represented by
a staggered field with a different bare mass.
But normally a single staggered field
generates four quark species, or ``tastes,'' in the continuum limit.
The four tastes do have equal renormalized masses
thanks to the lattice staggered-fermion symmetries \cite{mgjs}.\footnote{
  In principle, one could account for the up, down, strange, and charm quarks
  by the four tastes of a single staggered field
  with a general staggered mass term \cite{mgjs}.
  In practice, this is not done.
  See Ref.~\cite{Steve} for a discussion of the reasons.
}
In order to remove the excessive degrees of freedom
one takes the fourth root of the staggered-fermion determinant.%
\footnote{
  In the isospin limit, the up--down sector is represented by a square root
  of a staggered determinant with the common light quark mass.
}
The fourth-root recipe defines a renormalizable theory which,
to all orders in perturbation theory,  reproduces a local, unitary theory
with the correct number of light quarks
in the limit of a vanishing lattice spacing.
(It is assumed \cite{PQ} that the staggered theory
without the fourth root
behaves as expected in perturbation theory, which
is very plausible \cite{JG,Steve}.)
Non-perturbatively, the validity of the fourth-root recipe is a
non-trivial issue which has been the subject of much debate \cite{doubt}.
For a recent review, see Ref.~\cite{Steve}.

In a formal expansion in the lattice spacing,
the massless staggered action splits into marginal terms
that have a $U(4)$ taste symmetry and irrelevant terms
that break the symmetry explicitly.
Because of the absence of an exact four-fold degeneracy in the spectrum
of the staggered Dirac operator,
the fourth-root theory is non-local, and not unitary, at any
non-zero lattice spacing \cite{BGS}.
The question that must be addressed is whether the non-locality
disappears, and unitarity is recovered, in the continuum limit.

First, a degree of control over the infra-red behavior
must be maintained, and I will assume that the quark masses
are all positive \cite{SV,DH,CBma,BGSS}.
In a nut-shell, the following tentative reasoning summarizes how the
fourth-root recipe might be valid in the continuum limit:
The taste-violating effects of staggered fermions arise from
irrelevant operators.
These effects should scale to zero in the continuum limit
like a positive power of the lattice spacing.
Hence exact four-fold taste degeneracy is recovered
in the continuum limit. The effective low-energy
Dirac operator attains the form $\tD  \otimes {\bf 1}$,
where $\tD$ is a local operator that carries no taste index,
and where ${\bf 1}$ is the four-by-four identity matrix in taste space.
The fourth root of $\det(\tD  \otimes {\bf 1})$ is $\det(\tD)$.
This fourth root is analytic, leading to
locality and unitarity.\footnote{
  The analytic continuation to Minkowski space must be performed
  after the continuum limit.
}

Before making any concrete claims, a framework is needed
where statements about the continuum limit
will be well defined to begin with. The natural tool for the task
is renormalization-group (RG) block transformations,
which have already been used in the free theory \cite{rg}.
A coarse-lattice theory is obtained via $n$ blocking steps,
starting from a fine-lattice theory which, in the case at hand,
contains staggered fermions.  The process is repeated with more and more
blocking steps, but the coarse-lattice spacing (in physical units)
is held fixed.  With each additional blocking step
the initial fine-lattice spacing gets smaller,
while the bare parameters are adjusted to maintain constant physics.
In the limit $n\to\infty$, one obtains a (well-defined!) coarse-lattice theory
that generates a set of continuum-limit observables.
By setting the coarse-lattice spacing
to be small enough, we ensure that the observables
are rich enough to extract all the QCD physics.

Using this blocking framework, I make the following two-pronged argument.
I first derive the non-controversial claim that exact taste symmetry
is recovered in the continuum limit of the ordinary, local staggered theory.
The concrete physical properties on which this conclusion rests
are itemized.  Like many fundamental properties,
no rigorous proofs of these physical properties are available.
Yet, they are more than just plausible.  It is
difficult to imagine how they could be spoiled
without grossly affecting the continuum limit of lattice QCD as we know it.

I next examine to what extent the same
physical properties remain valid in the fourth-root theory.
Properties that have to do purely with the short-distance behavior,
and generalize relatively easily to the fourth-root theory,
include power-counting renormalizability \cite{PQ,JG,Steve},
and the locality of the contributions to the gauge-field effective action
generated by the integration over the ultra-violet fermion modes.

A crucial ingredient is the scaling of taste-breaking effects.
Because of the absence of a local fermion action,
it is unclear whether a scaling analysis is at all possible.
In other words, it is unclear what is the merit of the observation that
the taste-breaking effects arise, formally, from an irrelevant operator.
I bypass this difficulty by building a representation of the RG-blocked theory
where all the taste violations can be traced back to a local operator,
whose scaling can be computed by developing the appropriate perturbative
expansion.  I argue that the result should indeed reproduce the scaling law
of an irrelevant operator.

If the taste violations vanish in the limit of infinitely many
blocking steps, then, after sufficiently many blocking steps, it should
be possible to find local theories in the correct universality class
that provide a good approximation of the fourth-root theory.
I construct such local theories explicitly via a \textit{reweighting}
of the blocked fourth-root theory that amounts to discarding
all the taste-breaking terms.
Starting from the reweighted theories and working
back towards the fourth-root theory, I then conclude
that exact taste symmetry is recovered
in the continuum limit of the fourth-root staggered theory as well.
As I have explained above, this implies the validity of the fourth-root
recipe.

This paper is organized as follows.
The RG-blocking framework is introduced in Sec.~\ref{overview}.
Technical details have been kept to a minimum, and are mostly relegated
to several appendices.  The beginning of Sec.~\ref{overview} contains
a brief summary both of the section itself
and of the content of the appendices.
Sec.~\ref{overview} ends with the introduction of the reweighted theories,
at which point I give a description of the key steps of the argument,
the details of which will be presented in subsequent sections.

The recovery of exact taste symmetry in the continuum limit of
the ordinary, local staggered theory is discussed in Sec.~\ref{unroot}.
The reweighted theories are discussed in detail in Sec.~\ref{uv}.
Finally, in Sec.~\ref{rooted}, the reweighted theories are used to establish
the recovery of exact taste symmetry in the continuum limit of
the fourth-root theory. My conclusions are given in Sec.~\ref{conclusion}.

This paper is long, and addresses both conceptual and technical questions.
In order to help the reader find his/her way through,
I have organized the paper
such that the essentials are summarized in the following (sub)sections:
Sec.~\ref{plan} explains the technical layout of the argument;
Sec.~\ref{rewcap} gives a summary of key properties of the reweighted theories;
and Sec.~\ref{conclusion} contains the conclusions.
For a brief account of this work, see Ref.~\cite{CMY}.

\section{\label{overview} The RG blocking framework}
In this section I introduce the RG blocking framework,
referring mostly to the ordinary staggered theory for pedagogical reasons.
This section provides a bird's eye overview.
It contains only the minimum needed to follow the logic of the main argument,
as given in subsequent sections.
Much of what goes into the construction has been relegated to several
appendices.  In the appropriate places, I refer to the relevant
appendices for a more elaborate discussion.  The summary below serves
as a ``Table of Contents'' both for this section and for the appendices.

The blocking transformations are introduced in Sec.~\ref{blockn}
which also serves to set the notation.
The first blocking step is special,
since it is used to make the transition from the standard
one-component formalism of staggered fermions to a taste-basis
representation.  This special step is discussed in App.~\ref{tst1}.
The same appendix also provides technical details on the fermion blocking
kernels for subsequent blocking steps (App.~\ref{tst2}),
and contains a proof of the positivity
of fermion determinants encountered in the blocking process  (App.~\ref{tst3}).
Next, Sec.~\ref{conv} casts the partition function of the resulting blocked theory
in a form that will be repeatedly used below.
Sec.~\ref{pullb} introduces the pull-back mapping of operators
from the coarse lattice back to the original fine lattice,
as well as its main uses.
A more elaborated discussion of the pull-back mapping
may be found in App.~\ref{pullback}, alongside with some details
on the gauge-field blocking kernels.
Also relegated to appendices are the generation of blocked gauge-field
ensembles (App.~\ref{ensemble}), a general discussion of
lattice symmetries under the blocking transformation (App.~\ref{sym}),
as well as a more specific discussion of the hypercubic and chiral
symmetries (App.~\ref{disorder}).

The reasons why the interacting fourth-root theory cannot be local
for any finite lattice spacing have been spelled out in Ref.~\cite{BGS},
which also contains further details on various Dirac operators
encountered within the blocking framework.  For completeness, a brief
review of these reasons is given in App.~\ref{nnlcl}.
The remaining two appendices (App.~\ref{mbound} and App.~\ref{iroprep})
deal with scaling issues.

In this section, I continue in Sec.~\ref{free} with
a summary of the main lessons from the
free theory \cite{rg}.  Finally, I introduce the fourth-root
theory and the various reweighted theories in Sec.~\ref{plan},
and give an overview
of the argument, to be presented in detail in subsequent sections,
the conclusion of which is that the fourth-root recipe is
valid in the continuum limit.

\subsection{\label{blockn} The blocking transformations}
Originally, the partition function of the ordinary staggered theory is
\begin{equation}
  Z = \int \cd\cu \cd\c \cd\bc\; \exp[-S_g(\cu) -\bc \Dstag(\cu) \c]\,,
\label{Zstag}
\end{equation}
where $\Dstag(\cu)=\Dstag(x,y;\cu)$ is the staggered Dirac operator,
and $\bc(x),\c(y)$ is the staggered field.
The fine-lattice coordinates are denoted $x,y$,
and the lattice spacing is $a_f$.
The link variables are $U_{\m,x}$, and the gauge field as a whole
will be denoted $\cu=\{U_{\m,x}\}$.
The gauge-field action is $S_g(\cu)$.
Summations over all lattice sites will
be suppressed.

As already mentioned, I assume that all the quark masses
are positive.\footnote{
  See Refs.~\cite{DH,BGSS} for a discussion of how to implement the physical
  theory with a negative quark-mass using the fourth-root recipe.
}
In order to avoid unnecessarily cluttered notation I will consider
the ordinary and fourth-root theories with a single flavor of
staggered fermions. The generalization to more than one flavor is obvious.

I will perform $n+1$ blocking steps, labeled as $k=0,1,\ldots,n$.
The first, $k=0$ step is special; it transforms the staggered field
from its usual one-component basis to a taste basis,
which is then retained in all subsequent blocking steps.
The $k=0$ step maintains the number of fermionic degrees of freedom.
It is described in detail in App.~\ref{tasteinv}.
In the subsequent blocking steps,
thinning out of all degrees of freedom (fermions and gauge field) occurs.
I have chosen to block (and thin out)
the gauge field in the special $k=0$ step as well,
essentially for no better reason than making the notation more tractable.

In every blocking step the lattice spacing is increased by a factor of two.
Thus, $a_k = 2^{k+1} a_f$ for  $k=0,\ldots,n$.  When speaking of the
coarse-lattice theory I will refer to the theory obtained
at the last, $k=n$ step.  Its lattice spacing will also be denoted
$a_c \equiv a_n$.  When I increase the number of blocking steps,
the coarse-lattice spacing $a_c$ will be held fixed in physical units,
and the fine-lattice spacing will decrease as $a_f = 2^{-n-1} a_c$.
The bare parameters on the fine lattice are adjusted
to maintain constant physics.  I will assume that the (fixed) length
of each dimension of the coarse lattice is finite.
Since, by assumption, all quarks are massive, no subtlety should arise
in taking the thermodynamical limit.

For $k=0,\ldots,n$, the blocked fermion and anti-fermion fields
on the $k^{\rm th}$ lattice will be denoted as $\j^{(k)}_{\a i}(\tx^{(k)})$
and $\bj^{(k)}_{\a i}(\tx^{(k)})$ respectively,
where $\tx^{(k)}$ are the coordinates on the $k^{\rm th}$ lattice.
The indices $\a$ and $i$, both ranging from one to four,
are the Dirac and the taste index respectively.
The blocked link variables will be denoted $V^{(k)}_{\m,\tx^{(k)}}$.
The $k^{\rm th}$-step blocked gauge field
as a whole is denoted $\cv^{(k)}=\{V^{(k)}_{\m,\tx^{(k)}}\}$.

Each blocking step is performed by multiplying the integrand of the
partition function of the previous step by one, written in a sophisticated
form \cite{bellw} (for reviews of the renormalization group,
see Refs.~\cite{rgrev,fp}).
Since integrations are over a compact space or else involve
Grassmann variables, the order of integrations can be chosen at will.
The result of the first, special blocking step, and the subsequent
$n$ ordinary blocking steps, is summarized by the following equation:
\begin{subequations}
\label{nlocal}
\begin{eqnarray}
  Z
  &=&
  \int \cd\cu \cd\c \cd\bc\;
   \exp[-S_g(\cu) -\bc \Dstag(\cu) \c]
\label{nlocala}
\\
  && \times
  \int \prod_{k=0}^{n} \Big[ \cd\cv^{(k)} \cd\j^{(k)} \cd\bj^{(k)} \Big]\;
  \exp\Big[-\ck\Big(\cu,\c,\bc,
  \{\cv^{(k)},\j^{(k)},\bj^{(k)}\} \Big)\Big]
\NON
  &=&
  \int \cd\cv^{(n)} \cd\j^{(n)} \cd\bj^{(n)}\;
  \exp\Big[-S_n\Big(\cv^{(n)},\j^{(n)},\bj^{(n)}\Big)\Big] \,,
\label{nlocalb}
\end{eqnarray}
\end{subequations}
where $S_n$ is the final coarse-lattice action.
In Eq.~(\ref{nlocala}), $\ck$ represents the sum of all the blocking kernels.
The notation $\{\cv^{(k)},\ldots\}$
signifies dependence on the listed fields for all $0 \le k \le n$.
Itemizing the blocking kernels,
\begin{equation}
\label{K}
  \ck = \sum_{k=0}^n \Big( \ck_{B}^{(k)} + \ck_{F}^{(k)} \Big) \,,
\end{equation}
in which the subscripts $B,F$ refer to bosons (\ie the gauge field)
and fermions respectively, we have
\begin{subequations}
\label{KB}
\begin{eqnarray}
  \ck_{B}^{(0)} &=& \cb_0\left(\cv^{(0)},\cu\right) + \cn_0(\cu) \,,
\label{KB0}
\\
  \ck_{B}^{(k)} &=& \cb_k\Big(\cv^{(k)},\cv^{(k-1)}\Big)
   + \cn_k\Big(\cv^{(k-1)}\Big) \,,
\label{KBk}
\end{eqnarray}
\end{subequations}
where
\begin{subequations}
\label{N}
\begin{eqnarray}
  \exp[+\cn_0(\cu)] &=&
  \int \cd\cv^{(0)} \exp\Big[-\cb_0\Big(\cv^{(0)},\cu\Big)\Big]\,,
\label{Na}
\\
  \exp\Big[+\cn_k\Big(\cv^{(k-1)}\Big)\Big] &=&
  \int \cd\cv^{(k)} \exp\Big[-\cb_k\Big(\cv^{(k)},\cv^{(k-1)}\Big)\Big]\,,
\label{Nb}
\end{eqnarray}
\end{subequations}
and
\begin{subequations}
\label{KF}
\begin{eqnarray}
  \ck_{F}^{(0)} &=& \a_0
  \Big(\bj^{(0)} - \bc\, Q^{(0)\dagger}(\cu)\Big)
     \Big(\j^{(0)} - Q^{(0)}(\cu) \c\Big)\,,
\label{KF0}
\\
  \ck_{F}^{(k)} &=& \a_k
  \Big[\bj^{(k)} - \bj^{(k-1)} Q^{(k)\dagger}\Big(\cv^{(k-1)}\Big)\Big]
     \Big[\j^{(k)} - Q^{(k)}\Big(\cv^{(k-1)}\Big) \j^{(k-1)}\Big]\,.
\label{KFk}
\end{eqnarray}
\end{subequations}
In Eqs.~(\ref{KBk}), (\ref{Nb}) and (\ref{KFk}), the $k$-range is $1\le k \le n$.
The fermion blocking kernel $Q^{(k)}(\cv^{(k-1)})$
depends on the gauge field on the $(k-1)^{\rm th}$ lattice only.
It is an ultra-local and gauge covariant
linear transformation that maps each $2^4$ hypercube of the
$(k-1)^{\rm th}$ lattice to a single site of the $k^{\rm th}$ lattice.
The blocking parameter $\a_k$ has mass-dimension one.
In this paper I will assume that $\a_k$ is chosen to be of order
$a_k^{-1} = 2^{n-k} a_c^{-1}$. Note that the gaussian integral
\begin{equation}
  \int \cd\j^{(k)} \cd\bj^{(k)} \exp\Big(-\ck_{F}^{(k)}\Big)\,,
\label{grass}
\end{equation}
yields a trivial constant; in Eq.~(\ref{nlocala}), this constant
was absorbed into the definition of the Grassmann measure.
For more details on the fermion blocking kernels,
see App.~\ref{tasteinv}. For the gauge-boson blocking kernels,
see App.~\ref{pullback}.

\subsection{\label{conv} A convenient representation}
Keeping track of the taste-symmetry violations is evidently important.
For this purpose,
the action $S_n$ that results from the blocking transformations
is not very useful because it contains multi-fermion interactions.
In the fourth-root theory, that action would furthermore be non-local,
and completely intractable.
The problem can be circumvented by noting that
the fermion blocking transformations are gaussian,
as can be seen from Eqs.~(\ref{nlocala}) and (\ref{KF}).
Let us integrate out the original fermion variables as well as all the blocked
fermion variables, except for those that live on the coarsest lattice.
But let us not integrate out the original gauge field nor any
of the blocked gauge fields.  All the integrations we explicitly do are,
thus, gaussian.  The result is
\begin{equation}
  Z
  =
  \int \cd\cu
  \prod_{k=0}^{n} \Big[ \cd\cv^{(k)} \Big]\;
  \bltz_n\Big(1;\cu,\{\cv^{(k)}\}\Big)
  \int \cd\j^{(n)} \cd\bj^{(n)}
  \exp\Big(-\bj^{(n)} D_n\, \j^{(n)}\Big)\,,
\label{gauss}
\end{equation}
where
\begin{eqnarray}
  \bltz_n\Big(\g\,;\cu,\{\cv^{(k)}\}\Big)
  &=&
  \exp\Big(-S_g - \sum_{k=0}^n \ck_{B}^{(k)}\Big)
\label{bltz}
\\
  && \times \;
  \det^\g\Big[\Big(\a_0 G_0\Big)^{-1}\Big]\,
  \prod_{k=1}^{n}
  \det^\g\Big[\Big(\a_k^{1/16}\, G_k\Big)^{-1}\Big] \,.
\nonumber
\end{eqnarray}
In the ordinary staggered theory one has $\g=1$.
In the fourth-root theory we will have $\g=\quart$.
Here
\begin{subequations}
\label{Dn}
\begin{eqnarray}
  D_0^{-1} &=& \a_0^{-1} + Q^{(0)} \Dstag^{-1} Q^{(0)\dagger} \,,
\label{Dna}
\\
  D_k^{-1} &=& \a_k^{-1} + Q^{(k)} D_{k-1}^{-1} Q^{(k)\dagger} \,,
  \qquad k=1,\ldots,n\,,
\label{Dnb}
\end{eqnarray}
\end{subequations}
\vspace{-4ex} 
\begin{subequations}
\label{Gn}
\begin{eqnarray}
  G_0^{-1} &=& \Dstag + \a_0 Q^{(0)\dagger} Q^{(0)} \,,
\label{Gna}
\\
  G_k^{-1} &=& D_{k-1} + \a_k Q^{(k)\dagger} Q^{(k)} \,.
  \qquad k=1,\ldots,n.
\label{Gnb}
\end{eqnarray}
\end{subequations}
Equation (\ref{bltz}) includes the constant resulting from
the Grassmann integration (\ref{grass}), for each $k$.
The different powers of $\a_k$ arise because the $(k+1)^{\rm th}$ lattice
has sixteen times fewer fermionic degrees of freedom compared to
the  $k^{\rm th}$ lattice,
except for the $k=0$ step which does not reduce the number
of fermionic degrees of freedom.

For fixed values of the original as well as all the blocked gauge fields,
we have the factorization formula \cite{bellw}
\begin{equation}
  \det(\Dstag)
  = \det(D_n)\,
  \det\Big[\Big(\a_0 G_0\Big)^{-1}\Big]\;
  \prod_{k=1}^{n}
  \det\Big[\Big(\a_k^{1/16}\, G_k\Big)^{-1}\Big] \,.
\label{fctrn}
\end{equation}
Equation (\ref{fctrn}) shows how the fermionic short-distance fluctuations
are gradually removed from the theory.
Each factor of $\det(G_k^{-1})$ results from integrating out
fermionic degrees of freedom during the $k$-th blocking step,
and generates an effective action for the collection of
gauge fields $\cu,\cv^{(0)},\cdots,\cv^{(k-1)}$,
\begin{subequations}
\label{SeffG}
\begin{eqnarray}
  \Seff^0 &=&  \log \det(\a_0\, G_0)\,,
\label{SeffGa}
\\
  \Seff^k &=&  \log \det\Big(\a_k^{1/16}\, G_k\Big)\,,
  \qquad k=1,\ldots,n\,.
\label{SeffGb}
\end{eqnarray}
\end{subequations}

All the long-distance physics is contained in the RG-blocked
Dirac operator $D_n$. That $D_n$ faithfully reproduces
the long-distance physics can be seen as follows.
By successive applications of Eq.~(\ref{Dn}),
the blocked-field propagator $D_n^{-1}$ can be expressed in terms of
the original propagator $\Dstag^{-1}$ between a special type of smeared sources
built up from the product of the fermion blocking kernels (cf.\ Eq.~(\ref{DQn})).
As I explain below, this is but a special case of a more general mechanism.

For $m>0$, one can show that $\det(D_k)$ is  strictly positive
and $\det(G_k^{-1})$ is positive. See App.~\ref{tst3} for the proof.

\subsection{\label{pullb} Pull-back mapping}
The RG-blocking transformations start off at the cutoff scale and proceed
gradually towards lower energy scales.
But the blocking transformations facilitate
another operation that works in the reverse direction.
Suppose that we want to calculate the expectation value of an operator
$\co^{(n)}=\co^{(n)}(\cv^{(n)},\j^{(n)},\bj^{(n)})$
defined explicitly in terms of the fields of the $n^{\rm th}$ lattice.
Using Eq.~(\ref{nlocal}), the expectation value may be calculated as follows.
We begin by integrating over the $n^{\rm th}$-lattice fields,
then over the $(n-1)^{\rm th}$-lattice fields, and so on.
If, however, we stop at any intermediate step $k$,
the result of the integrations we have done so far will be expressible
solely in terms of the fields of the $k^{\rm th}$ lattice.
This procedure defines a
{\it pull-back} mapping of any operator from the $n^{\rm th}$ to the
$k^{\rm th}$ lattice, that by construction preserves expectation values.

Explicitly, for any $-1 \le j \le n-1$,
the pull-back mapping $\ct^{(j,n)}: \co^{(n)} \to \co^{(j)}$
is defined by
\begin{equation}
  \ct^{(j,n)} \co^{(n)}
  =
  \int \prod_{k=j+1}^{n} \Big[ \cd\cv^{(k)} \cd\j^{(k)} \cd\bj^{(k)} \Big]\;
  \exp\Big[-\sum_{k=j+1}^{n} \Big( \ck_{B}^{(k)} + \ck_{F}^{(k)} \Big) \Big]\;
  \co^{(n)}\,.
\label{pull}
\end{equation}
As promised, by construction the pull-back mapping preserves
the value of observables,
\begin{equation}
  \svev{\ct^{(j,n)} \co^{(n)}}_j
  =
  \svev{\co^{(n)}}_n\,.
\label{observe}
\end{equation}
Here the expectation value $\svev{\cdots}_n$
is defined by the representation
of the partition function in Eq.~(\ref{nlocala}).
Taken together, these equations merely say that we may perform
the integrations in Eq.~(\ref{nlocal}) by first integrating over the blocked
fields labeled by $j+1\le k \le n$, and then integrating over
the remaining blocked fields as well as over the original fields.
The value $j=-1$ accounts for the original fine-lattice theory,
and $\ct^{(-1,n)}$ is the pull-back  from the last-step coarse lattice
all the way to the original staggered theory on the fine lattice.

The pull-back mapping is ultra-local if and only if the blocking kernels are.
An operator supported on a compact subset of the $n^{\rm th}$ lattice
is mapped by $\ct^{(j,n)}$ to an operator supported on a corresponding,
only somewhat bigger, subset of the $j^{\rm th}$ lattice.

An immediate corollary is that the coarse-lattice observables
form a proper subset of the fine-lattice observables.
The coarse-lattice expectation value of the operator $\co^{(n)}$
is equal to the fine-lattice expectation value of
the operator $\ct^{(-1,n)} \co^{(n)}$.
As alluded to earlier, the reconstruction of the blocked fermion
propagator $D_n^{-1}$ from its predecessors is in fact an example
of the pull-back mapping in action.

This innocuous corollary leads to another, all important, result.
Every coarse-lattice observable,
being simultaneously a fine-lattice observable via the pull-back mapping,
is constrained by all the fine-lattice symmetries.
In this sense, the (physical) consequences of the exact lattice
symmetries cannot ``be lost'' by the blocking process.

More can be said on the role of specific
lattice symmetries within the blocking framework.
The interested reader is referred to App.~\ref{pullback}
for a more detailed discussion of the pull-back mapping.
Effects of the blocking transformations on the fine-lattice symmetries,
including the relevance of the pull-back mapping,
are discussed in App.~\ref{sym} and App.~\ref{disorder}.

\subsection{\label{free} Lessons from the free theory}
In this subsection I
review the main results of RG-blocking in the free theory \cite{rg}.
With no gauge fields, Eq.~(\ref{fctrn}) takes a somewhat special form.
By a unitary change of variables one can switch back and forth between
the Dirac operators $D_{stag}$, in the one-component basis, and
$D_{taste}$, in the taste basis \cite{taste,saclay},
and\footnote{\label{Gnotation}
  Equation (\ref{fctrn}) reduces to Eq.~(\ref{fctrFree})
  in the limit $\a_0 \to\infty$,
  where $\det((\a_0 G_0)^{-1}) \to 1$ and $D_0 \to D_{taste}$ \cite{BGS}.
  What I denote as $G_k$ in this paper was denoted $\G_k$ in Ref.~\cite{rg},
  see Eq.~(6) therein. The notation $G_k$ has a related, but different meaning
  in Ref.~\cite{rg}.  Also, for compatibility with Ref.~\cite{cmp},
  in Ref.~\cite{rg} the blocking parameters $\a_k$ were all chosen
  to have a fixed, $O(1)$ value in units of the coarse-lattice spacing,
  whereas here I make the more natural assumption $\a_k=O(a_k^{-1})$.
}
\begin{equation}
  \det(D_{stag})
  = \det(D_{taste})
  = \det(D_n)\, \prod_{k=1}^{n}
  \det\Big[\Big(\a_k^{1/16}\, G_k\Big)^{-1}\Big] \,.
\label{fctrFree}
\end{equation}

An explicit expression for the free RG-blocked Dirac operator $D_n$
may be written down.
Its taste-violating part $\D_n$ (see Eq.~(\ref{Dinv}) below)
has a norm bounded by\footnote{
  This bound is rigorous in the free theory \cite{rg,cmp}.
}
\begin{equation}
  \| a_c\, \D_n \| = O(2^{-n}) = O(a_0/a_c)\,.
\label{boundD}
\end{equation}
In the limit $n\to\infty$, all the taste-violating terms
go to zero, and
\begin{subequations}
\label{Drg}
\begin{eqnarray}
  \lim_{n\to\infty} D_n &=& D_{rg}\otimes {\bf 1}  \,,
\label{Drga}
\\
  \lim_{n\to\infty}
  {\det(D_{taste}) \over \prod_{k=1}^{n} \det(\a_k^{-1/16} G_k^{-1})}
  &=&  \rule{0ex}{4ex} \det^4(D_{rg})
  \,. \qquad
\label{Drgb}
\end{eqnarray}
\end{subequations}
Again, ${\bf 1}$ is the identity matrix in taste space.
The ``one-taste'' operator $D_{rg}$ is local,
and $\det(D_{rg})$ qualifies as a fourth root
of $\det(D_{taste})=\det(D_{stag})$ in the sense of Eq.~(\ref{Drgb}).

By repeatedly integrating out the short-distance fluctuations
we thus obtain a coarse-lattice operator with an exact four-fold degeneracy
in the limit $n\to\infty$.  The power-law scaling
of the taste-breaking terms is clearly as dictated by their origin:
irrelevant operators with mass-dimension equal to five.
  Intuitively this can be understood as follows.
  The dimension-five taste-violating terms in $D_{taste}$
  are multiplied by an explicit factor of $a_0$,
  the initial taste-basis lattice spacing.
  But the momentum flowing through the fermion line
  is in effect of order $|p|\sim a_c^{-1}$ at most.
  The relative size of the taste-violating terms is therefore
  at most of order $a_0/a_c$.

As a simple corollary of the rigorous work of Ref.~\cite{cmp}, one can prove
that $D_k$, $G_k^{-1}$, and its inverse $G_k$, are all local,
bounded operators. Mathematical rigor set aside,
one can understand how $G_k^{-1}$
develops an $O(\a_k)$ gap directly from Eq.~(\ref{Gn}).
Since the fermion mass is very small in any lattice units,
it can be ignored for this purpose.  Also, from now on,
I assume that $\a_0$ in Eq.~(\ref{Dna}) has a finite, $O(a_0^{-1})$ value.
Hence Eq.~(\ref{fctrn}) (and not Eq.~(\ref{fctrFree})) must be used also in the free
theory.  For further details on the $k=0$ step see Ref.~\cite{BGS}
and App.~\ref{tasteinv}.

For $k\ge 1$, the massless Dirac operator $D_{k-1}$ satisfies a
Ginsparg-Wilson (GW) relation \cite{gw}, in which $[\g_5\otimes \x_5]$
takes the role usually played by $\g_5$.
Here $\x_5$ is the representation of $\g_5$ that acts
on the taste index (see App.~\ref{tasteinv}).\footnote{
  In the massless limit $D_{\rm rg}$ (Eq.~(\ref{Drg}))
  satisfies the usual Ginsparg-Wilson relation.
}
The eigenvalues of $D_{k-1}$
thus lie on a circle in the right half of the complex plane,
with the imaginary axis tangent to the eigenvalue circle
on the left \cite{rg,BGS}.
In order to obtain $G_k^{-1}$ from $D_{k-1}$, we add the blocking-kernel
term $\a_k Q^{(k)\dagger} Q^{(k)}$.
This new term is positive semi-definite. It affects mostly the small-momentum
modes located near the origin of the Brillouin zone,
and pushes their eigenvalues to the right
by an amount proportional to $\a_k$.  The result is that no eigenvalue
remains in an $O(\a_k)$ neighborhood of the origin.  In other words,
$G_k^{-1}$ has developed a gap of order $\a_k$.\footnote{
  When the eigenvalues $\l_i$ are complex we may define the gap
  as $\min|\l_i|$.
}
Its inverse $G_k$ will thus have a decay
rate of order $\a_k$ as well.
Furthermore, since one may also obtain the blocked
Dirac operator as
\begin{equation}
  D_k = \a_k -\a_k^2\, Q^{(k)} G_k Q^{(k)\dagger} \,,
\label{DG}
\end{equation}
it follows that the decay rates of the kernels of $D_k$ and of $G_{k+1}^{-1}$
should be $O(\a_k)$ too.  The argument may now be repeated for the
$(k+1)^{\rm th}$ step.

\subsection{\label{plan} Overview of the argument}
When no roots are taken,
the local lattice theory with $N_f$ staggered fields belongs
to the \textit{universality class} of QCD with $4N_f$ quarks.
The usual universality classification is, however,
inapplicable to the fourth-root theory, because of its non-locality.
There is evidence
from staggered chiral perturbation theory
that the outcome of taking the fourth root may be
described in terms of an extended Hilbert space
containing unphysical states with non-zero taste charges:
the contributions of these taste-charged states as
intermediate states are thus also unphysical.
A unitary subspace with the correct number of quarks,
one per staggered field, will exist only in the continuum limit,
and only provided that exact taste symmetry is recovered
\cite{BGS,BGSS,CB4f,SP}.

The range of the non-locality present
in the fourth-root theory appears to be set by infra-red scales of the theory:
the masses of the various staggered pions  \cite{BGS,CMY}.
The taste-breaking terms driving the non-locality
are lattice artifacts, and
would naively be expected to vanish in the continuum limit.
If indeed exact taste symmetry is recovered
in the continuum limit of the fourth-root theory, then,
when the lattice spacing has become small enough,
it is logically necessary that local lattice theories in the desired
universality class exist which provide a good approximation
of the fourth-root theory.  Such local theories could be constructed by
simply discarding the taste-violating terms from the blocked
fourth-root theory. This is the idea behind the introduction
of \textit{reweighted} theories.

Consider first the ordinary, local staggered theory.
With the help of Eqs.~(\ref{gauss}) and (\ref{bltz}),
the original staggered partition function
in Eq.~(\ref{Zstag}) can be re-expressed as
\begin{equation}
  Z = Z_n
  \equiv
  \int \cd\cu
  \prod_{k=0}^{n} \Big[ \cd\cv^{(k)} \Big]\;
  \bltz_n\Big(1;\cu,\{\cv^{(k)}\}\Big)\,
  \det(D_n) \,.
\label{Zn}
\end{equation}
In short, this expression results from $n+1$ blocking steps,
after which the fermion fields have been integrated out altogether,
while retaining explicitly the integral over the original,
fine-lattice gauge field $\cu$, as well as over the
gauge fields $\cv^{(k)}$ of all the blocking steps.  Recall that
the special $k=0$ step facilitates the transition
from the usual one-component staggered basis to a taste basis.

In order to keep track of taste-symmetry violations, let us
split the blocked Dirac operator $D_n$
into its taste-invariant and non-invariant parts:
\begin{subequations}
\label{Dinv}
\begin{eqnarray}
  D_n &=& D_{inv,n} + \D_n \,,
\label{Dinva}
\\
  D_{inv,n} &=& \tD_{inv,n} \otimes {\bf 1}\,,
\label{Dinvb}
\\
  \tD_{inv,n} &=& {1\over 4}\; \ts(D_{n}) \,,
\label{Dinvc}
\end{eqnarray}
\end{subequations}
where $\ts$ denotes tracing over the taste index only.
The taste non-invariant part $\D_n$ is traceless on the taste index.

By construction, $D_n$ accounts for physics over distances on
the order of the coarse-lattice spacing $a_c$ or longer.
In particular, its taste violating part $\D_n$ accounts
for all taste-symmetry violations at the energy scale $a_c^{-1}$ and below.
The question is how big are the taste-symmetry violating effects
in the spectrum of $D_n$.

Originally, the staggered Dirac operator $D_{stag}$
exhibits taste-symmetry violations
on all scales. In the low-lying spectrum they are small \cite{evs};
but they grow gradually with the energy scale,
until they become $O(1)$ at the scale of the original
lattice cutoff $a_f^{-1}=2^{n+1}a_c^{-1}$.
I will argue that, nevertheless,  by choosing
$n$ large enough the taste-violating effects
in the spectrum of $D_n$ can be made arbitrarily small.
RG blocking removes the ultra-violet fermionic modes.
Their remnant is, {\it mutatis mutandis}, the effective action $\Seff^k$
(Eq.~(\ref{SeffG})).
This effective action is a sum of (products of) Wilson loops
and, I will claim, it is local on both the ordinary and
fourth-root staggered ensembles.  Being a local functional of the
(original and blocked) gauge fields,
but not a functional of the fermion fields,
it cannot give rise to any taste-symmetry violations at large distances.
In this sense, the ultra-violet taste violations have been eliminated.
I will further claim that, with every additional blocking step,
the remaining taste-violations in the entire eigenvalue spectrum of
the blocked Dirac operator get smaller, uniformly,
basically because this spectrum consists of only ``low-energy'' modes
with respect to the fine-lattice scale.
In the limit of infinitely many blocking steps,
taste symmetry is fully recovered.

We may discard all the taste-breaking effects from the staggered theory,
by hand, after only a finite number of blocking steps.
Truncating the blocked Dirac operator in Eq.~(\ref{Zn}) to its taste-invariant
part $D_{inv,n}$ gives rise to the following reweighted theory
\begin{subequations}
\label{Zninv}
\begin{equation}
  Z_{inv,n} =
  \int \cd\cu
  \prod_{k=0}^{n} \Big[ \cd\cv^{(k)} \Big]\;
  \bltz_n\Big(1;\cu,\{\cv^{(k)}\}\Big)\,
  \det(D_{inv,n})\,.
\label{Zninva}
\end{equation}
A more conventional looking path-integral representation may be
obtained by rewriting $\det(D_{inv,n})$ as a path integral over
(four-taste) coarse-lattice fermion fields $\j^{(n)},\bj^{(n)}$,
as in Eq.~(\ref{gauss}), and then integrating out the ``tower'' of
gauge fields except for the coarse-lattice gauge field $\cv^{(n)}$.
This gives
\begin{equation}
  Z_{inv,n} =
  \int \cd\cv^{(n)} \cd\j^{(n)} \cd\bj^{(n)}\;
  \exp\Big[-S_{inv,n}\Big(\cv^{(n)},\j^{(n)},\bj^{(n)}\Big) \Big] \,,
\label{Zninvb}
\end{equation}
\end{subequations}
which is to be compared with the path integral representation (\ref{nlocalb})
of the blocked staggered theory.  Unlike the staggered theory,
the reweighted theory has no shift invariance \cite{SvdD,mgjs}.
Instead, it has exact taste-$U(4)$ invariance by construction.
Another difference is that the above constructed reweighted theory
does not have an exact chiral symmetry in the massless limit.\footnote{
  More sophisticated reweighted theories may be constructed.
  See Ref.~\cite{BGS} for a construction that maintains
  the exact chiral symmetry of the $m\to 0$ limit.
\label{invov}
}

Reweighting at blocking level $n$ generates a sequence of theories
$Z_{inv,n}$ which are different from each other, as well as from the
staggered theory.  But I will argue in Sec.~\ref{unroot} that,
because $\D_n$ is an irrelevant operator,
the (sequence of) reweighted theories
has the same continuum limit as the (blocked) staggered theory.
Each reweighted theory enjoys exact taste symmetry by
construction, and this implies (the uncontroversial result) that exact
taste symmetry is recovered in the continuum limit
of the ordinary staggered theory.
The proof works by establishing the existence of a convergent expansion
relating the staggered and reweighted theories when $n$ is large enough.
One must require that all the quark masses be non-zero,
consistent with the fact that the chiral and continuum limits
do not always commute \cite{SV,DH,CBma,BGSS}.

Moving on to the fourth-root theory, its partition function
cannot be represented as an ordinary path integral
with a local fermion action.  Rather, it is given by \cite{trick}
\begin{equation}
  \Zroot = \int \cd\cu \exp(-S_g)\; {\rm det}^{1/4}(\Dstag)\,,
\label{Z4}
\end{equation}
where the positive fourth root is taken.
As in the ordinary staggered theory,
this may be re-expressed in an $n$-step RG-blocked form as
\begin{equation}
  \Zroot =
  \Zroot_n \equiv
  \int \cd\cu
  \prod_{k=0}^{n} \Big[ \cd\cv^{(k)} \Big]\;
  \bltz_n\Big(\quart;\cu,\{\cv^{(k)}\}\Big)\,
  \det^{1/4}(D_n) \,.
\label{Z4n}
\end{equation}
Again let us remove the taste-breaking terms by hand,
which gives rise to a new family of reweighted theories
\begin{equation}
  \Zroot_{inv,n}
  =
  \int \cd\cu
  \prod_{k=0}^{n} \Big[ \cd\cv^{(k)} \Big]\;
  \bltz_n\Big(\quart;\cu,\{\cv^{(k)}\}\Big)\,
  \det(\tD_{inv,n}) \,.
\label{Z4ninv}
\end{equation}
Here I have used the exact taste symmetry of
$D_{inv,n} =  \tD_{inv,n} \otimes {\bf 1}$ to take
the analytic fourth root of its determinant.\footnote{
  For large enough $n$, $\det(D_{inv,n})$ and $\det(\tD_{inv,n})$
  are positive, see Sec.~\ref{rooted}.
}

One can represent $\det(\tD_{inv,n})$ as a fermion path integral.
This suggests that the validity of the continuum limit
of the fourth-root theory could be established by, once again,
showing that the sequence of reweighted theories has the
same continuum limit as the blocked fourth-root theory.
But we must now face two hurdles that were not encountered
in the local, ordinary staggered theory.

The new hurdles are addressed in Sec.~\ref{uv}.
First, we must show that the reweighted theories derived from
the fourth-root theory are local.  This is done in Sec.~\ref{lcl} by showing
that, on the basis of plausible assumptions,
the effective action $\Seff^k$ and the blocked Dirac operator $D_k$
are local on the $k^{\rm th}$ lattice scale,
on both the ordinary and the fourth-root staggered ensembles.
Because $\tD_{inv,n}$ is defined by
a trace projection, $\tD_{inv,n}$ and $\D_n$ are then separately local.

Introducing coarse-lattice Dirac fields $q^{(n)},\bq^{(n)}$
that, this time, carry no taste index,
and once again integrating out the ``tower'' of
gauge fields except for the coarse-lattice gauge field $\cv^{(n)}$
the reweighted fourth-root partition function then takes the form
(compare Eq.~(\ref{Zninvb})),
\begin{subequations}
\label{Z4q}
\begin{eqnarray}
  \Zroot_{inv,n}
  &=&
  \int \cd\cu
  \prod_{k=0}^{n} \Big[ \cd\cv^{(k)} \Big]\;
  \bltz_n\Big(\quart;\cu,\{\cv^{(k)}\}\Big)
\label{Z4qa}
\\
  && \hspace{10ex} \times
  \int \cd q^{(n)} \cd\bq^{(n)}\;
  \exp\Big(-\bq^{(n)} \tD_{inv,n}\, q^{(n)}\Big)
\NON
  &=&  \rule{0ex}{4ex}
  \int \cd\cv^{(n)} \cd q^{(n)} \cd\bq^{(n)}\;
  \exp\Big[-S^{root}_{inv,n}\Big(\cv^{(n)},q^{(n)},\bq^{(n)}\Big) \Big]
  \,. \qquad
\label{Z4qb}
\end{eqnarray}
\end{subequations}
The ``one-taste'' action
$S^{root}_{inv,n}=S^{root}_{inv,n}(\cv^{(n)},q^{(n)},\bq^{(n)})$
is complicated, and contains many multi-fermion interactions,
just like $S_n$ and $S_{inv,n}$ encountered earlier.
What matters, however, is that $S^{root}_{inv,n}$ too is local
on the coarse-lattice scale if, in particular, $\tD_{inv,n}$ is;
the ``surgery'' of removing the taste violations
has also removed the non-localities of the blocked fourth-root theory!

I stress that the argument for locality of $\Seff^k$ and $D_k$
does \textit{not} require that the underlying theory be local.
This non-perturbative argument is very general, and only makes mild use of the
\textit{renormalizability} of all theories including
the fourth-root theory \cite{PQ,Steve} to establish the existence
of a weak-coupling regime.

At this point we expect
that the reweighted theories $\Zroot_{inv,n}$
derived from the fourth-root theory are local,
and belong to the desired universality class.  Now comes the second hurdle.
Convergence of the reweighted and the staggered theories
to the same continuum limit depends on the scaling of the taste-breaking
effects.  But the fourth-root theory does not have a local action in
the first place, so how are we to perform any scaling analysis?

The scaling of taste-breaking effects in gauge-invariant observables
can, of course, be studied numerically, and what one finds is in
agreement with what one would naively expect \cite{milc,SP,FM}.
But, while no undesirable effects have been encountered
at presently accessible values of the lattice spacing and the quark masses,
numerical results alone cannot alleviate the concern that
closer to the continuum and/or the chiral limit
the scaling of taste-breaking effects might eventually be altered
in undesirable ways due to the lack of a local lattice action.

On the lattice, a scaling analysis rests on two pillars.
On the non-perturbative side,
we define a theory, construct local operators within it, and set up an
RG transformation. On the perturbative side, if the theory
is power-counting renormalizable, we
can compute the scaling of any local operator.  An important special case
is to take the local operator to be the action itself, or individual terms
within it.

In this paper I show how to generalize the scaling analysis
to the fourth-root theory.
First, the fourth-root theory \textit{is} renormalizable.
The multi-gauge-field representation of the blocked fourth-root theory,
Eq.~(\ref{Z4n}), then allows us to bypass the lack of a local fermion action.
Instead, we may study the scaling of the blocked Dirac operator $D_k$
and the effective action $\Seff^k$.
Having first shown by non-perturbative considerations
that both of them are local operators,
their scaling can be computed by setting up
the appropriate perturbative expansion, which is done in Sec.~\ref{rewpt}.
I find that any local operator scales in the same way
in the staggered theory and in the corresponding reweighted theories
(with or without the fourth root).  In particular, the taste-breaking
part of the blocked Dirac operator, $\D_n$, indeed scales as an
irrelevant operator.

Finally, in Sec.~\ref{rooted} I reconstruct the rooted theory
from the corresponding reweighted theories, and establish the validity
of its continuum limit.

My conclusion rests on renormalizability of the fourth-root theory,
concerning which there is little doubt, and the fact
that renormalizability is ``inherited'' by the reweighted theories
(Sec.~\ref{pcr}).
My conclusion also rests on two additional key features that have to do
with locality (Sec.~\ref{lcl}), and scaling (Sec.~\ref{rewpt}).
I give plausible arguments for each of them,
but confirmation must await more detailed future investigations.
Where, then, do we stand today?
One can draw an indirect but important lesson
from the ordinary staggered theory.
In that case we (believe we) know what is the continuum limit.
Moreover, the reweighted theories are tightly constrained by
the convergent expansion relating them to the -- local -- staggered theory.
This leaves little doubt that all the key properties are valid in this case.
But the argumentation of Sec.~\ref{uv} makes very little
discrimination between the ordinary and the fourth-root cases.
As I explain in more detail later,
this increases confidence that nothing essential has been overlooked,
and that the claimed properties are valid in the fourth-root case as well.

\section{\label{unroot} Continuum limit of the ordinary staggered theory}
In this section I discuss the continuum limit of the ordinary
staggered-fermion theory.
Continuum-limit observables that can be computed within the coarse-lattice
theory are introduced in Sec.~\ref{CL}.  In Sec.~\ref{scaling}, I list scaling
relations that follow from a standard RG analysis in the
ordinary staggered theory.  These scaling
relations imply the recovery of taste symmetry in the continuum limit.
This is inferred in Sec.~\ref{reweigh} by comparing the blocked
staggered theory to the reweighted theory at each blocking level $n$.
Provided that the renormalized quark mass is non-zero,
I show that the two theories are connected by a convergent expansion
when $n$ is large enough, and that any difference between them vanishes in
the limit $n\to\infty$.

\subsection{\label{CL} Continuum-limit observables}
The continuum limit corresponds to the limit of infinitely many
blocking steps, which is taken while holding fixed
the coarse-lattice spacing $a_c$ in physical units.
The fine-lattice spacing $a_f$ of the staggered theory
goes to zero, $a_f\L = 2^{-n-1}a_c\L\to 0$,
where $\L$ is the QCD scale.
Constant physics is maintained by adjusting the bare parameters
such that the renormalized gauge coupling $g_r(a_c)$ and quark mass $m_r(a_c)$
are kept fixed.  The coarse-lattice spacing plays the role
of the renormalization scale.
The fermion's wave-function renormalization is controlled by
the parameter $z^{(k)}$ in Eq.~(\ref{covQ}), that fixes the overall normalization of
the gauge-covariant blocking kernel $Q^{(k)}$ in the interacting theory.
I will assume that the $z^{(k)}$'s have been
adjusted so as to impose a wave-function
renormalization condition on the blocked fermion fields
at the renormalization scale $a_c$.

For simplicity I will restrict the discussion to meson observables.
Sources for mesons are added
into the fermion action on the $n^{\rm th}$ lattice as follows:\footnote{
  For general sources, see appendix B of Ref.~\cite{Steve}.
}
\begin{equation}
  S^{(n)}_{source}(J)
  = \bj^{(n)} J\cdot\cs^{(n)}\, \j^{(n)} \,,
\label{SFJ}
\end{equation}
where
\begin{eqnarray}
  \bj^{(n)} J\cdot\cs^{(n)}\, \j^{(n)}
  &\equiv& \sum_{\tx^{(n)}} \sum_{i} J_i(\tx^{(n)})\,
  O^{(n)}_i(\tx^{(n)}) \,,
\label{JM}
\\
  \co^{(n)}_i(\tx^{(n)})
  &=& \sum_{\ty^{(n)},\tz^{(n)}} \bj^{(n)}(\ty^{(n)})\,
  \cs^{(n)}_i(\tx^{(n)},\ty^{(n)},\tz^{(n)};\cv^{(n)})\,
  \j^{(n)}(\tz^{(n)}) \,. \qquad
\label{Oi}
\end{eqnarray}
The kernels $\cs^{(n)}_i$ are gauge-covariant and ultra-local.\footnote{
  See App.~\ref{tst2} for the renormalization of composite
  coarse-lattice operators.
}
Augmenting the fermion action in Eq.~(\ref{gauss}) by the source term (\ref{JM})
and performing the Grassmann integration we obtain the partition function
with sources (compare Eq.~(\ref{Zn})):
\begin{equation}
  Z_n(a_c;J) =
  \int \cd\cu
  \prod_{k=0}^{n} \Big[ \cd\cv^{(k)} \Big]\;
  \bltz_n\Big(1;\cu,\{\cv^{(k)}\}\Big)\,
  \det\Big(D_n+J\cdot\cs^{(n)}\Big) \,,
\label{ZJ}
\end{equation}
as well as the normalized version
\begin{equation}
  \cz_n(a_c;J) = Z_n(a_c;J)/Z_n(a_c;0) \,.
\label{ZJnorm}
\end{equation}
Meson correlation functions, renormalized at the scale $a_c$,
are generated by functional differentiation of $\cz_n(a_c;J)$.
The (assumed) existence of the continuum limit
means that the $n\to\infty$ limit of the normalized generating functional
$\cz_n(a_c;J)$ is smooth:
\begin{equation}
  \cz_\infty(a_c;J) = \lim_{n\to\infty} \cz_n(a_c;J) \,.
\label{Zlim}
\end{equation}
By differentiation of $\cz_\infty(a_c;J)$ one generates
continuum-limit meson correlators in euclidean space.

Before moving on let me comment on the set of
coarse-lattice observables.  In Eq.~(\ref{observe}),
which states the equality of observables under the pull-back mapping,
let us choose $j=n-1$.
Instead of using the number of blocking steps as a label,
we may use the corresponding lattice spacing for this purpose.
The equation then takes the form
\begin{equation}
  \svev{\ct(a_c/2,a_c)\, \co(a_c)}_{a_c/2}
  =
  \svev{\co(a_c)}_{a_c}\,,
\label{acobserve}
\end{equation}
in self-explanatory notation.
Equation (\ref{acobserve}) remains valid in the continuum limit,
where the pull-back mapping $\ct(a_c/2,a_c)$ remains well defined.
This equation identifies the observables
of the coarse-lattice theory $\cz_\infty(a_c;J)$ with a proper subset
of the observables of $\cz_\infty(a_c/2;J)$,
the coarse-lattice theory with half the lattice spacing.
Because observables are lost in the blocking, we need to
take $a_c \Lambda$ to be small enough in the first place to ensure that
the observables derived from $\cz_\infty(a_c;J)$
are rich enough to extract all the QCD physics.
Additional reasons for choosing $a_c \Lambda$ small will be encountered
in Sec.~\ref{uv}.

\subsection{\label{scaling} Scaling of irrelevant operators}
I now list the scaling laws needed to establish
the recovery of exact taste symmetry in the continuum limit of the ordinary
staggered theory, as they apply within the RG-blocking
framework of this paper.

First, the very existence of the continuum limit in QCD derives from
asymptotic freedom,\footnote{
  The argument may be extended to other theories such as lattice QED
  with staggered fermions, if a finite but ``beyond the Planck scale''
  lattice cutoff is acceptable.
\label{QED}
}
or, in other words, from the scaling properties
of the running coupling $g_r(a_c)$ as a function of the bare coupling.

The remaining scaling laws pertain to the fermion sector.
For the restoration of taste symmetry in all observables,
two scaling laws will be necessary:
\begin{eqnarray}
  \left\| D_n^{-1} \right\| & \leqx & {1\over m_r(a_c)} \,,
\label{mscale}
\\
  \left\| \D_n \right\|  & \leqx &  {a_f\over a_c^2}
  = {2^{-(n+1)}\over a_c}\,.
\label{irrscale}
\end{eqnarray}
These scaling laws are assumed to hold in an \textit{ensemble average}
sense;  they do not hold on all configurations,
but they (are assumed to) hold after averaging over the configurations
in the ensemble. Each configuration is generated as described
in App.~\ref{ensemble}: one complete configuration consists of
a ``mother'' configuration $\cu_i$ of the fine-lattice gauge field,
as well as of ``daughter'' configurations
$\cv^{(0)}_i,\cv^{(1)}_i,\ldots,\cv^{(n)}_i$ of blocked gauge fields.

The bound (\ref{mscale}) relates the lowest eigenvalues of the
blocked Dirac operator $D_n$ to the running quark mass $m_r(a_c)$.
This inequality says that the value of $m_r(a_c)$
measured on an ensemble of configurations is
set by the lowest eigenvalues of $D_n$, provided that
the wave-function renormalization condition was imposed at the
coarse-lattice scale.
The bound (\ref{mscale}) is needed to tame the infra-red
behavior in the fermion sector.
We will see in the next subsection how it enters.
For some further comments on the bound (\ref{mscale})
and its parallel in the fourth-root theory,
see App.~\ref{mbound}.

The crucial scaling law (\ref{irrscale}) determines the scaling of
$\D_n$, the taste-breaking part of the blocked Dirac operator $D_n$
(see Eq.~(\ref{Dinv})).
In the taste representation,
the leading taste violations arise from dimension-five irrelevant
operators \cite{taste,saclay,rg}, and the
right-hand side of Eq.~(\ref{irrscale}) gives the anticipated scaling
based on this engineering dimension
(compare Eq.~(\ref{boundD}) for the free theory).
The ``$\leqx$'' sign means that in the interacting theory
the inequality holds up to logarithmic corrections, that is,
powers of $\log(a_c/a_f)$ or, equivalently, powers of $n$.

In a standard RG application the fermions and the gauge field
are both blocked at each step,
and the scaling laws apply to (the parameters of) the blocked action.
Here, the scaling laws are assumed within the representation
(\ref{gauss}), which is superficially different in that
none of the gauge fields have been integrated over explicitly.
However, because the sources couple to coarse-lattice fields only
(cf.\  Eq.~(\ref{Oi})),
we may imagine that we first integrate over $\cu,\cv^{(1)},\ldots,\cv^{(n-1)}$
and only later over the coarse-lattice (fermions and) gauge field.
This is equivalent to inserting the source term into Eq.~(\ref{nlocalb})
which contains the blocked-lattice action $S_n$.
Thus, in the context of the ordinary staggered theory,
the above scaling laws are on the same footing as the scaling laws
used in a completely standard RG-blocking context.

Because of the scaling of the taste-breaking part as given by Eq.~(\ref{irrscale}),
the bound (\ref{mscale}) practically applies to both $D_n^{-1}$
and $D_{inv,n}^{-1}$.  When using Eq.~(\ref{mscale}) I will disregard the
difference between $\|D_n^{-1}\|$ and $\|D_{inv,n}^{-1}\|$.

\subsection{\label{reweigh} Recovery of taste symmetry in the continuum limit}
Assuming the existence of the continuum limit and the scaling laws
(\ref{mscale}) and (\ref{irrscale}), I will now prove that
exact taste symmetry is recovered in this limit
for all the coarse-lattice observables of the ordinary
staggered theory, provided $m_r(a_c) > 0$.
The proof makes use of the reweighted theories introduced in Sec.~\ref{plan},
and reveals why it is necessary
to avoid an exactly massless lattice theory.

I first add the source term (\ref{SFJ}) to the $n^{\rm th}$ reweighted theory
(cf.\ Eq.~(\ref{Zninva})):
\begin{equation}
  Z_{inv,n}(a_c;J)
  =
  \int \cd\cu
  \prod_{k=0}^{n} \Big[ \cd\cv^{(k)} \Big]\;
  \bltz_n\Big(1;\cu,\{\cv^{(k)}\}\Big)\,
  \det\Big(D_{inv,n}+J\cdot\cs^{(n)}\Big)
\label{Zinv}
\end{equation}
I also introduce a family of partition functions\footnote{
  For positivity of the determinants see Sec.~\ref{rooted}.
}
(with sources) in which
$D_n$ is replaced by $D_{inv,n}+t\D_n$, where
$t$ takes values in the interval $0\le t \le 1$.  Explicitly,
\begin{subequations}
\label{Zinter}
\begin{eqnarray}
  Z_{inter,n}(t,a_c;J) &=&
  \int \cd\cu
  \prod_{k=0}^{n} \Big[ \cd\cv^{(k)} \Big]\;
  \bltz_n\Big(1;\cu,\{\cv^{(k)}\}\Big)\,
\label{Zintera}
\\
  && \rule{0ex}{3ex} \times
  \det\Big(D_n+J\cdot\cs^{(n)}\Big)\;
  \det\left[1 + (t-1)\D_n \Big( D_n+J\cdot\cs^{(n)} \Big)^{-1} \right]
\NON
  &=&  \rule{0ex}{4ex}
  \int \cd\cu
  \prod_{k=0}^{n} \Big[ \cd\cv^{(k)} \Big]\;
  \bltz_n\Big(1;\cu,\{\cv^{(k)}\}\Big)\,
\label{Zinterb}
\\
  && \rule{0ex}{3ex} \times
  \det\Big(D_{inv,n}+J\cdot\cs^{(n)}\Big)\;
  \det\left[1 + t\D_n \Big(D_{inv,n}+J\cdot\cs^{(n)}\Big)^{-1} \right] \,.
\nonumber
\end{eqnarray}
\end{subequations}
These partition functions interpolate between
the reweighted theory for $t=0$, and the staggered theory for $t=1$.
Normalized generating functionals are defined in analogy with Eq.~(\ref{ZJnorm}),
and the $n\to\infty$ limits in analogy with Eq.~(\ref{Zlim}).

I will now show that the $n\to\infty$ limit does not depend on $t$,
{\it viz.},
\begin{equation}
  \cz_\infty(a_c;J)
  = \cz_{inter,\infty}(t,a_c;J)
  = \cz_{inv,\infty}(a_c;J)\,.
\label{universal}
\end{equation}
The key step is to bound the last factor in Eq.~(\ref{Zintera}) as
\begin{equation}
  \exp\tr\log\left(1 + (t-1)\D_n D_n^{-1} \right)
  = 1 + O(\e_n^2)\,,
\label{trlog}
\end{equation}
where
\begin{equation}
  \e_n = \Big\| D_n^{-1} \Big\|\, \Big\| \D_n \Big\| \,.
\label{bound}
\end{equation}
I have used that $\D_n$ is traceless on the taste index, as well
as the geometric series for the staggered propagator
$D_n^{-1}= D_{inv,n}^{-1} - D_{inv,n}^{-1} \D_nD_{inv,n}^{-1}
+ \cdots$.
  The sources are infinitesimal and do not interfere with any bound
  valid for $J=0$.
The (unnormalized) partition function
of any interpolating theory, $Z_{inter,n}(t,a_c;J)$,
is obtained from the staggered partition function by reweighting
with the left-hand side of Eq.~(\ref{trlog}).  Because the scaling
laws  (\ref{mscale}) and  (\ref{irrscale}) hold by assumption
on the staggered ensemble, it follows that
\begin{equation}
  \e_n \,\leqx\, {a_f\over a_c^2\, m_r(a_c)}
    = {2^{-(n+1)}\over a_c\, m_r(a_c)}\,.
\label{expnd}
\end{equation}
We arrive at several important conclusions.
First, each term in the Taylor expansion of the logarithm in
Eq.~(\ref{trlog}) is bounded by the corresponding power of $\e_n$.
Second, because the product $a_c\, m_r(a_c)$ is held fixed,
there will be an $n_0$ such that, for any $n\ge n_0$, one has $\e_n < 1$
and the Taylor expansion converges. It follows that
the change in any (meson) observable over the interval $0\le t \le 1$
is $O(a_f^2)$.  Finally, since $\e_n\to 0$ for $n\to\infty$,
we readily arrive at Eq.~(\ref{universal}).

The reweighted theories $\cz_{inv,n}(a_c;J)$ have exact $U(4)$ taste symmetry
by construction, and the same is true for the limiting theory
$\cz_{inv,\infty}(a_c;J)$.  But the limit is independent of $t$, and so
the staggered generating functional $\cz_\infty(a_c;J)$
has exact taste symmetry as well; the RG-blocked staggered theory
becomes taste-invariant in the limit of infinitely many blocking steps
if, in particular, Eq.~(\ref{irrscale}) holds.
As far as the rate of restoration of taste symmetry is concerned,
this is recognized as the familiar result that
the discretization errors of staggered fermions
are proportional to the (fine) lattice-spacing squared
\cite{Suss,SinP,Luo,asq,LS}.

In the massless staggered theory one has $m_r(a_c)=0$,
and the bound (\ref{expnd}) becomes an empty statement.
Therefore it is not possible to infer the recovery of full taste symmetry
in the exactly-massless case.  This is consistent with the established
fact that the continuum and the chiral limits of staggered fermions
do not always commute \cite{SV,DH,CBma,BGSS}.

The result I have established
readily generalizes to all the coarse-lattice observables.
Given a coarse-lattice operator
$\co^{(n)}=\co^{(n)}(\j^{(n)},\bj^{(n)},\cv^{(n)})$,
I introduce the notation $\svev{\co^{(n)}_{t'}}_{t}$
for a ``mixed,'' unnormalized expectation value where the sea quarks have
Dirac operator $D_{int,n}(t)=D_{inv,n}+t\D_n$,
while the valence quarks have Dirac operator $D_{int,n}(t')$.
To be precise, the Boltzmann weight is given by Eq.~(\ref{Zinter})
(with $J=0$), while the valence propagators
are given by $D_{int,n}^{-1}(t')$.
We then have
\begin{subequations}
\label{sv}
\begin{eqnarray}
  \svev{\co^{(n)}_t}_t
  &=&
  \svev{ \co^{(n)}_t \; \exp \left[\tr\!\log
  \Big( 1 + (t-1)\D_n D_{n}^{-1} \Big) \right]\, }_{1}
\label{sva}
\\
  &=& \rule{0ex}{4ex}
  \svev{\co^{(n)}_t}_{1} \left( 1 + O(\e_n^2)\right)
\label{svb}
\\
  &=& \rule{0ex}{4ex}
  \svev{\co^{(n)}_{1}}_{1} \left( 1 + O(\e_n)\right) \,.
\label{svc}
\end{eqnarray}
\end{subequations}
Equation (\ref{sva}) re-expresses an unnormalized observable
of $Z_{inter,n}(t)$
as a correlation function computed on the staggered ensemble
with a reweighting factor.
Equation (\ref{svb}) follows from the validity of the scaling laws
on the staggered ensemble.
Last, Eq.~(\ref{svc}) replaces any valence propagator $D_{int,n}^{-1}(t')$
by the staggered propagator $D_n^{-1}$.
The expansion of $D_{int,n}^{-1}(t')$
as a power series in $\D_n D_n^{-1}$ has the same convergence
properties as that of the logarithm in Eq.~(\ref{trlog}).\footnote{
  I have allowed for an $O(\e_n)$ mismatch in the observable,
  though presumably for any
  physical quantity of interest one could construct
  a coarse-lattice observable that would have only $O(a_f^2)$
  discretization errors, in which case the additional mismatch
  incurred in Eq.~(\ref{svc}) is likely to stay  $O(\e_n^2)$ as well.
}
As a special case, $\svev{1}_t = \svev{1}_1 (1 + O(\e_n^2))$,
and Eq.~(\ref{sv}) readily generalizes to normalized expectation values.
It follows that all the staggered and the reweighted ($t=0$)
observables have the same $n\to\infty$ limit, which now establishes the
exact taste symmetry of all observables.

Equality of all the observables implies the equality of the ``fixed-point''
coarse-lattice actions obtained in the limit $n\to\infty$.
Comparing once again the staggered and the reweighted theories, this means
(cf.\ Eqs.~(\ref{nlocalb}) and (\ref{Zninvb}))
\begin{eqnarray}
  S_\infty(a_c;\cv,\j,\bj\,)
  &\equiv& \lim_{n\to\infty} S_n(a_c;\cv,\j,\bj\,)
\NON
  &=& \lim_{n\to\infty} S_{inv,n}(a_c;\cv,\j,\bj\,)
  \equiv S_{inv,\infty}(a_c;\cv,\j,\bj\,) \,.
\label{Slim}
\end{eqnarray}
In analogy with Eq.~(\ref{acobserve}), I have dropped the
blocking-step label attached to the coarse-lattice fields,
and traded it with the coarse-lattice spacing.
Adding in the source term (\ref{SFJ})
then gives
\begin{equation}
  \cz_\infty(a_c;J)
  = \int \cd\cv \cd \j \cd\bj\;
  \exp\left[-S_\infty(a_c;\cv,\j,\bj) - \bj J\cdot\cs(a_c;\cv)\, \j\right] \,.
\label{SZlim}
\end{equation}
By Eq.~(\ref{Slim}), the action $S_\infty(a_c;\cv,\j,\bj)$ of the limiting
RG-blocked staggered theory has acquired exact taste-$U(4)$ invariance.
Equation (\ref{SZlim}) can be used to derive taste-$SU(4)$ Ward-Takahashi
identities that will be exactly satisfied in the limiting theory.
The corresponding result in the fourth-root theory
(see discussion below Eq.~(\ref{Z4invns}))
will put on a firm basis the observation made in Refs.~\cite{BGSS,Steve}
that no paradoxes can be derived from the extended taste symmetry
of the continuum-limit fourth-root theory.

\section{\label{uv} Reweighted theories at weak coupling}
In the previous section I have shown that, for large enough blocking
level $n$, the reweighted theory can be reached from the local
staggered theory by means of a convergent expansion.
The converse is also true: Setting $t=1$ in Eq.~(\ref{Zinterb}),
we can reconstruct the staggered theory from
the reweighted theory.
The convergence of the ($t$-)expansion in Eq.~(\ref{Zinterb})
is controlled by $\e_{inv,n}$, where
\begin{equation}
  \e_{inv,n}
  = \Big\| D_{inv,n}^{-1} \Big\|_{inv}\, \Big\| \D_n \Big\|_{inv} \,.
\label{rebound}
\end{equation}
The notation $\| \cdot \|_{inv}$ means that the norms are now to be evaluated
on the \textit{reweighted} ensemble.  In view of the established
scaling of $\D_n$ on the staggered ensemble, Eq.~(\ref{irrscale}),
the difference between any staggered-ensemble expectation value
and the corresponding reweighted-ensemble expectation value must be very small.
Indeed we must have $\e_{inv,n} \approx \e_n$,
up to corrections which are of higher order (in either of them).

The ability to go back and forth between the staggered and reweighted theories
implies that the reweighted theory associated with the local staggered theory
must have the following key properties:
\begin{itemize}

\item A suitable notion of renormalizability;

\item Locality on the coarse-lattice scale;

\item Validity of the scaling laws of Sec.~\ref{scaling}, including
  in particular the scaling of $\D_n$ as in Eq.~(\ref{irrscale}),
  on the reweighted ensemble.
\end{itemize}
In this section I explore direct evidence for these key physical properties.
The main output is that, step by step, every argument about
the (four taste) reweighted theory derived from the local staggered theory
generalizes straightforwardly to the (one taste) reweighted version
of the fourth-root theory.
In Sec.~\ref{rooted}, this will allow me to
establish the validity of the continuum limit of the fourth-root theory.
The $n\to\infty$ limit of the blocked fourth-root theory
will be reached via the corresponding limit of the sequence of reweighted
theories.  Since the reweighted theories are all local on the coarse-lattice
scale, the same will be true for the (common) limiting theory.

Power-counting renormalizability of the ordinary and fourth-root
staggered theories, alongside with the derived reweighted theories,
is discussed in Sec.~\ref{pcr}. Locality of the reweighted
theories is addressed in Sec.~\ref{lcl}.
The scaling laws are discussed in Sec.~\ref{rewpt},
relegating some further issues to App.~\ref{iroprep}.
I summarize the emerging physical picture in Sec.~\ref{rewcap}.

In the rest of this paper I will assume that the coarse-lattice
scale has been chosen to satisfy $a_c \ll \L^{-1}$.
This has the following implications. (1) Because of asymptotic freedom,
the running coupling constant $g_r(a_c)$ is weak at the coarse-lattice scale
as well as on all shorter distance scales. (2) One can {\it define}
lattice-regularized QCD to be local if it is local at the coarse-lattice scale.
(3) The coarse-lattice observables are rich enough to extract all of
the QCD physics.

\subsection{\label{pcr} Renormalizability}
I begin with a brief account of what is known about the renormalizability
of the ordinary and fourth-root staggered theories (Sec.~\ref{rstag}).
I then offer a natural definition of renormalizability for reweighted theories,
from which it follows that a reweighted theory
is automatically renormalizable if the underlying staggered theory is
(Sec.~\ref{rrew}).

\subsubsection{\label{rstag} Staggered theory}
Unlike Wilson fermions \cite{TR},
the task of deriving lattice power-counting theorems and
all-orders renormalizability remains to be completed for staggered fermions
(for recent progress, see Refs.~\cite{JG,grunstg}).  Still,
it is widely believed that the ordinary staggered theory is renormalizable
to all orders.  The main evidence comes from a one-loop calculation
accompanied by the observation that the staggered-fermion symmetries forbid
the generation of any relevant or marginal terms not already
present in the staggered action \cite{mgjs}.  In particular,
the taste-breaking terms remain irrelevant to all orders.

As first noted in Ref.~\cite{PQ}, all-orders renormalizability should
extend from the ordinary staggered theory to the fourth-root theory.
The argument relies on the familiar replica trick.\footnote{
  The relevance of the replica trick \cite{PK} for the low-energy pion sector
  of the fourth-root theory has been discussed in Refs.~\cite{AB,CBma,CB4f}.
}
Consider an ordinary staggered theory
with $n_r$ copies, or replicas, of equal-mass staggered fields.
At this stage, $n_r$ is a positive integer.
At each order in perturbation theory, the counter-terms needed
to renormalize the lattice theory will be polynomials in $n_r$,
because $n_r$ only enters as an overall
multiplicative factor attached to every closed fermion loop.
Next, we consider the analytic continuation in $n_r$ to arbitrary real values,
which corresponds to raising the staggered determinant to
a (possibly fractional) power $n_r$.
This continuation is unique, because the (polynomial!) $n_r$-dependence
of the diagrammatic expansion at each order, including the counter-terms,
is already known.  Thus,
the counter-terms derived for integer $n_r$ will be just enough to
renormalize the fractional-power theory for {\it any}\ {value} of $n_r$.
(While this captures the essence of the argument, it amounts
to an over-simplification.
For a more thorough discussion, see Ref.~\cite{Steve}.)

Thus, while the fourth-root theory is non-local \cite{BGS},
renormalizability is \textit{not} lost.  Retaining renormalizability
turns out to be the absolutely essential starting point from
which everything else follows.

\subsubsection{\label{rrew} Reweighted theory}
Renormalizability of a lattice theory means that, by adjusting the bare
parameters, the correlation functions
of the renormalized fields have a finite limit
when the lattice spacing goes to zero
and momenta in physical units are kept fixed,
to any order in perturbation theory.
In the present RG-blocking context I will assume that the correlation
functions under study are constructed from coarse-lattice fields.
The external momenta all belong to the Brillouin zone of the coarse lattice.

While the reweighted theories depend on both the fine- and coarse-lattice
scales, I will adopt exactly the same criterion to define when
they are renormalizable.  The implication is that renormalizability carries
over automatically from the staggered theory to the derived, reweighted
theories.  The reason is that, to leading order (in the fine-lattice spacing),
the difference between a given coarse-lattice correlation
function in the staggered and in the reweighted theory is equal to
the taste-breaking part the staggered correlation function.
In Sec.~\ref{rewpt} below I will argue that the taste-breaking part of any diagram
vanishes in the continuum ($n\to\infty$) limit,
and so the staggered and the reweighted
theories assign the same continuum-limit value for every
correlation function,\footnote{
  The notion of renormalizability is often assumed to include the requirement
  that the continuum-limit value of each diagram can be made equal
  to the value computed using some other regularization method by
  a finite renormalization \cite{TR}.  To the extent that this is true for
  the staggered theory, this will readily generalize to the (sequence of)
  reweighted theories as well.
}
to any order in perturbation theory.\footnote{
  Extending this claim to the non-perturbative level is
  the subject of Sec.~\ref{rooted}.
}
This prediction applies to both the ordinary
and the fourth-root staggered theories,
in fact to any real value of the number of replicas $n_r$.

In order to avoid unrelated non-perturbative complications,
as well as to ensure the existence of a weak-coupling regime,
I will further restrict $n_r$ to positive values
where the one-loop beta function (which depends linearly on $n_r$)
remains asymptotically free.  The main results
of Sec.~\ref{lcl} below are valid under these mild restrictions only.
As a preparation for the next stage, let me write down the staggered-fermion
RG-blocked partition function (without sources) for a general number of
replicas $n_r$,
\begin{equation}
  Z_n(t,n_r,a_c) =
  \int \cd\cu
  \prod_{k=0}^{n} \Big[ \cd\cv^{(k)} \Big]\;
  \bltz_n\Big(n_r;\cu,\{\cv^{(k)}\}\Big)\,
  \det^{n_r}\Big(D_{inv,n}+t\D_n\Big) \,,
\label{Znr}
\end{equation}
in which I have also kept the interpolating parameter $0\le t \le 1$
of Eq.~(\ref{Zinter}).
All the partition functions studied in this paper are special cases
of $Z_n(t,n_r,a_c)$.  As claimed above, they are all renormalizable.

\subsection{\label{lcl} Locality of $\Seff^k$ and $D_k$}
The main result of this subsection is that both the effective action
$\Seff^k$ (Eq.~(\ref{SeffG})) and the blocked Dirac operator $D_k$ (Eq.~(\ref{Dn}))
are local on the $k^{\rm th}$ lattice scale.  In more detail, integrating out
fermionic degrees of freedom at the $k^{\rm th}$ blocking step
generates a local effective action $\Seff^k$ for the gauge field,
and the Dirac operator governing the remaining fermionic degrees
of freedom is local too.  Because $\tD_{inv,n}$ is defined by
trace projection, $\tD_{inv,n}$ and $\D_n$ are separately local.
I will argue that this is true on the ensemble generated by
$Z_n(t,n_r,a_c)$ defined above, for any $t$ and $n_r$
(in the indicated ranges).  The argument relies on the renormalizability
of the lattice theory defined by $Z_n(t,n_r,a_c)$,
but it does not require that that theory be local in itself.

As noted in Sec.~\ref{plan} (see Eq.~(\ref{Z4q})),
a corollary of crucial importance is that
the theory defined by $Z_n(t,n_r,a_c)$ turns out to have a
local coarse-lattice action whenever raising of the fermion determinant
to the $n_r^{\rm th}$ power is an analytic operation.
A local coarse-lattice action \textit{defines} when
a reweighted or an interpolated theory is local.
A local coarse-lattice action is obtained for
the ordinary ($n_r=1$) staggered theory, as well as
for any theory derived from it by varying $t$;  and for the $n_r=1/4$
theory at $t=0$, which is recognized as the ($n^{\rm th}$)
reweighted theory derived from the fourth-root
theory, cf.\  Eq.~(\ref{Z4q}).\footnote{
  The same is true at $t=0$ for $n_r=n_s/4$, where $n_s$ is a positive
  integer, interpreted as the number of equal-mass sea quarks.
  All other (fractional!) values of $n_r$ fail to yield a local action,
  even at $t=0$, because there is no local Dirac operator $\tD$ such that
  $\det^{1/n_r}(\tD)=\det(D_{inv,n})$.  Notice that, in Sec.~\ref{unroot},
  it was not necessary to use the locality of the reweighted theories
  because the ordinary staggered theory is by itself local.
}

Let me begin with RG-blocking in a pure Yang-Mills theory.
While again no rigorous proofs exist,
it is widely accepted (see \eg Ref.~\cite{fp}) that the pure Yang-Mills lattice
action obtained after $n$ blocking steps is local.
The coarse-lattice action can be approximated by
the continuum form
\begin{equation}
  S_n(a_c) \approx {1\over g_r^2(a_c)} \int d^4x\, F_{\m\n}^2
  + \mbox{discretization errors.}
\label{contYM}
\end{equation}
The ensemble of coarse-lattice
configurations generated by the Boltzmann weight $B_n(a_c)=\exp(-S_n(a_c))$
will correspond to the correct running coupling $g_r(a_c)$.
(See App.~\ref{ensemble} for the generation of blocked-lattice gauge field
configurations.)

Next let us consider lattice QCD.  After integrating over all fields
except for the coarse-lattice gauge field,
one arrives at a Boltzmann weight of the general form
(again using continuum notation)
\begin{equation}
  B_n(a_c) \approx
  \exp\bigg[-{1\over g_r^2(a_c)} \int d^4x\, F_{\m\n}^2
  + O(1)
  + \mbox{discretization errors} \bigg] \,.
\label{contB}
\end{equation}
Here ``$O(1)$'' stands for terms occurring at zeroth or higher  order
in the expansion in powers of $g_r(a_c)$,
which are also of zeroth order in the expansion in powers
of the lattice spacing (\ie terms that survive the continuum limit).
The form (\ref{contB}) depends on renormalizability,
because the ultra-violet divergent part of the fermion determinant
that renormalizes the coupling constant has been separated out explicitly.
All other effects of the integration over the fermion fields
are contained in the last two terms in Eq.~(\ref{contB}).  These terms
obviously include the effects of virtual quark loops, and they are not local.
For the considerations in this subsection, the only thing
that matters is that the renormalized Yang-Mills action inside
of the Boltzmann weight is parametrically larger by $1/g_r^2(a_c)$.
Equation (\ref{contB})
applies to the partition function $Z_n(t,n_r,a_c)$ as well
(after all but the integration over $\cv^{(n)}$ has been done),
because this partition function too defines a renormalizable theory.

With Eq.~(\ref{contB}) in hand, we are ready to discuss the locality
properties of $\Seff^k$ and $D_k$.
Let me summarize the relevant discussion from Sec.~\ref{overview}
(in particular around Eq.~(\ref{DG})), but now in terms of
the hermitian \cite{BGS} operator
\begin{equation}
  H_k = [\g_5\otimes \x_5]\, G_k^{-1}\,.
\label{Hk}
\end{equation}
First, functional differentiation of $\Seff^k$ with respect to the
(original or blocked)
link variables generates expressions that depend on both $H_k$ and $H_k^{-1}$.
Locality of $\Seff^k$ then follows provided that $H_k$ and $H_k^{-1}$
are both local on the $k^{\rm th}$ lattice scale.
The locality of $D_k$, $H_k$ and $H_k^{-1}$ is established iteratively
using Eqs.~(\ref{Gnb}) and (\ref{DG}).  (For the $k=0$ step, see Ref.~\cite{BGS}.)
The only non-trivial step is to demonstrate
the short-range nature of $H_k^{-1}$.
In the free theory, this follows because $H_k$ has an $O(\a_k)$ gap.

In the interacting theory one has to replace the notion of a spectral gap
by the notion of a mobility edge.
The properties established in the free theory
will carry over to any smooth gauge field,
and, more technically, to any order in lattice perturbation theory.
This leaves open the following question.
In the presence of very rough, lattice-size structures in the gauge field,
or ``dislocations,'' could $H_k$ develop
much smaller eigenvalues, which
in turn would spoil the short-range character of its inverse?
I will now argue that the answer is negative.

Before coming to the main argument I should note that it
is logically possible that all eigenvalues $\l^{(k)}_i$
of $H_k$ may always satisfy $|\l^{(k)}_i| \ge \l^{(k)}_{min}>0$,
for some $\l^{(k)}_{min}=O(\a_k)$.
In other words, $H_k$ might have an $O(\a_k)$ gap
for {\it all}\ {gauge} fields.  If this is true, we are done.
Because it is unknown if this is true, I will disregard this possibility.

According to the theory of disordered systems
(see Ref.~\cite{mobt,mobn} and references therein),
the right question becomes what is the {\it mobility edge} of $H_k$.
In general, the spectrum of $H_k$ will consist of
localized eigenstates (at the scale $a_k$)
with eigenvalues $0 \le |\l^{(k)}_i| \le \l^{(k)}_c$.
Above the critical value $\l^{(k)}_c$ the eigenstates become extended.
By definition, $\l^{(k)}_c \ge 0$ is the mobility edge of $H_k$.
The value of $\l^{(k)}_c$ is a property of the ensemble.

I will argue that the mobility edge of $H_k$ is $O(\a_k)$ in the
weak-coupling regime.
On general grounds, any localized eigenmodes lying below the mobility edge,
be their eigenvalues as small as they may, will not spoil
the short-range character of the inverse $H_k^{-1}$ \cite{HJL,mobt}.
The decay length of $H_k^{-1}$ is thus $O(a_k)$ as required.

At scales where lattice QCD is weakly coupled,
the physics that goes into the mobility edge is simple.
Because it may be unfamiliar, and since $H_k$ itself has not been studied
numerically yet, let me digress to describe results obtained
in the study of the mobility edge of the hermitian Wilson operator $H_W$.
In Ref.~\cite{mobn} the mobility edge of $H_W$ was determined
for the super-critical bare mass $am_0=-1.5$, and for a range of values
of $\beta=6/g_0^2$ on quenched ensembles.  This example is relevant
for the following reason.  First, for the chosen parameters,
$H_W$ and $H_k$ both have $O(a^{-1})$ gaps in the free theory,
where $a$ is the relevant lattice spacing.  Moreover,
at any super-critical bare mass the spectrum of $H_W$ can,
and does, reach zero.  Therefore, the analogy will become relevant in case
that future numerical work will demonstrate the existence of
low-lying eigenvalues of $H_k$.  According to the argument below,
the corresponding eigenmodes will necessarily be localized.

For the case at hand,
the most interesting finding of Ref.~\cite{mobn} was that
the mobility edge was very close to the free-theory gap
for several different gauge actions, even at the not-very-large
cutoff scale $a^{-1} \sim 2$ GeV.
At stronger coupling (lower cutoffs) the mobility edge did go down,
eventually reaching zero when the Aoki phase was entered.

Returning to the RG-blocking context, we set the coarse-lattice spacing
such that $g_r(a_c)$ is as weak as we like.  Therefore we are
interested in values of the mobility edge at weaker couplings
than any of those already studied.  The results of Ref.~\cite{mobn} suggest that,
on a very weakly-coupled pure Yang-Mills ensemble,
the mobility edge of any operator,
including the super-critical $H_W$ and $H_k$,
will be very close to the free-theory gap;
for $\beta\to\infty$ the mobility edge will continuously
approach the free-theory gap.

It remains to consider the inclusion of a fermion determinant
raised to some positive (but not necessarily integer!) power
at weak coupling, as described by Eq.~(\ref{contB}).
This should have little effect on
mobility edges which are already $O(a^{-1})$.
A different power of a fermion determinant
does change the beta-function and the running of the gauge coupling,
so we should be a bit careful in what we are comparing.
Consider first the operator $H_n$ of the last blocking step.  We may
compare the value of its mobility edge on a pure Yang-Mills ensemble
to the corresponding value on a dynamical-fermion ensemble
that has the same coarse-lattice running coupling $g_r(a_c)$.
Once the coarse-lattice couplings are equal,
the remaining contribution of the fermion determinant in Eq.~(\ref{contB})
is parametrically smaller by a factor of $g_r^2(a_c)$
compared to the Yang-Mills action which is the leading term.
Changes to the spectrum near a mobility edge which is already $O(a_c^{-1})$
and, therefore, any further deviations of the mobility edge itself
from the free-theory gap, are expected to be very small.

We actually need to know something about the mobility edges
of $H_k$ for all $k \le n$.  For each $k$, we may compare the blocked
staggered ensemble to a new Yang-Mills ensemble chosen such
that the running coupling $g_r(a_k)$ at the $k^{\rm th}$ lattice scale
is the same in the two theories.  Again a similar conclusion will follow.

What can, and will, be significantly affected
by the inclusion of fermion determinants is the small-eigenvalue localized
spectrum (if there were any near-zero eigenvalues to begin with;
see for example Ref.~\cite{mobt}).
Since the Boltzmann weight contains $\det(G_k^{-1})=\det(H_k)$
raised to a positive power, the transition from the pure Yang-Mills
ensemble to dynamical staggered-fermion ensembles will lead to fewer
near-zero eigenvalues of $H_k$, for all $k$.

Let me summarize the anticipated physical situation.
In the free theory, $H_k$ has an $O(\a_k)$ gap.  On weakly-coupled
pure Yang-Mills ensembles, the mobility edge of $H_k$ is expected
to be very close to the free-theory gap
and, thus, $O(\a_k)$ by itself.  Now, starting at $n_r=0$, let us
gradually increase the power of the staggered-fermion
determinant in Eq.~(\ref{Znr}), while maintaining
a fixed renormalized coupling $g_r(a_c)$ at the coarse-lattice scale.
Any near-zero eigenvalues of $H_k$ will be gradually suppressed,
but otherwise nothing much should change in the spectrum of $H_k$,
for all $k$.  In particular, any further change in the mobility edge will be
even smaller than the,
by itself small, deviation from the free-theory gap on
the pure Yang-Mills ensemble.\footnote{
  I use asymptotic freedom to bound $g_r(a_k)$ by $g_r(a_c)$.
  This restricts my conclusions to the range of $n_r$-values where
  the one-loop beta-function is negative. See also footnote \ref{QED}.
}
As a result, the mobility edge of $H_k$ will remain $O(\a_k)$, and
the decay length of $H_k^{-1}$ will remain $O(a_k)$.
As explained above, this implies the locality of
$D_k$, $H_k$, $H_k^{-1}$, and $\Seff^k$, for all $k\le n$.

\subsection{\label{rewpt} Scaling of $\D_n$}
In Sec.~\ref{pcr} I have explained why
the partition function (\ref{Znr}) defines a renormalizable theory
for any $n_r$ and $t$.  For generic values of these parameters, this theory
is non-local.  Nevertheless, thanks to its power-counting renormalizability
we may study the scaling
of any local operator constructed within such a theory.
This includes in particular the operators listed at the end of
the previous subsection, whose locality I have just established
by non-perturbative considerations.

I will be mostly interested in the scaling of $\D_n$,
the taste-breaking part of the blocked Dirac operator,
in the staggered ($t=1$) and in the reweighted ($t=0$) theory.
In this subsection I argue that Eq.~(\ref{irrscale}) correctly describes
the leading power-law scaling of $\D_n$.
I furthermore find that the logarithmic corrections to the scaling
of $\D_n$ depend on $n_r$ but not on $t$.
The argument is heuristic, and will have to be confirmed by future
calculations.

Setting up perturbation theory is in principle straightforward.
In reality, perturbation theory for a reweighted theory is unfamiliar,
and technically rather different from ordinary staggered
perturbation theory.
It can be gradually built up in several steps:
\begin{quotation}
\noindent\vspace{-4ex}
\begin{description}
\item[\textit{Step 1.}]
Ordinary staggered perturbation theory;

\item[\textit{Step 2'.}]
Staggered perturbation theory with the fermions in a taste basis obtained
via a unitary change of variables \cite{saclay,Luo};

\item[\textit{Step 2.}]
Staggered perturbation theory with the fermions in a taste basis obtained
via a gaussian smearing RG-like step \cite{BGS};

\item[\textit{Step 3.}]
Multi-gauge-field perturbation theory for the blocked staggered
theories of Eqs.~(\ref{gauss}), (\ref{Zn}) or (\ref{Z4n});

\item[\textit{Step 4.}]
Multi-gauge-field perturbation theory for a reweighted theory
(or, more generally, for $Z_n(t,n_r,a_c)$).

\end{description}

\end{quotation}
As I will explain, computing the scaling of $\D_n$ within the fully developed
perturbative setup of Step~4 can be reduced,
via Steps~3 and~2 (skipping Step~2'), to a calculation
in ordinary staggered perturbation theory (Step~1).
Focusing on the fourth-root theory, the outcome is that in spite
of the lack of a local action, we nevertheless have at our disposal
a local operator $\D_n$ that, on the one hand, accounts for
all the taste violations in blocked observables,
and, on the other hand, is controlled
by staggered perturbation theory.

Ordinary staggered perturbation theory is a well-developed technique
and there is no need to discuss it here
(see \eg Ref.~\cite{Steve} and references therein).
I now discuss all the other steps listed above.

In taste-basis perturbation theory the fermion
momentum ranges over a reduced Brillouin zone,
and the sixteen sites of each $2^4$ hypercube are accounted
for by the Dirac and the taste degrees of freedom.
A taste-basis perturbative expansion is usually not used for two reasons.
The unitary transformation to a taste basis is not unique,
and complicates the form of many symmetries \cite{saclay,Luo}
(see Apps.~\ref{tasteinv}, \ref{sym} and \ref{disorder} for more details).
Also, in a taste-basis diagrammatic expansion, taste violations
occur in both the propagator and the vertices, as can immediately be seen
by inspection of the free taste-basis Dirac operator (see App.~\ref{iroprep}
for further discussion).  This is to be contrasted
with the usual staggered perturbation theory,
where the momentum-space propagator is taste-symmetric and taste violations
reside in the vertices only \cite{mgjs,Steve}.

Next, consider the staggered theory obtained by
performing the special $k=0$ ``blocking'' step introduced in
Sec.~\ref{blockn} for the fermions \textit{only}.
No blocking is applied to the gauge field $\cu$.
The resulting Dirac operator is $D_0$ of Eq.~(\ref{Dna}).
Following the notation of Ref.~\cite{BGS} where its explicit form was derived,
in this subsection I denote it as $D_{taste}=D_{taste}(\a_0,m)$.
In the free theory one can write $D_{taste}=i(\ca+\cb)+\cm$,
where $\ca$, $\cb$ and $\cm$ are all hermitian,
and correspond to the $Dirac\, \otimes\, taste$ structures
$\g_\m \otimes \id$, $\g_5 \otimes i\x_\m \x_5$, and $\id \otimes \id$
respectively.  As usual, $D_{taste}(\a_0,m)$ satisfies
a GW relation in the massless limit.  A key feature of $D_{taste}(\a_0,m)$
is that, provided $\a_0<\infty$, dropping the taste-breaking part $\cb$
does \textit{not} introduce any new doublers into the theory \cite{BGS}.
The gaussian-smearing transformation reduces to the previous
unitary transformation in the limit $\a_0\to\infty$, where $\cm(\infty,m)=m$.
Here, as in Sec.~\ref{overview}, I assume that $\a_0=O(a_f^{-1})$.

With its extra technical hassles, staggered
perturbation theory with the Dirac operator $D_{taste}$
is, clearly, highly relevant to the present blocking framework.
For the purpose of the discussion below, I only need to draw attention
to one fact.  By using the general procedure of Ref.~\cite{HNA}
one can prove that $D_{taste}$
retains all the staggered symmetries, albeit in a complicated form.\footnote{
  The underlying reason is that, on top of the previously applied unitary
  transformation, gaussian smearing changes the propagator
  by a contact term only.  The long-distance propagator is unchanged,
  and therefore all the symmetry constraints on long-distance correlation
  functions must hold, as can be shown using the pull-back
  mapping (Sec.~\ref{pullb} and App.~\ref{sym}).
  The modified, Ginsparg-Wilson-L\"uscher (GWL) chiral symmetry \cite{ML}
  associated with the GW relation
  is a special case of the general construction of Ref.~\cite{HNA}.
}
These symmetries forbid the appearance of
taste-breaking relevant or marginal terms
through loop corrections \cite{mgjs}.
The taste-violating part of the fermion self-energy,
denoted $\G_{t.v.}$, is therefore $O(p^2)$ in lattice units.
But since taste violations occur now
in both vertices and propagators, this result necessarily represent
a delicate cancellation. Schematically: $\G_{t.v.}^v-\G_{t.v.}^l=O(p^2)$
where both $\G_{t.v.}^v$ and $\G_{t.v.}^l$ are $O(1)$,
and where $\G_{t.v.}^v$ accounts for the contribution of diagrams
with at least one taste-breaking vertex, while $\G_{t.v.}^l$ accounts
for the remaining contributions, in which taste-breaking arises
from the fermion lines only.

The thing to notice is that this delicate cancellation would be hampered
had we truncated the free propagator
$(-i(\ca+\cb)+\cm)/(\ca^2+\cb^2+\cm^2)$
to the ``linearized'' form  $(-i(\ca+\cb)+\cm)/(\ca^2+\cm^2)$
obtained keeping only the first taste-breaking term
in the geometric-series expansion of the free propagator
about the taste-symmetric $(-i\ca+\cm)/(\ca^2+\cm^2)$.
This truncation is not entirely foolish,
because it does not introduce doublers into the theory.
Yet, it mutilates
shift symmetry (see App.~\ref{iroprep} for a related discussion).
Dynamically, the reason is that, for lattice-scale
momenta that contribute to the loop integrals, the taste-breaking
part of the propagator is not small relative to the taste symmetric part.
Therefore, truncating the propagator
will give rise to large, imbalanced changes in
$\G_{t.v.}^v$ and $\G_{t.v.}^l$.

I will now argue that the main change brought about by (many!) iterations of
the RG transformation
is that the taste-breaking term $\D_n$ becomes \textit{uniformly}
small, for all the coarse-lattice momenta.  As a result, for large $n$
the scaling of taste-breaking effects will be controlled by
diagrams with a \textit{single} insertion of $\D_n$.

Before we do any RG-blocked diagrammatic calculation, we must
first set up the appropriate perturbative expansion.
Usually lattice perturbation theory is based on the expansion
of the link variables as $U_{\mu,x}=\exp(iga A_{\mu,x})$.
When the definition of the lattice theory makes use of the representation
(\ref{gauss}), a similar expansion will have to be applied to the
entire tower of gauge fields $\cu,\cv^{(0)},\cv^{(1)},\ldots,\cv^{(n)}$
simultaneously.  With this, one can in principle set up
the perturbative expansion for every theory that can be cast in the form
of Eq.~(\ref{Znr}), because the closed-form expressions for $D_k$, $\tD_k$,
$\D_k$ and $S_{eff}^k$ as functionals of all the gauge fields are known.

The next stage is to consider
the multi-gauge-field perturbative expansion of the RG-blocked
staggered theory ($t=1$ in Eq.~(\ref{Znr})).  This perturbation theory will
reproduce all the scaling laws derived using ordinary
staggered perturbation theory.
The reason is simply that, as already noted in Sec.~\ref{scaling},
one can first integrate out $\cu,\cv^{(0)},\cv^{(1)},\ldots,\cv^{(n-1)}$.
At this point, one has effectively recovered the diagrammatic expansion
derived from the action $S_n$ (Eq.~(\ref{nlocalb})) of the $n^{\rm th}$ level
blocked theory. The scaling of the parameters of $S_n$,
in turn, must agree with the predictions of ordinary lattice
perturbation theory.\footnote{
  Alternatively, we may consider the pull-back of $\D_n$ to the original
  fine-lattice staggered theory, where we may again study its scaling
  as a function of $n$.
}
While this description is, strictly speaking, applicable
for integer values of $n_r$, all other values can be
reached via the replica trick, cf.~Sec.~\ref{pcr}.\footnote{
  As long as we stay within the confines of perturbation theory, this procedure
  gives meaning to the blocked action $S_n$ for any real value of $n_r$.
  Conceptually, this is similar to the way the diagrammatic expansion
  gives meaning to the dimensionally-continued action in
  dimensional regularization.
}

The last step is to show that $\D_n$ must scale in the same way
in the (blocked) staggered theory and in the corresponding
reweighted theory.\footnote{
  The argument can be generalized to $t\ne 0$, with the same conclusion.
}
This result is established by isolating the diagrams that
determine the scaling of $\D_n$ in each theory, and showing
that they amount to exactly the same set of diagrams.

In the blocked staggered theory, $\D_n$ gives rise to taste breaking effects
either through vertices or though the expansion of the free blocked propagator
as $D_n^{-1}= D_{inv,n}^{-1} - D_{inv,n}^{-1} \D_n D_{inv,n}^{-1} + \cdots$.
Next comes the main observation.  The momentum that flows though any fermion
line is, in all cases, a coarse lattice momentum
$p\, \leqx\, 1/a_c$.  After sufficiently many blocking steps,
any coarse-lattice momentum will be very small in fine-lattice units:
$p a_f \sim a_f/a_c \ll 1$.
(This is true whether the fermion line forms a closed loop or connects
to an external leg.)  In contrast, $\D_n$ embodies taste-breaking
effects coming from all higher momentum scales
up to the fine-lattice cutoff $1/a_f$.
This means that any mechanism needed to ensure the smallness of
all taste breaking effects on the coarse lattice,
such as cancellations based on symmetries,
must be built into the functional form of
$\D_n=\D_n(\cu,\cv^{(0)},\ldots,\cv^{(n)})$ itself.
Said differently, most of the needed cancellations must occur
over distance scales much smaller than $a_c$. Therefore,
they will not occur in an expectation value with multiple insertions
of the operator $\D_n$, unless they already occurred in every
expectation value with a single insertion of $\D_n$.

Within the multi-gauge-field diagrammatic expansion,
this translates into the statement that
any insertion of $\D_n$ in any diagram must scale as $a_f/a_c$
in coarse-lattice units.
Given a diagram of the blocked staggered theory, let us now drop
every contribution where the total number of insertions of $\D_n$
(coming from both propagators and vertices) is bigger than one.
In the remaining taste-violating diagrams,
the fermion propagator is $D_{inv,n}^{-1}$,
and they contain exactly one insertion of $\D_n$.  These diagrams
determine the scaling of $\D_n$ in the blocked staggered theory,
and, as argued above, will reproduce the taste-breaking scaling laws
of the original staggered theory.

But, clearly, the very same set of diagrams is what determines
the scaling of $\D_n$ in the \textit{reweighted} theory!
I conclude that, in any reweighted theory,
$\D_n$ must scale in the same way as in the original staggered theory.
While this will not be needed, the diagrammatic correspondence
is clearly tight enough to encompass the logarithmic corrections
as well.  Thus, the logarithmic corrections depend on $n_r$,
but not on $t$.

More generally, the scaling behavior of any local operator
must be the same in the staggered and in the reweighted theory,
simply because the scaling will be insensitive to
dropping $\D_n$ from the blocked Dirac operator.  (The special case
considered above amounts to taking the local operator to be $\D_n$ itself.)
This implies that the physical observables of the reweighted theory should
have scaling violations proportional to $a_f^2$, as predicted by
(perturbation theory for) the original staggered theory.
The $n\to\infty$ limit therefore yields a ``perfect action'' theory \cite{fp}.
(See, however, App.~\ref{sym} for a discussion of related technical issues.)

In summary, we learn two important lessons.  After many blocking steps,
$\D_n$ will be
small on any staggered or reweighted ensemble.
We may thus compute its scaling behavior
on either ensemble by appealing to perturbation theory for
the multi-gauge-field representation of the blocked staggered theory ($t=1$).
Also, as long as we allow for coarse-lattice observables only,
this calculation further reduces
to a conventional scaling calculation in staggered perturbation theory,
augmented by the replica trick (for non-integer $n_r$).

In particular, I find that the power-law scaling (\ref{irrscale}) is valid
in the reweighted theory derived from the fourth-root theory.
Notice that I have assumed that $\D_n$ scales like
a dimension-five (and not like a dimension-six) operator.
But, as explained in Sec.~\ref{reweigh},
thanks to taste-tracelessness of $\D_n$ this assumption is consistent
with $O(a_f^2)$ scaling of the taste-violating effects
in all the physical observables.
In Sec.~\ref{rooted}, the scaling of $\D_n$ will be used
to establish the validity of the continuum limit of the fourth-root theory.

In this subsection I
gave only a very minimal discussion of the multi-gauge-field
diagrammatic expansion.  In  App.~\ref{iroprep}, I illustrate some further
aspects of this expansion by considering a few examples of terms
which are expected to occur in $\D_n$.

\subsection{\label{rewcap} Summary and future work}
We are almost done.  In the next section, the
reweighted theories will be used to establish
the validity of the fourth-root theory in the continuum limit.
This conclusion is a straightforward corollary of the emerging
physical picture of the reweighted theories.  I therefore pause to summarize
what has been learned.

All the reweighted theories introduced in Sec.~\ref{plan} share
the following key features:
(1) renormalizability, (2) locality, and (3) the same scaling
laws as the underlying staggered theory.
At the starting point is the all-essential observation that
the fourth-root theory is renormalizable if the ordinary staggered theory
is.  From this point on, basically the same reasoning
was applied in both the ordinary and fourth-root cases.
In fact, for the most part the arguments generalize to
any real number of replicas $n_r$ within the range
specified above Eq.~(\ref{Znr}).

As briefly discussed in Sec.~\ref{rstag}, renormalizability of both
the ordinary and fourth-root staggered theories is not as solidly
established as in other cases (notably Wilson fermions).
But there is no real reason to doubt it either.  For a recent, more
thorough discussion, see Ref.~\cite{Steve}.  As explained in Sec.~\ref{rrew},
the reweighted theories ``inherit'' their renormalizability from
the underlying staggered theory, in a rather trivial way.

Locality of the reweighted theories at the coarse-lattice scale rests on the
locality of the effective action $\Seff^k$ and
the blocked Dirac operator $D_k$, on the relevant ensemble.
Those locality properties, in turn, are set by the range of $H_k^{-1}$,
where the hermitian operator $H_k$ (Eq.~(\ref{Hk}))
accounts for the short-distance fermion
modes integrated out at the $k^{\rm th}$ step.
I have argued in Sec.~\ref{lcl} that
what is needed is that the \textit{mobility edge}
of $H_k$ be $O(1)$ in units of the $k^{\rm th}$ lattice scale.
I have drawn an analogy to a recent application of the theory of localization
to lattice QCD, specifically, to a study of
the mobility edge of the super-critical Wilson operator \cite{mobt,mobn}.
I concluded that, thanks to the existence of a weak-coupling regime
(which in turn is a consequence of renormalizability),
both the mobility edge of $H_k$
and the range of $H_k^{-1}$ will be $O(1)$ in $k^{\rm th}$ lattice
units, as required. Obviously, it will be necessary
to confirm the claims by numerical investigations of $H_k$ itself.
The first non-trivial instance
is provided by the $k=1$ blocking step.\footnote{
  Because of special features of the $k=0$ step,
  the operator $H_0$ is guaranteed to have a gap
  in the interacting theory too.
  The same is not true for $k\ge 1$.
}

Our knowledge about the ordinary staggered theory strengthens the claims
I have made. In the ordinary staggered theory,
based on standard RG considerations one assumes
that the blocked action $S_n$ (Eq.~(\ref{nlocalb}))
will be local on the coarse-lattice scale.  For this to be true,
the locality properties of $\Seff^k$ and $D_k$ must be as claimed.
But my reasoning in Sec.~\ref{lcl} did not discriminate between
the ordinary and the fourth-root ensembles.  This lends higher
credibility to the proposed physical picture in the fourth-root case as well.

My claims are on stronger footing for $n_r=0$ as well:
this quenched limit is closer to the actual
setup of the work reported in Ref.~\cite{mobn}.  The fourth-root value
$n_r=1/4$ may thus be reached by interpolation, starting either from
$n_r=0$ or $n_r=1$.  Once again, this supports the claims
made in the fourth-root case.

I now turn to the scaling of the taste-breaking effects
represented by $\D_n$.  The basic difficulty is simply that,
in the fourth-root theory, there is no local fermion action.
Thus, it is unclear if the taste violations that reside in the fermion
sector are amenable to a scaling analysis.

First, the fourth-root theory is renormalizable.
Therefore, even though the theory is non-local, we have a power counting
and we can study the scaling of any local operator.\footnote{
  See Ref.~\cite{Steve} for similar examples taken from Condensed Matter physics.
}
Specifically, I have shown that a scaling analysis in the fermion sector
is made possible thanks to the multi-gauge-field representation
introduced in Sec.~\ref{overview}.  This gives us access to the
operator $\D_n$ that accounts for all the taste-symmetry violations
in blocked observables.  According to the
discussion of Sec.~\ref{lcl},
$\D_n$ is a \textit{local} operator; therefore its scaling
can be computed using the appropriate (multi-gauge-field) perturbative
expansion.  Finally, I have argued that the needed scaling calculation
ultimately reduces to a calculation in ordinary staggered perturbation
theory augmented by the replica trick,
and that $\D_n$ indeed scales as an irrelevant operator should.

My arguments in Sec.~\ref{rewpt} were heuristic,
and it is clearly necessary to confirm them by performing the
appropriate perturbative calculations.
The actual scaling of $\D_n$ can also be investigated numerically,
at least on the fourth-root ensembles provided by MILC \cite{sugar}.
A first study was performed last year \cite{FM}.
Most of the arguments of this section rely on being in a (sufficiently)
weak-coupling regime, and it is important to understand how
close are we to this region in practice.
Numerically reweighting is clearly a challenge,
which, if successfully tackled,
could further strengthen confidence in the entire framework.
Another challenging project is to perform an accurate comparison of the
predictions of
the various perturbative expansions to numerical results obtained e.g.\
by measuring Wilson loops \cite{grunstg} or by adapting
the Schr\"odinger-functional technique  \cite{grun,sf,mrun,stagrun}.

\section{\label{rooted} Continuum limit of the fourth-root theory}
Assuming the properties of the reweighted
theories discussed in Sec.~\ref{uv}, in this section I prove the validity
of the fourth-root theory in the continuum limit.
As explained earlier,
when the blocking level $n$ is high enough, one can either reach
the reweighted theory from the staggered theory via a convergent
expansion, or work the other way around.  I find it appealing to
reconstruct the fourth-root theory from the reweighted theory,
because the latter is local, and is already expected to be in the correct
universality class.  The argument, that otherwise follows the same logic as in
Sec.~\ref{unroot}, is given in Sec.~\ref{cl}.  I add several comments on the
scaling analysis in Sec.~\ref{ns}.

\subsection{\label{cl} Recovery of locality in the continuum limit}
In the fourth-root theory I will make use of the scaling laws
(compare Eqs.~(\ref{mscale}) and (\ref{irrscale}))
\begin{equation}
  \left\| \D_n \right\|_{inv} \,\leqx\; {a_f\over a_c^2}\,,
\label{irr4scale}
\end{equation}
where the  subscript ``$inv$'' refers to the reweighted ensemble,
and
\begin{equation}
   \left\| D_{inv,n}^{-1} \right\|_{inv} \,\leqx\,  {1\over m_{inv,n}} \,,
\label{m4scale}
\end{equation}
where
\begin{equation}
  m_{inv,n} = m_r(a_c) + O(a_f/a_c^2) \,.
\label{minv}
\end{equation}
The scaling of $\D_n$ was discussed  in Sec.~\ref{rewpt}.
The leading power-law behavior of $\D_n$ is robust. It is unchanged by taking
the fourth root, and it is also independent of reweighting.
Turning to  Eq.~(\ref{minv}),
the origin of the rightmost term is simply that the transition
from $D_n$ to $D_{inv,n}$ amounts to dropping $\D_n$.  The latter is
$O(a_f/a_c^2)$ which, therefore, could entail similar changes in the
eigenvalues. See App.~\ref{mbound} for some further comments on
the bound (\ref{minv}).
Similar considerations show that the effective coupling constant
of the reweighted theory, denoted $g_{inv,n}$, satisfies
\begin{equation}
  g^2_{inv,n} = g^2_r(a_c) \left[1 + O\left((a_f/a_c)^2\right)\right] \,,
\label{ginv}
\end{equation}
where I have used Eq.~(\ref{contB}) and the taste-tracelessness of $\D_n$.
This implies that the reweighted theory is in a weak-coupling regime
if the coarse-lattice staggered theory is, and vice versa.

I will restrict the present discussion to meson observables
of the fourth-root theory.  The generalization to all other observables
requires additional technical steps which are discussed in
Appendix~B of Ref.~\cite{Steve}.
The physical meson observables are taste singlets.  They are probed
by restricting the source term of Sec.~\ref{CL}
to the form $\tJ\cdot\cs^{(n)}$,
where now $\cs_i^{(n)}=[\tcs^{(n)}_i \otimes {\bf 1}]$.
Here $\tcs^{(n)}_i$ carries no taste index and,
as usual, ${\bf 1}$ is the identity matrix in taste space.
Switching notation from $J$ to $\tJ$ is meant to remind us
that the sources now couple to taste singlets only.
The blocked fourth-root partition function with these sources is given by
\begin{equation}
  \Zroot_n(a_c;\tJ\,) =
  \int \cd\cu
  \prod_{k=0}^{n} \Big[ \cd\cv^{(k)} \Big]\;
  \bltz_n\Big(\quart;\cu,\{\cv^{(k)}\}\Big)\,
  \det^{1/4}\Big(D_n+\tJ\cdot\cs^{(n)}\Big) \,.
\label{Z4J}
\end{equation}
Equation (\ref{Z4J}) means that observables are constructed as follows
\cite{CB4f}.
Fermion -- anti-fermion contractions are done in the same way as
in the ordinary staggered theory;
then one applies the extra ``replica'' rule that a factor of
$\quart$ is to be attached to every closed fermion loop
occurring in the observable itself (in other words, to every valence
staggered-fermion loop).
With this replica rule in place,
the pull-back mapping defined in Eqs.~(\ref{pull}) and (\ref{pullF})
remains valid, and the same is true for Eq.~(\ref{observe}).
Thus, the ultra-local nature of the pull-back mapping is preserved,
even though the lattice action itself is not local.

I add the same source term to the reweighted theories, which gives rise to
\begin{equation}
  \Zroot_{inv,n}(a_c;\tJ\,)
  =
  \int \cd\cu
  \prod_{k=0}^{n} \Big[ \cd\cv^{(k)} \Big]\;
  \bltz_n\Big(\quart;\cu,\{\cv^{(k)}\}\Big)\,
  \det\Big(\tD_{inv,n}+\tJ\cdot\tcs^{(n)}\Big) \,.
\label{Z4inv}
\end{equation}
Here I have used the exact taste invariance of the reweighted
theories and the taste-singlet nature of the sources to take the analytic
fourth root.  In analogy with Sec.~\ref{unroot}, I also introduce
interpolating theories (with the same source), whose partition functions
can be expressed as:
\begin{eqnarray}
  \Zroot_{inter,n}(t,a_c;\tJ\,) &=&
  \int \cd\cu
  \prod_{k=0}^{n} \Big[ \cd\cv^{(k)} \Big]\;
  \bltz_n\Big(\quart;\cu,\{\cv^{(k)}\}\Big)\,
\label{Z4inter}
\\
  && \times
  \det\Big(\tD_{inv,n}+\tJ\cdot\tcs^{(n)}\Big)\;
  \det^{1/4}\bigg[1 + t\D_n\Big(D_{inv,n}+\tJ\cdot\cs^{(n)}\Big)^{-1} \bigg]\,.
\nonumber
\end{eqnarray}
Normalized varieties of all partition functions
are defined in analogy with Eq.~(\ref{ZJnorm}), \eg
\begin{equation}
  \czr_{inv,n}(a_c;\tJ\,) = \Zroot_{inv,n}(a_c;\tJ\,)/\Zroot_{inv,n}(a_c;0) \,.
\label{Z4Jnorm}
\end{equation}
I will assume that the continuum limit of the
(sequence of) reweighted theories exists:
\begin{equation}
  \czr_{inv,\infty}(a_c;\tJ\,)
  =  \lim_{n\to\infty}  \czr_{inv,n}(a_c;\tJ\,) \,.
\label{Z4lim}
\end{equation}
As usual, this is based on the scaling of the
coupling constant itself, which, in turn, is only negligibly
affected by reweighting (cf.\ Eq.~(\ref{ginv})).

We are now ready to reconstruct the observables of the fourth-root staggered
theory from those of the reweighted theory.  To this end I use that,
on the reweighted ensemble,
\begin{equation}
  \exp\left[{1\over 4}\tr\log\left(1 + t\D_n D_{inv,n}^{-1} \right)\right]
  = 1 + O\left((t\e_{inv,n})^2\right)\,.
\label{trlog4}
\end{equation}
The definition of $\e_{inv,n}$ is the same as in Eq.~(\ref{rebound}),
except that this is now in the context of the fourth-root theory of course.
The similarity between Eqs.~(\ref{trlog}) and (\ref{trlog4}) is clear.
It follows that
\begin{equation}
  \Zroot_{inter,n}(t,a_c;\tJ\,)
  =  \Zroot_{inv,n}(a_c;\tJ\,)
  \left[ 1 + O\left((t\e_{inv,n})^2\right)\right] \,.
\label{resum4}
\end{equation}
Because $m_r(a_c)$ scales logarithmically while
$\D_n$ is suppressed by a power of the fine-lattice cutoff
(cf.\ Eq.~(\ref{irr4scale})), it is guaranteed that, for $n$ above a certain value,
we will have $\e_{inv,n}<1$ and, with it, convergence of the $t$-expansion
in Eq.~(\ref{trlog4}).  Once again, $\e_{inv,n}\to 0$
for $n\to\infty$, and the continuum limit is independent of $t$,
\begin{equation}
  \czr_\infty(a_c;J)
  = \czr_{inter,\infty}(t,a_c;J)
  = \czr_{inv,\infty}(a_c;J)\,.
\label{universal4}
\end{equation}

We now recall that the reweighted theories are local on the coarse-lattice
scale, as can be seen from the path integral representation (\ref{Z4q}),
and that they belong to the correct universality class.
In the limit $n\to\infty$ we thus have
\begin{equation}
  \czr_{inv,\infty}(a_c;\tJ\,)
  = \int \cd\cv \cd q \cd\bq\;
  \exp\left[-S^{root}_{inv,\infty}(a_c;\cv,q,\bq)
  - \bq \tJ\cdot\tcs(a_c;\cv)\, q\right] \,,
\label{SZ4lim}
\end{equation}
where the limiting action
\begin{equation}
  S^{root}_{inv,\infty}(a_c;\cv,q,\bq\,)
  =  \lim_{n\to\infty} S^{root}_{inv,n}(a_c;\cv,q,\bq\,)\,,
\label{S4lim}
\end{equation}
is local too.
But, by Eq.~(\ref{universal4}), $\czr_{inv,\infty}(a_c;\tJ\,)$
accounts for the continuum-limit
observables of the (blocked) fourth-root theory as well.
This establishes the validity of the continuum limit of the
fourth-root theory.\footnote{
  As explained in Sec.~\ref{rewpt}, $S^{root}_{inv,\infty}$ is a
  ``perfect'' action.
}

I conclude with two additional observations.
My first comment concerns the physical consequences of the
continuum-limit taste symmetry of the fourth-root theory.
One can lift the restriction on the sources and consider (meson)
observables with a general taste structure. The reweighted partition
function then takes the form
\begin{equation}
  \Zroot_{inv,n}(a_c;J)
  =
  \int \cd\cu
  \prod_{k=0}^{n} \Big[ \cd\cv^{(k)} \Big]\;
  \bltz_n\Big(\quart;\cu,\{\cv^{(k)}\}\Big)\,
  \det^{1/4}\Big(\tD_{inv,n}\otimes\id + J\cdot\cs^{(n)}\Big) \,.
\label{Z4invns}
\end{equation}
Taste-$SU(4)$ Ward-Takahashi identities may now be derived by
varying $J$.  These identities will be exact in the reweighted theory
for every $n$, and will be true up to $O(\e_{inv,n}^2)$ corrections
in the blocked fourth-root theory.  In the $n\to\infty$ limit,
these identities will become exact in the fourth-root theory as well.
If, however, we reinstate the restriction to taste-singlet observables,
then Eq.~(\ref{Z4invns}) evidently reduces to Eq.~(\ref{Z4inv}).
This means that no paradoxes can be derived based on
the taste symmetry of the continuum-limit fourth-root theory
(as claimed in Ref.~\cite{Mike}).  The taste non-singlet states live in
an extended, non-unitary Hilbert space; but a unitary, physical
sub-space exists.  A more practical conclusion concerns deciding
when one is allowed to use taste non-singlet operators (such as those in
Eq.~(\ref{Gpion})), which are often advantageous numerically,
instead of taste-singlets ones.  For a detailed discussion of these issues,
see Refs.~\cite{BGSS,Steve,CMY}.

Another observation is that, as I have
assumed in Sec.~\ref{plan}, $\det(D_{inv,n})$ and $\det(\tD_{inv,n})$
will both be strictly positive when $\e_n,\e_{inv,n}<1$.
I begin by re-writing the blocked staggered determinant as
\begin{equation}
  \det(D_n) = \det(D_{inv,n}) \,
  \det\Big(1 + \D_n D_{inv,n}^{-1} \Big) \,.
\label{invrest}
\end{equation}
Because $[\g_5\otimes \x_5]\,D_n$ is hermitian \cite{BGS}, and
$D_{inv,n}=\tD_{inv,n}\otimes {\bf 1}$ accounts for its taste-invariant part,
it follows that $\g_5\tD_{inv,n}$ is hermitian too.
Therefore, $\det(D_{inv,n})$ and $\det(\tD_{inv,n})$ are real.
Moreover, by my assumption, the expansion of the determinant in Eq.~(\ref{invrest})
is convergent (on both the staggered and reweighted ensembles), and the
rightmost determinant in Eq.~(\ref{invrest}) is thus strictly positive.
Since $\det(D_n)$ is strictly positive too (see Sec.~\ref{conv}),
it follows that $\det(D_{inv,n})$ is strictly positive.
Next I consider $\det(\tD_{inv,n})$.
Because $\det^4(\tD_{inv,n})=\det(D_{inv,n})$, we know that
$\det(\tD_{inv,n})$ cannot be zero and cannot flip sign.
By considering the limit where the bare mass of the original staggered
theory goes to infinity,
it follows that $\det(\tD_{inv,n})$ is strictly positive.

Finally I should note that it is quite certain that
the bounds I have made use of in this section and in Sec.~\ref{unroot}
must represent over-estimations. I return to this point in Sec.~\ref{conclusion}.

\subsection{\label{ns} Scaling in the reweighted theories revisited}
A key result of this paper is that the fermion sector
of the fourth-root theory becomes amenable to a scaling analysis
by means of the multi-gauge-field representation of the blocked theory.
As explained in Sec.~\ref{rewpt}, the scaling analysis
in the fourth-root theory can be carried out by reducing
it to a calculation in ordinary staggered perturbation theory
augmented by the replica trick.

Interestingly, in the reweighted fourth-root theory, the needed scaling laws
may be found without making any reference
to the replica trick in staggered perturbation theory.
According to this alternative route,
the calculation of the scaling of $\D_n$
(which is still done as described in Sec.~\ref{rewpt})
proceeds by first considering
only reweighted theories with $n_s=4n_r$ quark species,
where $n_s$ is a multiple of four,
and therefore $n_r$ is integer.  This means that the complete
calculation, including the part done on the staggered-theory side,
involves local theories only.  The scaling of $\D_n$ for any other
number of quark species $n_s$ in the reweighted theory
can now be found without any further reference to the staggered theory.
We simply analytically continue the previous result to the desired value
of $n_s$.  As usual, because the $n_s$ dependence
is known in closed form, the analytic continuation is uniquely determined.
Unlike in the staggered theory, however, this analytic continuation
only relates local (reweighted) theories to other local (reweighted) theories!
In particular, the scaling in the one-taste reweighted theory is inferred from
the scaling in reweighted theories
where the number of quark species is a multiple of four,
without ever having to perform a scaling analysis in the fourth-root theory.

Thus,
this line of argument relates the needed scaling properties of the reweighted
fourth-root theory to the local, ordinary staggered theory,
while passing only through local theories at intermediate steps.
The fourth-root theory is then encountered only at the very
last stage, where we reconstruct it from the reweighted theory,
as was done in Sec.~\ref{cl}.

I comment in passing that, ``forgetting'' where they came from,
the reweighted theories $Z_{inv,n}$ or $Z_{inv,n}^{root}$ each constitute
a family of local theories defined on a lattice with spacing $a_c$,
which depend on an additional parameter $n$.  The role of this parameter
is similar to the fifth dimension $L_5$ of domain-wall fermions:
when either $n$ or $L_5$ are sent to infinity, a GWL chiral symmetry is
recovered.\footnote{
  Of course, in a one-flavor theory, a GWL symmetry exists only in the
  free theory.
}
The actual construction of the reweighted theories would amount to a gross
``overkill,'' if our only aim was to find solutions of the GW relation.
The merit of the construction is that the same local operator, $\D_n$,
controls both the violations of the GWL chiral symmetry (that originates
from the staggered $U(1)_\e$ symmetry) in the reweighted theory,
and the deviations of the latter
from the corresponding staggered theory.

\section{\label{conclusion} Conclusion}
Like a journey through a dark wood, when dealing with a difficult problem
in quantum field theory one can never be too sure which is the right way,
and where danger is lurking.
I have concluded that the fourth-root recipe is valid
in the continuum limit using plausible assumptions.
Plausibility is, at the end of the day, in the eye of the beholder.
In this concluding section, I give my personal perspective
on what has been gained.

In a way, this paper trades one set of questions for another.
But while at the starting point the questions were rather vague,
the new questions are focused, technical,
and testable.  The initial worries basically stem from our lack of
experience with non-local theories.
A formal expansion of the staggered action suggests that the taste-breaking
terms are irrelevant operators, that would naively be expected to
vanish in the continuum limit. But it is unclear how to perform
a scaling analysis when there is no local fermion action in the first place.
Related, one must also translate the (tentative)
claim ``locality is recovered in the continuum limit''
into a well-defined statement.

This paper offers a solution to these problems.  By first RG-blocking
the (fourth-root) staggered theory and then enforcing exact taste symmetry
by reweighting, we obtain local coarse-lattice theories in
the desired universality class, which provide a good approximation
of the (fourth-root) staggered theory once the number of blocking
steps is large enough.  That the reweighted and staggered theories are indeed
close to each other, follows from a scaling analysis,
which, in the fourth-root case, is made possible
by the multi-gauge-field representation of the blocked theory
introduced in Sec.~\ref{overview}.  Within this representation,
the taste-breaking effects all arise from the taste-breaking
part $\D_n$ of the local, blocked Dirac operator $D_n$.

The reasoning of this paper has been presented early on in Sec.~\ref{plan}
and I now recapitulate it:
All-orders renormalizability of the reweighted theories follows from that of
the (ordinary and fourth-root) staggered theories (Sec.~\ref{pcr});
making mild use of renormalizability to establish the existence of
a weak-coupling regime, a robust non-perturbative consideration shows that
the reweighted theories are local (Sec.~\ref{lcl});
the scaling of the local operator $\D_n$, which embodies all
the taste violations in the blocked theory, can be traced back to a
calculation in ordinary staggered perturbation theory
(augmented by the replica trick in the case of the fourth-root theory),
and the result is that $\D_n$ indeed scales
as an irrelevant operator (Sec.~\ref{rewpt});
the smallness of $\D_n$ on the reweighted ensemble enables the
reconstruction of the staggered theory from the reweighted theory
by means of a convergent expansion (Sec.~\ref{rooted});
in the continuum limit, the difference between the (blocked) staggered theory
and the reweighted theory, which is already known to be in the
correct universality class, vanishes.
For the ordinary, local staggered theory this implies that exact taste
symmetry has been recovered; for the fourth-root theory, this
implies that it has become local.  Thus the fourth-root theory provides a
valid regularization of QCD.

This conclusion depends on confirming the key properties of
the reweighted theories. This amounts to verifying their locality,
checking the actual predictions of their perturbation theory,
as well as testing these predictions non-perturbatively (by numerical methods).
A summary of what each of the above amounts to has been
given in Sec.~\ref{rewcap}.

A detailed-level comprehensive study of all the properties of the
reweighted theories would be a major endeavor.  Nevertheless, already now
there is good reason to believe that the fourth-root theory
is indeed a valid regularization of QCD.
This conclusion derives from the comparison to the local
four-taste staggered theory.
In short, our understanding of the local staggered theory
is on essentially the same footing as with any other local lattice fermion
method.  If the continuum limit of the four-taste staggered theory
is what we think it is, then it is difficult to see how the claimed properties
of the reweighted theories derived from it could go wrong.
But then, the arguments for the key properties of the reweighted theories
apply, basically unchanged, to the four-taste (derived from local staggered)
and one-taste (derived from fourth-root staggered)
cases, both of which constitute local theories.
Thus, it is also difficult to see how any key property
of a reweighted theory could go wrong in the one-taste case,
if this does not happen in the four-taste case.

By following this line of argument one can in fact
avoid any reference to the replica trick -- which is
the manifestation of the non-locality in staggered perturbation theory.
Instead, one first derives the scaling laws for reweighted theories
where the number of quark species is a multiple of four
(and, thus, the original staggered theory is local).  From this,
one infers the scaling laws
for reweighted theories with any other number of quark species (Sec.~\ref{ns}).
Thus, any reference to the non-local fourth-root theory is avoided
until the very last step where it is reconstructed from the reweighted
theory.

Taste-breaking effects in the spectrum of the
staggered Dirac operator are largest at the (fine-lattice) cutoff scale.
But the largest taste-breaking effects are \textit{not} a major source
of non-locality; in fact they entail basically no non-locality, because
RG blocking trades all ultra-violet fermion modes
with a local correction to the gauge-field action.
This observation is nothing but a (part of the standard) description
of how symmetries broken by the lattice regulator are recovered in
the continuum limit.
A key result is that this feature is not lost by the fourth-root theory.

The remaining non-local effects have been argued to be associated
with the dimensionless, small parameter $a_f\L$ \cite{BGS,CMY,CB4f,SP}.
It follows from the results of this paper that \textit{all}
the non-local effects should be controlled by (powers of) $a_f\L$.
In Sec.~\ref{unroot} and Sec.~\ref{rooted},
I have bounded the relative size of taste-breaking effects
in long-distance observables,
hence also the relative size of non-local effects,
by powers of $a_f/(a_c^2\, m_r(a_c))$.
But, because the coarse-lattice spacing $a_c$ is
basically arbitrary (apart from the restriction $a_c \ll \L^{-1}$),
this has got to be an over-estimation.
In all likelihood, the actual relative size of the taste-breaking
and the non-local effects is on the order of $a_f\L^2/m_{phys}$
(or powers thereof),
where $m_{phys}$ is the renormalized quark mass extract from
some low-energy observable.  This is based on the anticipation that,
on low-energy modes of the staggered Dirac operator, the actual magnitude
of taste-splittings among quartets of eigenvalues should scale like $a_f\L^2$.
For related theoretical discussions, see Refs.~\cite{BGS,BGSS,mu}.

In numerical simulations, the taste-symmetry violations are observed to
decrease rapidly as $a_f$ is decreased, and indeed to be roughly proportional
to $(\alpha_s(a_f)\, a_f \L)^2$ (with ``improved'' staggered quarks)
\cite{milc}.  The presence of a fixed physical scale,
and not some $a_c^{-1} \gg \L$, makes it
possible to extrapolate to the continuum limit
using present-day computer resources.

This work lends strong support to the physical picture advocated in
Refs.~\cite{CB4f,BGSS,CMY}:  The non-localities of the fourth-root theory
can be interpreted in terms of an extended Hilbert space containing
states with, in general, non-zero taste charges.  The physical subspace
consists of the taste-singlet states.  The exact taste symmetry, recovered
in the continuum limit, relates physical and unphysical states,
and its Ward-Takahashi identities play a crucial role in
establishing unitarity in the physical subspace.

It is interesting to consider a closely related problem,
namely the use of the fourth-root recipe for finite-density simulations
(see Ref.~\cite{mu} and references therein).
Here there is a new, three-fold difficulty.
First, there are all the general difficulties having to do with a
complex measure, that set in when $\Re(\m) \ne 0$,
where $\m$ is the chemical potential.  Second, when trying to apply
the fourth-root recipe one confronts phase ambiguities,
and the systematic error they introduce must be kept under control.\footnote{
  This difficulty never appears when  $\Re(\m)=0$.
  When the quark mass is strictly positive
  the staggered determinant is strictly
  positive too, and the positive, analytic fourth root can always be chosen.
}
Last, a non-zero quark mass is no longer an effective infra-red cutoff.
The eigenvalues reach the origin in the complex plane for
realistic values of the chemical potential.
Even in this case, it has been argued in Ref.~\cite{mu}
that everything is {\it in principle} under control,
provided that the continuum limit is taken before the thermodynamical
limit.  The crucial grouping of eigenvalues into quartets --
near the origin in the complex plane and beyond -- can still be done
when one is close enough to the continuum limit,
if the volume (in physical units) is finite. However,
the systematic error due to the phase ambiguities is
parametrically much larger, and grows with a positive power of the volume.
For more details, see Ref.~\cite{mu}.

Returning to zero density,
the up and down quark masses used in numerical
simulations are larger than their physical values.
Extrapolation of numerical results to the physical point
requires the appropriate low-energy effective theory.
For the development of staggered chiral perturbation theory (\scpt)
see Refs.~\cite{PQ,LS,AB}.  Recently, based on plausible assumptions within
the context of the chiral effective theory,
it has been argued that \scpt\
augmented by the replica trick
is indeed the correct low-energy description
of the pion sector of the fourth-root theory \cite{CB4f}.

It will be interesting to re-derive \scpt\
with the replica trick directly from
the underlying theory, the (RG-blocked) fourth-root theory.
The difficulty is that the effective theory depends on the
number of replicas $n_r$ both explicitly, as well as implicitly
through the $n_r$-dependence of its low-energy constants.
Normally, the dependence of low-energy constants on the parameters
of the underlying theory is non-perturbative, and is not known.
The challenge is to cast the (RG-blocked) underlying theory into a new form
where the necessary analytic continuation in the number of
fermion species of some type can be done in a closed form.
Work on this subject is in progress \cite{hybrid}.

\vspace{5ex}
\noindent {\bf Acknowledgements}
\vspace{3ex}

I am very grateful to Maarten Golterman for extensive discussions
and for his numerous insightful comments.  I thank Claude Bernard,
Steve Sharpe, and Ben Svetitsky for discussions, comments, and questions.
I also wish to mention discussions I have had with Aharon Casher
on the concept of renormalization-group blocking.
I thank the organizers of the \textit{Workshop on the Fourth Root of the
Staggered Fermion Determinant} for creating a timely opportunity
to discuss the issue; that workshop was held in the
Institute for Nuclear Theory at the University of Washington, Seattle,
which I thank for hospitality.
The current revision (hep-lat/0607007v3) represents significant progress
in the understanding of scaling issues, as well as of the role
of the local, four-taste staggered theory as an ``anchor.''
This progress was gained through extensive discussions with Claude Bernard,
Maarten Golterman and Steve Sharpe during and after the annual
lattice meeting, \textit{LATTICE 2006}, Tucson.
This research is
supported by the Israel Science Foundation under grant no.~173/05.

\appendix

\section{\label{tasteinv} The fermion blocking kernels}
In this appendix I describe the fermion blocking kernels in some
more detail.  The transition to a taste representation in
the special $k=0$ step is discussed in App.~\ref{tst1}.
The blocking kernels of all subsequent steps are introduced
in App.~\ref{tst2}.  In App.~\ref{tst3}, I prove the positivity
of $\det(D_n)$ and  $\det(G_n^{-1})$.

\subsection{\label{tst1} Taste representation in the interacting theory}
In the case of free staggered fermions there is a unitary transformation
between the one-component field $\c(x)$ and the taste-basis field
$\j^{(0)}_{\a i}(\tx^{(0)})$, given explicitly by \cite{taste,saclay}
\begin{equation}
  \j^{(0)}_{\a i}(\tx^{(0)})
  = \sum_{r_\m=0,1}
  (\g_1^{r_1} \g_2^{r_2} \g_3^{r_3} \g_4^{r_4})_{\a i}\, \c(2\tx^{(0)}+r)\,.
\label{Gm}
\end{equation}
For notation see Sec.~\ref{blockn}.
Writing Eq.~(\ref{Gm}) compactly as $\j = \G\,\c$,
the ``conjugate'' Grassmann variables
are related by $\bj = \bc\, \G^\dagger$.
Recall that the fine-lattice spacing $a_f$
of the one-component field, and the lattice spacing $a_0$ of
the taste-basis field, are related via $a_0=2a_f$.
Equation (\ref{Gm}) makes use of the embedding
of the taste-basis lattice into the fine lattice,
and the fact that each fine-lattice
site has a unique representation as $x_\m=2\tx^{(0)}_\m+r_\m$,
where $r_\m=0,1$.

In a free theory, RG blocking normally works by suppressing modes with
a lattice-scale momentum.  It is therefore natural to apply the blocking in
the taste basis \cite{rg}, where all the long-distance physics comes from
the vicinity of the origin in the Brillouin zone.
The one-component formalism would be inconvenient\footnote{
  RG blocking of \textit{free} one-component staggered fermions
  was discussed in Ref.~\cite{RGonecomp}.
}
because the long-distance
physics comes from all sixteen ``corners'' of the Brillouin zone.

Unlike the free theory,
an equal choice between a one-component basis and a taste basis
does not exist in the interacting theory.
Lattice QCD with staggered fermions must be defined in the one-component
formalism.  According to power-counting arguments
and explicit one-loop calculations,
only this formalism has enough symmetry to ensure the
multiplicative renormalization of the staggered-fermion mass term
and the recovery of full rotation and taste symmetries
in the continuum limit \cite{mgjs,MW}.
This state of affairs poses a difficulty for the RG program.
The question is how to accommodate
all the symmetries of the standard one-component formalism
in a taste-basis representation that will, in turn, provide the starting point
for the succession of RG-blocking steps.

The unitary transformation from the one-component basis to the taste basis
can be promoted to a gauge-covariant one,
\begin{eqnarray}
  \j^{(0)}_{\a i}(\tx^{(0)})
  &=&
  Q^{(0)}_{\a i}(\tx^{(0)}) \c
\label{tstQ}
\\
  &\equiv&
  \sum_{r_\m=0,1}
  (\g_1^{r_1} \g_2^{r_2} \g_3^{r_3} \g_4^{r_4})_{\a i}\,
  \cw\Big(2\tx^{(0)},2\tx^{(0)}+r;\cu\Big)\,
  \c(2\tx^{(0)}+r)\,.
\nonumber
\end{eqnarray}
Setting $x_0=2\tx^{(0)}$ for short,
an explicit choice for the parallel transporter is \cite{saclay}
\begin{equation}
  \cw\Big(x_0,x_0+r;\cu\Big)
  =
  U_{1,x_0}^{r_1}\,
  U_{2,x_0+\hat{1}r_1}^{r_2}\,
  U_{3,x_0+\hat{1}r_1+\hat{2}r_2}^{r_3}\,
  U_{4,x_0+\hat{1}r_1+\hat{2}r_2+\hat{3}r_3}^{r_4}\,,
\label{tstQfix}
\end{equation}
where $\hat\m$ is the fine-lattice unit vector in the $\m$ direction.
With the notation of Eq.~(\ref{tstQ}) we similarly have
$\bj^{(0)}_{\a i}(\tx^{(0)}) = \bc\, Q^{(0)\dagger}_{\a i}(\tx^{(0)})$,
where hermitian conjugation applies to the color matrices.
Parallel transporting the fine-lattice variables entails
well-defined transformation properties for the taste-basis variables
under fine-lattice gauge transformations. Thus, gauge invariance
is maintained by the blocking transformation.

The covariant blocking kernel (\ref{tstQ})
illustrates, however, an inherent problem.  Any concrete choice of
the gauge-covariant blocking kernel will transform
non-trivially under hypercubic rotations.  In Eq.~(\ref{tstQ}) this is seen
both in the special role of the hypercube's site with relative coordinates
$r_\m=0$, because these relative coordinates transform non-trivially
under hypercubic rotations; and, for a similar reason,
in the specific ordering of traversing the axes in Eq.~(\ref{tstQfix}).

The solution adopted in this paper is to perform the transition from the
original one-component formalism to a taste representation as a gaussian
RG-like transformation, in which no thinning out of the fermionic
degrees of freedom (but only of the gauge field) occurs.
For the taste-basis Dirac operator resulting from this transformation,
see Ref.~\cite{BGS}.
This comprises the special $k=0$ blocking step introduced in Sec.~\ref{blockn}.
Within the gaussian blocking transformation,
$\j(\tx)$ is loosely equal to $Q^{(0)}(\tx^{(0)};\cu)\, \c$,
and
$\bj(\tx)$ is loosely equal to $\bc\, Q^{(0)\dagger}(\tx^{(0)};\cu)$.
For a precise statement, see App.~\ref{pullback}.

Of course, replacing the unitary change of variables (\ref{tstQ})
by a gaussian transformation does not by itself solve the difficulty
with hypercubic rotations. But we may now overcome it by creating a coherent
superposition over a family of different blocking transformations.
This is explained in App.~\ref{disorder} (see also App.~\ref{sym}).

\subsection{\label{tst2} Fermion blocking kernels for $1\le k \le n$}
For completeness, let me specify the fermion blocking kernels of the
subsequent, $1\le k \le n$ blocking steps. In the free theory I take
\begin{equation}
   Q^{(k)}(\tx^{(k)})\, \j^{(k-1)}
  = {1\over 16}\sum_{r_\m=0,1} \j^{(k-1)}(2\tx^{(k)}+r)\,.
\label{freeQ}
\end{equation}
In analogy with Eq.~(\ref{tstQ}) we may define a covariant version,
\begin{equation}
  Q^{(k)}(\tx^{(k)};\cv^{(k-1)})\, \j^{(k-1)}
  = {z^{(k)}\over 16}
  \sum_{r_\m=0,1} \cw\Big(2\tx^{(k)},2\tx^{(k)}+r;\cv^{(k-1)}\Big)\,
  \j^{(k-1)}(2\tx^{(k)}+r)\,,
\label{covQ}
\end{equation}
where
the parallel transporters are defined
analogously to Eq.~(\ref{tstQfix}), but now in terms of the blocked gauge field
of the $(k-1)^{\rm th}$ lattice.
These definitions imply that the linear transformation
$Q^{(0)}$ is unitary, whereas for $1\le k \le n$, the product
$Q^{(k)}Q^{(k)\dagger}$ is equal to $(z^{(k)})^2/16$ times the identity
matrix on the $k^{\rm th}$ lattice.  The difficulty with hypercubic
symmetry recurs at every blocking step, and again it is solved in
a similar manner (see App.~\ref{disorder}).

The constants $z^{(k)}$ are adjusted to impose a wave-function
renormalization condition on the fermion fields at each blocking level.
Usually, lattice renormalization produces factors of $\log(a\m)$
where $\m$ is the renormalization scale.  But in an RG-blocking setup
one has $a \to a_{k-1}$, $\m \to 1/a_k$ at the $k^{\rm th}$ blocking step.
Whence $\log(a\m) \to \log(2)$, and
we may expect $z^{(k)}=1 + c_k\, g_r^2(a_k)/(16\p^2)$, where $c_k=O(1)$.
Of course, the product of all the $z^{(k)}$'s can diverge (or vanish)
in the limit $n\to\infty$, as dictated by the integrated anomalous
dimension of the fermion field.

The expectation value of any product of local composite operators constructed
from the coarse-lattice fields will always be finite.
Therefore, composite operators do not necessarily require a separate
renormalization.  (Once again this can be explained by the fact that
the ratio of the cutoff and renormalization scales is a finite, fixed number.)
One might, however, opt to impose specific renormalization conditions
for certain composite operators.  A renormalization condition
imposed on a composite operator at the coarse-lattice scale
will in general entail some finite renormalization.

\subsection{\label{tst3} Positivity of $\det(D_n)$ and $\det(G_n^{-1})$}
Here I prove that $\det(D_n)$ and  $\det(G_n^{-1})$
are positive for $m>0$.
In more detail, I will prove that, like $\det(D_{stag})$,
also $\det(D_n)$ is real and strictly positive for $m>0$.
It follows from Eq.~(\ref{fctrn}) that $\det(G_n^{-1})$ is real positive
(for the issue of zero eigenvalues of $G_n^{-1}$, see Sec.~\ref{lcl}).

I begin by noting \cite{BGS} that
$[\g_5\otimes \x_5] D_n [\g_5\otimes \x_5] = D_n^\dagger$.\footnote{
  The transformation (\ref{Gm}) implies $\x_\m=\g_\m^T$
  if the Dirac and the taste matrices both act from the left.
\label{transposeG}
}
It follows that $[\g_5\otimes \x_5] D_n$ is hermitian, and $\det(D_n)$
is real.  Moreover, complex eigenvalues of $D_n$ must occur in pairs
with conjugate values. Therefore $\det(D_n)$
will be (strictly) positive if all the real eigenvalues are
(strictly) positive.\footnote{
  If $D_n$ has no real eigenvalues, the strict positivity
  of $\det(D_n)$ follows trivially.
}

By ``undoing'' the all the blocking steps one can express
the blocked propagator as
\begin{equation}
  D_n^{-1}
  = R_n + Q_n \, D_{stag}^{-1} \, Q_n^\dagger \,,
\label{DQn}
\end{equation}
where $Q_n = Q^{(n)} Q^{(n-1)} \cdots Q^{(0)}$, and
$R_n>0$ is determined iteratively from $R_0=\a_0^{-1}$
and $R_k = \a_k^{-1} + ((z^{(k)})^2/16) R_{k-1} $
(for the free theory, see Eq.~(\ref{Rn})).
It is now straightforward to show that, if $\J$ is an eigenstate of $D_n$
with real eigenvalue $\l$, then
\begin{equation}
  {1\over \l}
  = R_n + \J^\dagger Q_n \, {m\over -D_{msls}^2+m^2} \, Q_n^\dagger \J\,.
\label{oneoverl}
\end{equation}
I have used that $D_{stag}=D_{msls}+m$, where the massless staggered
operator $D_{msls}$ is anti-hermitian.
For $m>0$, it follows that $R_n \le \l^{-1} < \infty$.
Hence all the real eigenvalues of $D_n$ are (finite and) strictly positive
for $m>0$.

\section{\label{pullback} More details on the pull-back mapping}
Here I discuss in more detail the pull-back mapping
introduced in Sec.~\ref{blockn}.
First, considering the original as well as all the blocked gauge fields
as a fixed background, let us discuss the pull-back mapping $\ct^{(j,n)}_F$ of
the fermions only.  In analogy with Eq.~(\ref{pull}) it is defined for
$-1 \le j\le n-1$ by
\begin{equation}
  \ct^{(j,n)}_F \co^{(n)}
  =
  \int \prod_{k=j+1}^{n} \Big[ \cd\j^{(k)} \cd\bj^{(k)} \Big]\;
  \exp\Big[-\sum_{k=j+1}^{n} \ck_{F}^{(k)}  \Big]\;
  \co^{(n)}\,.
\label{pullF}
\end{equation}
As an example, consider the action of $\ct_F^{(n-1,n)}$ on
a fermion bilinear. Using Eq.~(\ref{pullF}) one has
\begin{equation}
  \ct_F^{(n-1,n)} \Big[\j^{(n)}(\tx^{(n)})\;\; \bj^{(n)}(\ty^{(n)}) \Big]
  =  {\d_{\tx^{(n)},\ty^{(n)}} \over \a_n} +
     \Big[ Q^{(n)}(\tx^{(n)})\, \j^{(n-1)} \Big] \;
     \Big[ \bj^{(n-1)}\, Q^{(n)\dagger}(\ty^{(n)}) \Big] \,.
\label{mappf}
\end{equation}
The resemblance to Eq.~(\ref{Dnb}) is evident.
Notice that for $\tx^{(n)}\ne\ty^{(n)}$ there is no contact term.
This generalizes to the product of any number of fermion
and anti-fermion fields.
Thus,
the fermion pull-back mapping realizes the operator relation
\begin{subequations}
\label{Opr}
\begin{eqnarray}
  \ct_F^{(n-1,n)}\, \j^{(n)}(\tx^{(n)})
  &\simeq& Q^{(n)}(\tx^{(n)})\, \j^{(n-1)} \,,
\label{Opra}
\\
  \ct_F^{(n-1,n)}\, \bj^{(n)}(\tx^{(n)})
  &\simeq& \bj^{(n-1)}\, Q^{(n)\dagger}(\tx^{(n)}) \,,
\label{Oprb}
\end{eqnarray}
\end{subequations}
where the right-hand sides were defined in App.~\ref{tasteinv}, and where the
$\,\simeq\,$ sign means equality
up to the contact terms that arise when a fermion and an anti-fermion
reside on the same site of the coarse lattice.
Observe that, for the fermion kernels of App.~\ref{tasteinv},
if no fermion and anti-fermion reside on
the same site of the coarse lattice, then no contact terms
will arise under the pull-back $\ct_F^{(j,n)}$ for any $j$.

Next let us consider the action of the pull-back mapping on
the gauge fields as well.
First, a few more details on the gauge-field blocking kernels are needed.
The non-linear blocking kernel $\cb_k$ is constructed as a sum over
the links of the $k^{\rm th}$ lattice,
\begin{equation}
  \cb_k\Big(\cv^{(k)},\cv^{(k-1)}\Big)
  =
  \sum_{\m,\tx^{(k)}} \cf_k
  \Big[V^{(k)}_{\m,\tx^{(k)}},
  W^{(k)}_{\m,\tx^{(k)}}\Big(\cv^{(k-1)}\Big)\Big]\,.
\label{Bx}
\end{equation}
A simple choice, consistent with the gauge-transformation properties
of the fermion kernels (\ref{covQ}), is
\begin{subequations}
\label{BF}
\begin{eqnarray}
  \cf_k(V,W) &=& -\b_k\,\tr (V^\dagger W)\,,
\label{BFa}
\\ \rule{0ex}{3ex}
  W^{(k)}_{\m,\tx^{(k)}} &=&
  V^{(k-1)}_{\m,2\tx^{(k)}} V^{(k-1)}_{\m,2\tx^{(k)}+\hat\m} \,,
\label{BFb}
\end{eqnarray}
\end{subequations}
where $\b_k>0$ is a new blocking parameter.
Since $W\in SU(3)$, one can use the invariance of the Haar measure
to show that $\cn_k(\cv^{(k-1)})$ in Eq.~(\ref{KBk})
reduces to a numerical constant.
Many other choices of $\cb_k$ are possible, see \eg Ref.~\cite{FM}.

Considering the defining equation (\ref{nlocala}),
the gauge-field blocking kernel $\cb_k$ may be viewed as a generalized action.
This generalized action couples each $k^{\rm th}$-lattice link
$V^{(k)}_{\m,\tx^{(k)}}$ to the gauge field on the $(k-1)^{\rm th}$ lattice,
but it does not couple directly any two $k^{\rm th}$-lattice links.
As an example, let $F(g)$ denote some function of $g\in SU(3)$.
Considering an operator of the form $F(V^{(n)}_{\m,\tx^{(n)}})\, \co^{(n)}$,
where $\co^{(n)}$ does not depend on $V^{(n)}_{\m,\tx^{(n)}}$,
it follows that
\begin{equation}
  \ct^{(n-1,n)} \Big[ F\Big(V^{(n)}_{\m,\tx^{(n)}}\Big)\, \co^{(n)} \Big]
  =
  \Big[\ct^{(n-1,n)} F\Big(V^{(n)}_{\m,\tx^{(n)}}\Big)\Big]
  \Big[\ct^{(n-1,n)}  \co^{(n)}\Big]\,.
\label{Tb}
\end{equation}
Returning to the general case,
while the explicit expression gets more complicated with every pull-back step,
the pull-back mapping is ultra-local because the blocking kernels are.
If $\co^{(n)}$ has a compact support, then
the support of $\ct^{(j,n)} \co^{(n)}$ will only slightly increase
for all $j<n$.

\section{\label{ensemble} Ensembles of blocked configurations}
Having introduced all the blocking kernels
(see App.~\ref{tasteinv} and App.~\ref{pullback}), let us discuss
the generation of ensembles of blocked gauge fields.
This is necessary, for example, for the computation of the coarse-lattice
observables numerically.
The issue is how to generate blocked
gauge-field configurations from pre-existing fine-lattice configurations.
Using Eq.~(\ref{fctrn}) in Eq.~(\ref{Znr}) for $t=1$, we get
\begin{subequations}
\label{weight}
\begin{eqnarray}
  Z_n(1,n_r,a_c) &=& \int \cd\cu
  \prod_{k=0}^{n} \Big[ \cd\cv^{(k)} \Big]\;
  \exp\Big(-S_g - \sum_{k=0}^n \ck_{B}^{(k)}\Big)\;
  \det^{n_r}(\Dstag) \qquad
\label{weighta}
\\
  &=& \int \cd\cu \exp(-S_g)\; \det^{n_r}(\Dstag) \equiv Z(n_r) \,.
\label{weightb}
\end{eqnarray}
\end{subequations}
Equation (\ref{weightb}) reminds us that, in the original
staggered theory, the Boltzmann weight
of the fine-lattice gauge field $\cu$ has the form
$\exp(-S_g)\,\det^{n_r}(\Dstag)$ with $n_r=\quart,$ $\half$, or 1.
Equation (\ref{weighta})
provides a Boltzmann weight for all the blocked gauge fields as well.
In view of Eq.~(\ref{weightb}), the process begins with
an ensemble of fine-lattice configurations generated in the usual way.
Given a ``mother'' configuration $\cu_i$ in this ensemble, one can generate
a ``daughter'' configuration
of the once-blocked gauge field $\cv^{(0)}_i$ by a new Monte-Carlo process,
by taking $\exp[-\ck^{(0)}_B(\cv^{(0)},\cu_i)]$ as a Boltzmann weight
while holding the fine-lattice gauge field $\cu_i$ fixed.
As it should, the probability to obtain the pair $\{ \cu_i,\cv^{(0)}_i \}$
is given by the product of the original (normalized) Boltzmann weight
$Z^{-1}(n_r) \exp(-S_g(\cu_i))\, \det^{n_r}(\Dstag(\cu_i))$
and the new Boltzmann weight
$\exp[-\ck^{(0)}_B(\cv^{(0)}_i,\cu_i)]$.
The process may be repeated on further blocking steps $1 \le k \le n$,
each time generating a 
$k^{\rm th}$-lattice
daughter configuration from the existing $(k-1)^{\rm th}$-lattice
daughter configuration $\cv^{(k-1)}_i$
using the Boltzmann weight
$\exp[-\ck^{(k)}_B(\cv^{(k)},\cv^{(k-1)}_i)]$.

Once an ensemble of fine-lattice configurations
and of daughter configurations for all $0 \le k \le n$ has been generated,
it can be used to calculate any observable.
The blocked-lattice fermion propagator $D_n^{-1}$ can be calculated
by repeatedly applying Eq.~(\ref{Dn}) until an expression involving the fine-lattice
propagator $\Dstag^{-1}$ is obtained.
The blocking kernel $Q^{(k)}$ is an explicit functional of $\cv^{(k-1)}$ only,
therefore it should be evaluated using a $(k-1)^{\rm th}$-lattice
daughter configuration.\footnote{
  In the limit $\b_k\to\infty$ the blocking Boltzmann weight
  $\exp[-\ck^{(k)}_B(\cv^{(k)},\cv^{(k-1)})]$ collapses to a $\d$-function,
  and the daughter configurations reduce to well-defined functionals
  of the original fine-lattice gauge field.
  In the case of Eq.~(\ref{BF}), for example, one finds that
  that $V^{(0)}_{\m,\tx^{(0)}}$ is equal to
  $U_{\m,2\tx^{(0)}} U_{\m,2\tx^{(0)}+\hat\m}$,
  and $V^{(k)}_{\m,\tx^{(k)}}$ is equal to
  $V^{(k-1)}_{\m,2\tx^{(k)}} V^{(k-1)}_{\m,2\tx^{(k)}+\hat\m}$.
}

\section{\label{sym} Lattice symmetries under the blocking transformation}
As explained in Sec.~\ref{pullb},
thanks to the pull-back mapping each coarse-lattice observable
is at the same time a fine-lattice observable; as such, it is constrained
by all the staggered-fermion symmetries of the original theory.
In a more technical sense, a given fine-lattice symmetry may or may not
survive as a manifest symmetry on the coarse lattice.  I will now discuss the
fine-lattice symmetries one by one.

Translation and gauge invariance have a prominent role,
and they are secured by construction.  With the fermion and gauge-field
blocking kernels introduced above, the blocked-lattice action $S_n$
(cf.~Eq.~(\ref{nlocalb})) retains these symmetries manifestly.\footnote{
  Obviously, the size of the translation group gets smaller
  with each blocking step.
}

The situation is more subtle with respect to hypercubic symmetry.
As explained in App.~\ref{tasteinv}, the fermion blocking kernels
transform non-trivially under $90^0$ rotations.
As a result,  with the blocking transformations
as introduced in Eq.~(\ref{nlocal}), in fact $S_n$
is not invariant under hypercubic rotations.
On a closer look, the reason can also be understood as follows.
Consider the pull-back $\ct^{(-1,n)} \co^{(n)}$ of some operator
from the coarse to the fine lattice.
Under a fine-lattice rotation, $\ct^{(-1,n)} \co^{(n)}$ transforms
in the usual way.  But, because of the non-trivial transformation
properties of the blocking kernels, the rotated fine-lattice operator
cannot be obtained as the pull-back of any coarse-lattice operator!
In other words, the observables of the coarse-lattice theory
do not constitute complete representations of the hypercubic group.

This difficulty can be solved by allowing the blocking kernels
to depend on additional degrees of freedom, or {\it disorder fields}.
Each blocking step in Eq.~(\ref{nlocal}) is promoted to a
coherent superposition of block transformations summed over
all values of the disorder fields.  The details are given in App.~\ref{disorder}.
Briefly, a disorder field allows for parallel transporting of
the fermion variables of a given $2^4$ hypercube to any of its sixteen
sites in turn. Another disorder field allows for all possible orderings
for traversing the axes.
The gauge-field blocking kernels are similarly adapted.
With the disorder fields in place,
the blocked action becomes manifestly invariant under hypercubic rotations.

Next, the $U(1)_\e$ symmetry
of the (massless) one-component formalism
turns into a Ginsparg-Wilson-L\"uscher (GWL) chiral symmetry \cite{gw,ML}
in the blocked-lattice theories.
See Refs.~\cite{rg,BGS} for a detailed discussion of both
the massless and the massive cases.
For some further observations, see App.~\ref{disorder}.

The last symmetry of the (one-flavor) staggered theory
is shift symmetry.
It is generated by four anti-commuting elements $L_\m$.
The action of $L_\m$ involves a one-unit translation in the $\m$ direction,
as well as the multiplication of $\c(x)$ and $\bc(x)$ by sign factors.
In the low energy limit, shift symmetry reduces to a discrete subgroup
of taste-$U(4)$.  The importance of shift symmetry is that,
without it, a cutoff-scale mass term that breaks the $U(4)$
taste symmetry may be induced.
This unacceptable mass term is indeed generated if one couples the
free taste-basis Dirac operator directly to a gauge field \cite{BJ},
as was found by an explicit one-loop calculation in Ref.~\cite{MW}.

In contrast, the taste-basis representation constructed
in this paper avoids this problem.  To see this, consider the
expectation value of the pulled-back fermion propagator
(compare Eq.~(\ref{DQn}))\footnote{
  We may either consider the expectation value in Eq.~(\ref{prop})
  in a fixed gauge, or replace it by a gauge invariant one, obtained
  \eg by connecting the coarse-lattice fermion and anti-fermion
  by a coarse-lattice Wilson line.
}
\begin{equation}
  G(x^{(n)},y^{(n)})
  = \svev{\ct^{(-1,n)}
  \Big[\, \j^{(n)}(x^{(n)})\ \ \bj^{(n)}(y^{(n)})\, \Big]}_{-1} \,.
\label{prop}
\end{equation}
Here, cf.~Eq.~(\ref{observe}), the subscript ``$-1$'' refers to the expectation
value of the pulled-back operator in the original staggered theory.
It is straightforward to verify that the presence of the
offensive mass term in the coarse-lattice propagator would imply
that $G(x^{(n)},y^{(n)})$
is not invariant under shift symmetry \cite{mgjs,Luo}.
This is impossible, however, because $G(x^{(n)},y^{(n)})$ is
a correlation function of the original staggered theory.

As a last exercise that nicely exhibits how the coarse-lattice
observables are constrained by the fine-lattice symmetries,
let us examine the two-point function of a coarse-lattice operator with
the quantum numbers of an
exactly massless (taste non-singlet) Goldstone pion \cite{rg},
in a {\it two}-flavor theory.
As an interpolating field we may take
\begin{subequations}
\label{Gpion}
\begin{equation}
  \p^{(n)}_{ab}(\tx^{(n)})
  = \bj^{(n)}_a(\tx^{(n)})\,[\g_5 \otimes \x_5]\,\j^{(n)}_b(\tx^{(n)})\,,
\label{Gpiona}
\end{equation}
where again $\g_5$ and $\x_5$ act on the Dirac and the taste indices
respectively,
and $a,b=1,2,$ label the staggered flavor. Then
\begin{equation}
  \svev{\p^{(n)}_{12}(\tx^{(n)})\;\p^{(n)}_{21}(\ty^{(n)})}_n
  = \svev{\cg(\tx^{(n)},\ty^{(n)})} \,,
\label{Gpionb}
\end{equation}
where
\begin{equation}
  \cg(\tx^{(n)},\ty^{(n)})
  = \tr \Big([\g_5 \otimes \x_5] D_n^{-1}(\tx^{(n)},\ty^{(n)})
      [\g_5 \otimes \x_5] D_n^{-1}(\ty^{(n)},\tx^{(n)})\Big) .
\label{Gpionc}
\end{equation}
\end{subequations}
The expectation value on the left-hand side of Eq.~(\ref{Gpionb}) is
with respect to the partition function in Eq.~(\ref{nlocal}),
while on the right-hand side it is with respect to Eq.~(\ref{weighta}).\footnote{
  In other words, the right-hand side is to be evaluated on
  an ensemble of blocked configurations, cf.~App.~\ref{ensemble}.
}
Via the pull-back mapping, the interpolating coarse-lattice fields
we use represent specific smeared sources on the fine lattice.
Now, one can show that
$\cg(\tx^{(n)},\ty^{(n)})$ is strictly positive \cite{rg,BGS}.
This rules out the possibility of
destructive interference caused by these smeared sources;
the asymptotic decay rate of the correlator
must be dictated by the lightest excitation of the original staggered
theory in that channel, the Goldstone pion. Once again, this shows that
no fermion mass terms that contradict any of the symmetries
of the original staggered theory could be generated, because such
a mass term would completely change the long-distance behavior of
this correlator.

A feature that may be confusing on first acquaintance is that
the limiting $n\to\infty$ coarse-lattice
theory is invariant only under $90^0$ rotations, and not
under continuous rotations.
This can be understood as follows. Via the pull-back mapping,
even a nominally scalar (or pseudoscalar) operator on the coarse lattice
has in effect some internal structure for its support,
that ``remembers'' the orientations of the axes of the lattice.
The lack of manifest invariance under
continuous rotations in the coarse-lattice theory is, once again, because
its observables do not constitute complete
representations of this symmetry group.\footnote{
  In the free theory,
  one can check that the operator $D_{rg}$ obtained in the $n\to\infty$
  limit (Eq.~(\ref{Drg})) indeed has only
  hypercubic rotation invariance \cite{rg}.
}
In a formal sense, the continuum-limit observables accessible
by the coarse-lattice theory form a discrete subset of the set of
``all'' continuum-limit observables.
If we would keep decreasing the coarse-lattice spacing, we expect that
the breaking of continuous rotational invariance
should go to zero like some positive power of $a_c\L$.

In more detail, consider as an example the $n\to\infty$ limit
of the pion two-point function in Eq.~(\ref{Gpionb}),
\begin{equation}
  \bar\cg(x,y;a_c)
  = \lim_{n\to\infty}
  \svev{\p^{(n)}_{12}(\tx)\;\p^{(n)}_{21}(\ty)}_n \,.
\label{sep}
\end{equation}
At large (euclidean) distances, where the correlator is dominated
by the Goldstone-pion intermediate state, one expects the factorization
\begin{equation}
  \bar\cg(x,y;a_c) \approx e^{-m_\p |x-y|}\; \cf^2(n_\m,a_c m_\p)\,.
\label{Psep}
\end{equation}
Here $|x-y|$ is the usual euclidean distance, and $n_\m$ is
the unit vector pointing in the direction of $x-y$.
Power corrections that depend on $|x-y|$ have been suppressed.
The (direction dependent!) form factor $\cf(n_\m,a_c m_\p)$
accounts for the coupling of our coarse-lattice interpolating field
to the pion intermediate state it creates.
The smeared fine-lattice operator that corresponds
(via the pull-back mapping) to the  coarse-lattice interpolating field
is manifestly invariant under $90^0$ rotations only.
As noted above, in the limit $a_c m_\p \to 0$, the form factor
$\cf(n_\m,a_c m_\p)$ should approach an ($n_\m$-independent) constant.

\section{\label{disorder} Hypercubic symmetry and disorder fields}
Here I discuss how manifest hypercubic invariance
is recovered by summing over disorder fields at each blocking step.
I will discuss mainly the $k=0$ step, which produces the transition from
the one-component basis to the taste basis.
Subsequent blocking steps work essentially in
the same way, except that they are somewhat simpler
because the subsequent blocking kernels act trivially on the Dirac and
taste indices.  Within this Appendix, I will usually
use the term ``coarse lattice'' for the taste-basis
lattice obtained via the $k=0$ blocking step,
in which case I will drop the corresponding superscript-label
of the fields and the coordinates.

\begin{figure}[thb]
\vspace*{3ex}
\begin{center}
\includegraphics*[height=5cm]{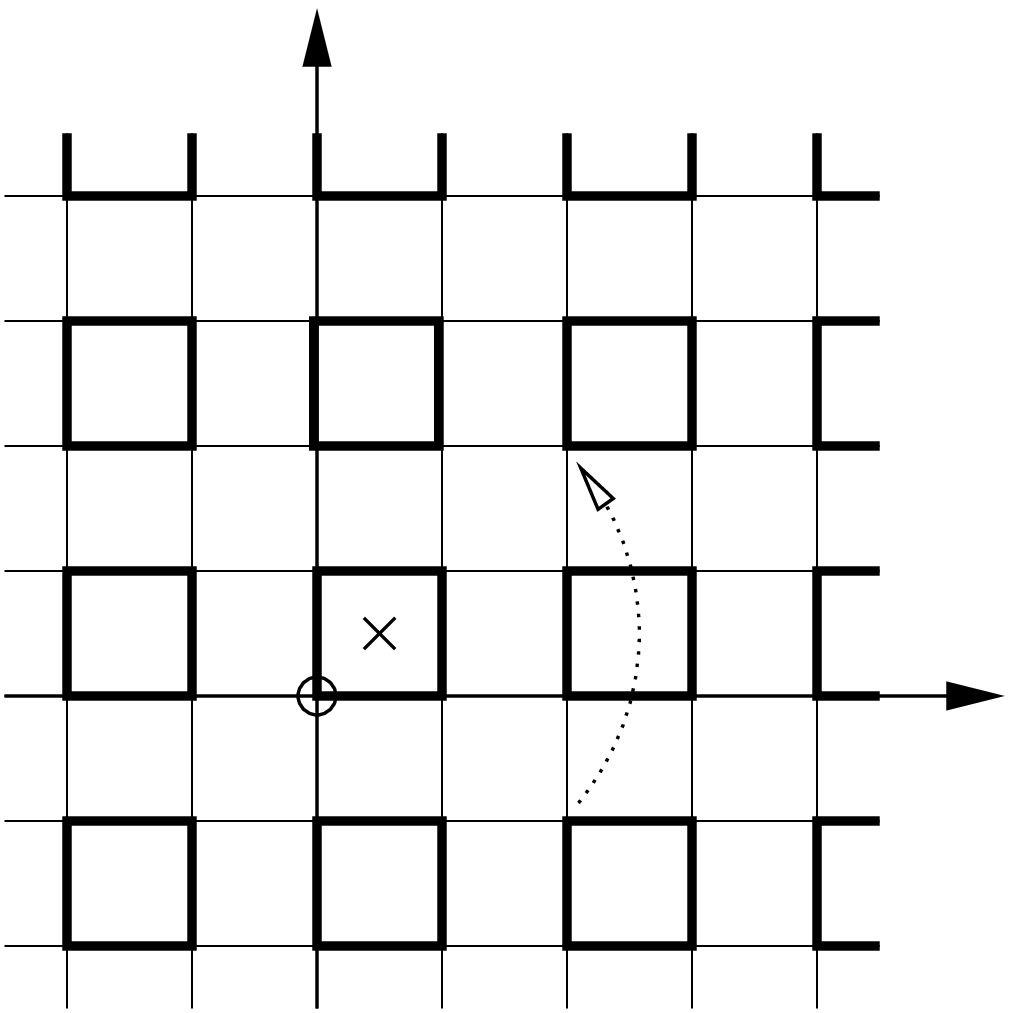}
\hspace{10ex}
\includegraphics*[height=5cm]{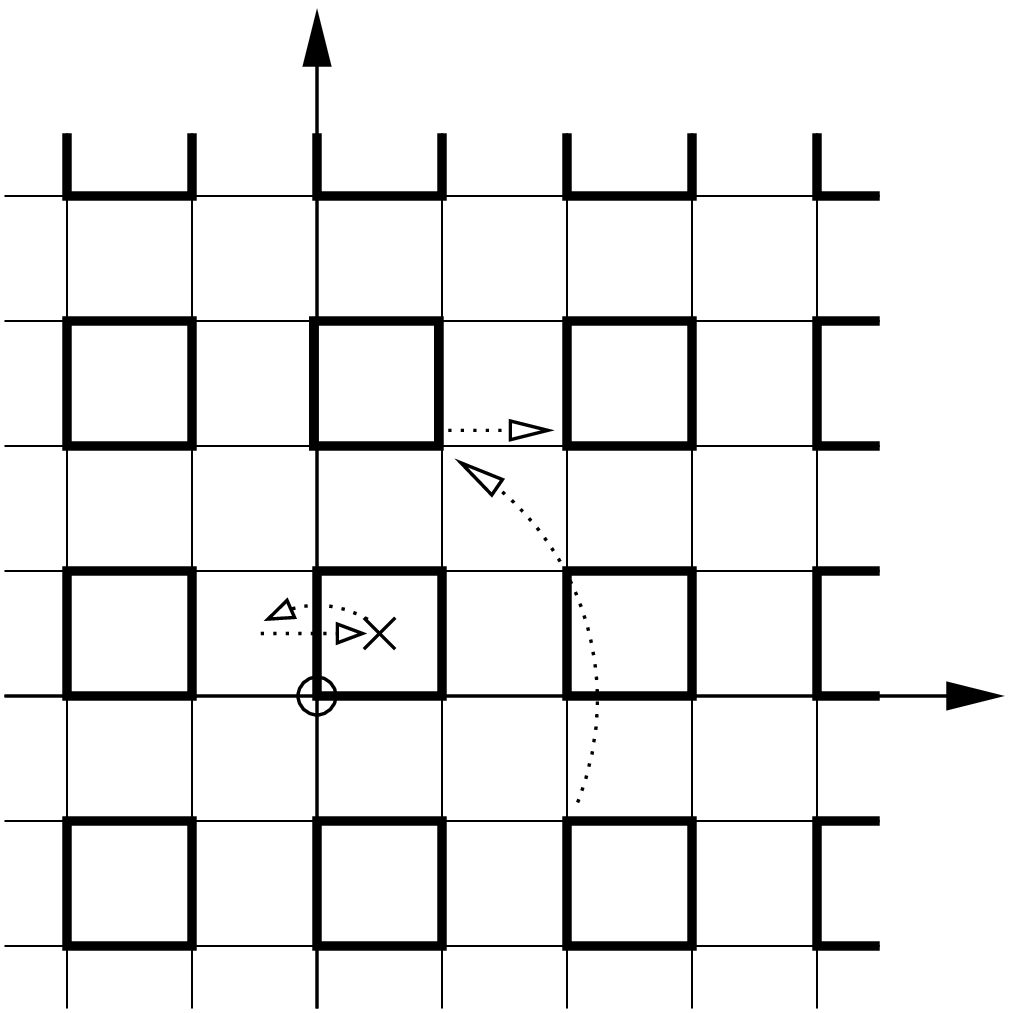}
\end{center}
\begin{quotation}
\caption{{\it Rotations.}
The two-dimensional example shows how fine- and coarse-lattice
rotations are related.
The small circle marks the origin. The point
$X=(\half,\half)$ is marked by a cross.
Thick squares show the blocking pattern.
{\it Left panel}: Counter-clockwise $90^0$
rotation about the point $X$. The blocked squares
are mapped onto themselves: centers are mapped to centers;
corners undergo a rotation with respect to
the square's center.
{\it Right panel}: The same effect is achieved by a rotation about the origin,
followed by a translation that brings the point $X$ back into its
original position.
\label{fig1}}
\end{quotation}
\vspace*{-3ex}
\end{figure}

The guiding principle is that we want to preserve the embedding of the
coarse (taste-basis) lattice into the original fine lattice,
\ie we want to maintain the same breakup of the fine lattice
into $2^4$ hypercubes.
On the coarse lattice, the $90^0$ rotation will be around the origin.
As can be seen from Fig.~\ref{fig1}, this corresponds
to a fine-lattice rotation around the origin which is either followed by,
or preceded with, a one-unit translation.  Recall that, for staggered fermions,
the fine-lattice translation group is generated by the four
anti-commuting shifts $L_\m$ that involve a one-unit translation and the
multiplication of the staggered fields by sign factors.
Let us denote by $\car$ the matrix that produces a $90^0$ rotation around
the origin.  This defines a linear, homogeneous mapping of the coordinates,
as well as of four-vectors.
The coarse- and fine-lattice coordinates rotations are given by
\begin{subequations}
\label{rot}
\begin{eqnarray}
  \tx \to \tx' &=&  \car \tx\,,
\label{rota}
\\
  x \to x' &=& \Rf(x) \equiv \car x + \D  \,,
\label{rotb}
\end{eqnarray}
\end{subequations}
where $\D$ is the one-unit translation that follows the rotation
(Fig.~\ref{fig1}).
The inverse of Eq.~(\ref{rotb}) is
\begin{equation}
  \Rf^{-1}(x') = \car^{-1}(x'-\D)  = \car^{-1}x' - \car^{-1}\D \,.
\label{invrot}
\end{equation}
Making the vector index explicit, the fine-lattice rotation is given by
$x'_\m = \car_{(\s\t)\m\n}\, x_\n + \d_{\s\m}$.
The rotation matrix
$\car_{(\s\t)\m\n} = \d_{\s\n}\d_{\t\m}-\d_{\s\m}\d_{\t\n} +P_{(\s\t)\m\n}\,$
produces the ``counter-clockwise'' rotation in the $(\s,\t)$ plane,
whereby $x'_\t=x_\s$ and $x'_\s=-x_\t$. Here
$P_{(\s\t)\m\n} = \d_{\m\n}-\d_{\s\m}\d_{\s\n}-\d_{\t\m}\d_{\t\n}$
is the projector on the $d-2$ invariant coordinates.
For the same rotation
one has $\D=\hat\s$ in Eq.~(\ref{rotb}), namely, the follow-up
translation is in the positive $\hat\s$ direction.

Space-time transformations act on fields by prescribing their value
at a point in terms of their value at the source of that point
(under the ``active'' coordinates transformation).
For a coarse-lattice rotation around the origin,
the value of the transformed staggered field at a fine-lattice point $x'$
will be determined in terms of its value at $\Rf^{-1}(x')$.
Therefore, the transformation applied to the staggered
field is first a shift from $x'$ back to $x'-\D$, and then a fine-lattice
rotation (around the origin) back to the original orientation,
cf.~Eq.~(\ref{invrot}).
Performing this combined transformation using the rules given in Ref.~\cite{mgjs}
and plugging the result into the right-hand side of Eq.~(\ref{Gm})
yields the taste-basis transformation rule
for hypercubic rotations \cite{MW}
\begin{equation}
  \j(\tx) \to \j'(\tx)
  = [R\otimes T^T]\, \j(\car^{-1} \tx) \,.
\label{transtst}
\end{equation}
Here $R_{(\s\t)}= 2^{-1/2} (1-\g_\s \g_\t)$ is the usual Dirac rotation,
while $T_{(\s\t)}= 2^{-1/2}(\g_\s-\g_\t) = T_{(\s\t)}^\dagger$
produces the rotation on the taste index.\footnote{
  The transformation rule in Ref.~\cite{MW} looks slightly different due to
  a further change of basis.  See also footnote~\ref{transposeG}.
}

The fermion blocking kernels (\ref{tstQ}) and (\ref{covQ}) are already
gauge covariant.
I now discuss how to ``covariantize'' their transformation properties
under hypercubic rotations.  The idea is to simply let any element of
the fine lattice, be it a site or a link, transform as it should
under the above fine-lattice rotation.  For the fermion blocking kernels
we need to make two choices.
What choice is being made will be prescribed by a set of discrete-valued
``disorder fields,'' that reside on suitable elements of the coarse lattice.
In detail, we have to decide to which one of the hypercube's
sixteen sites will all the fermion variables be parallel transported.
The chosen site will be determined by a vector field $\rel_\m$.
The possible values of $\rel_\m(\tx)$ are zero or one, and
the fine-lattice coordinates of the chosen site will be $2\tx+\rel(\tx)$.
We must also decide in which order to traverse the axes.
The ordering will be determined by another coarse-lattice field
$\perm=\perm(\tx)$
that takes values in $S_4$, the permutation group of four elements.

Let us next establish the transformation rules of these fields.
For the chosen-site field we demand that, if $x=2\tx+\rel(\tx)$,
then this relation will be respected by the rotation.
With $x'$ and $\tx'$ given by  (\ref{rot}), we must require
$x'=2\tx'+\rel'(\tx')$.  It is straightforward to show that the required
transformation rule is
\begin{equation}
  \rel'(\tx') = \car\,\rel(\car^{-1}\tx') + \D \,.
\label{transrel}
\end{equation}
Let us verify that Eq.~(\ref{transrel}) is a consistent transformation
on this field.  We must verify that for $\rel_\m=0,1$,
also $\rel'_\m$ takes only these two values.
Let us again consider the 
counter-clockwise rotation in the $(\s,\t)$ plane.
Only the $\s$ and $\t$ components undergo a non-trivial transformation,
which reads explicitly $\rel'_\t(\car_{(\s\t)}\tx)=\rel_\s(\tx)$
and $\rel'_\s(\car_{(\s\t)}\tx)=-\rel_\t(\tx)+1$.
We see that the translation by the unit vector $+\hat\s$
acts precisely to bring $\rel'_\s(\car_{(\s\t)}\tx)$ back
into the allowed range.

In order to write down the transformation rule for the axes-ordering field,
let us use the defining representation of the permutation group $S_4$
in terms of four-by-four orthogonal matrices,
each of which has one entry equal to one
and the rest equal to zero on every raw or column.
The axes ordering is then given by letting this matrix act on
the constant four-vector $v=(1234)$, that is, act on
the four-vector whose entries are given by $v_\m=\m$.
With this, the transformation rule is
\begin{equation}
  \perm'(\tx') = \perm(\car^{-1}\tx') \, \hat\p(\car) \,,
\label{transperm}
\end{equation}
where the permutations $\hat\p(\car_{(\s\t)}) \in S_4$
is represented by the four-by-four matrix
$\hat\p(\car_{(\s\t)})_{\m\n}
= \d_{\s\n}\d_{\t\m}+\d_{\s\m}\d_{\t\n} +P_{(\s\t)\m\n}\,$.

For completeness, recall the  transformation rule of the fine-lattice
gauge field, which is conveniently expressed as \cite{mgjs}
\begin{equation}
  U(x,y) \to U'(x,y) = U(\car^{-1} x,\car^{-1} y)\,,
\label{transU}
\end{equation}
where
\begin{equation}
  U(x,y) = \left\{
  \begin{array}{ll}
    U_{\m,x}\,, & y=x+\hat\m\,, \\
    U^\dagger_{\m,y}\,, & y=x-\hat\m\,, \\
    0\,, & {\rm otherwise}\,.
  \end{array}\right.
\label{Uxy}
\end{equation}

We are now ready to introduce new parallel transporters
that transform covariantly under rotations. Let
\begin{equation}
  \ww(x,y,\hat\p;\cu)\,,
\label{newW}
\end{equation}
be the parallel transporter from $y$ back to $x$, which
traverses the axes in the order determined by $\hat\p \in S_4$,
as follows.  With the constant four vector $v$ introduced above, we let
$v_{\hat\p} = \hat\p v$.  Starting at $y$ and letting $\n=v_{\hat\p}(4)$,
we first go along the $\n^{\rm th}$ axis until
the $\n^{\rm th}$ coordinate is equal to $x_\n$.
The direction is determined by the sign of $y_\n-x_\n$.
Then we go along the
axis specified by $v_{\hat\p}(3)$ and so on. In four steps, each
involving a straight line, we go from $y$ back to $x$.
Note that the parallel transporter in Eq.~(\ref{tstQfix})
corresponds to the special case of choosing $\hat\p$ as the identity element.

Armed with the more general parallel transporter (\ref{newW}),
we modify the fermion kernel of the $k=0$ step by replacing
$\cw$ of Eq.~(\ref{tstQfix}) with
$\ww(2\tx+\rel(\tx),2\tx+r(\tx),\perm(\tx);\cu)$.
For convenience, the dummy summation variable of Eq.~(\ref{tstQ})
has been promoted to a field; its transformation properties are,
obviously, the same as those of $\rel(\tx)$.
The so-constructed parallel transporter transforms as
\begin{eqnarray}
  && \ww(2\tx+\rel(\tx),2\tx+r(\tx),\perm(\tx);\cu) \to
\label{transW}
\\
  && \hspace{10ex}\to \; \ww(2\tx+\rel'(\tx),2\tx+r'(\tx),\perm'(\tx);\cu')
\NON
  && \hspace{10ex} = \;  \ww(2\car^{-1}\tx+\rel(\car^{-1}\tx),
      2\car^{-1}\tx+r(\car^{-1}\tx),\perm(\car^{-1}\tx);\cu)\,.
\nonumber
\end{eqnarray}
With this, the right-hand side of Eq.~(\ref{tstQ})
attains the same hypercubic transformation properties as
the taste-basis field, cf.~Eq.~(\ref{transtst}).

Now that the fermions are each time parallel transported to a different
hypercube's site, we must also modify the gauge-field blocking kernels,
so as to maintain gauge invariance of the coarse-lattice theory.
For the $k=0$ step, the new gauge-field blocking kernel
is obtained by replacing Eq.~(\ref{BFb}) with
\begin{equation}
  W_{\m,\tx}
  = \ww(2\tx+\rel(\tx),2(\tx+\hat\mu)+\rel(\tx+\hat\m),\Perm_\m(\tx);\cu) \,,
\label{BWx}
\end{equation}
where again $\ww$ is defined by Eq.~(\ref{newW}).
In Eq.~(\ref{BWx}), the axes ordering is chosen
independently for each coarse-lattice link,
according to a new disorder field $\Perm_\m(\tx)$
taking values in the permutation group $S_4$.
Introducing notation analogous to Eq.~(\ref{Uxy}),
\begin{equation}
  \Perm(\tx,\ty) = \left\{
  \begin{array}{ll}
    \Perm_\m(\tx)\,, & \ty=\tx+\hat\m\,, \\
    \Perm_\m^\dagger(\ty)\,, & \ty=\tx-\hat\m\,, \\
    0\,, & {\rm otherwise}\,,
  \end{array}\right.
\label{Oxy}
\end{equation}
its transformation rule is (compare  Eq.~(\ref{transperm}))
\begin{equation}
  \Perm(x,y) \to \Perm'(x,y)
  = \Perm(\car^{-1} \tx,\car^{-1} \ty)\, \hat\p(\car) \,.
\label{Otrans}
\end{equation}

We are now ready for the implementation.
A complete set of disorder fields is introduced at each blocking step.
Re-instating the blocking-step label,
the ``measure'' for the disorder fields is
\begin{equation}
  \sumx{k}
  \equiv \prod_{\tx^{(k)}}
  \slide{0}{\Bigg[}
  {1\over 24} \ \sum_{\perm^{(k)}(\tx^{(k)}) \in S_4} \
  \slide{0}{\Bigg]}\;
  \prod_{\m,\tx^{(k)}}
  \slide{0}{\Bigg[}
  {1\over 16\cdot 24} \ \sum_{\rel^{(k)}_\m(\tx^{(k)})=0,1} \ \
  \sum_{\Perm^{(k)}_\m(\tx^{(k)}) \in S_4} \
  \slide{0}{\Bigg]} \,.
\label{Ddis}
\end{equation}
The $k=0$ blocking step takes the form
\begin{subequations}
\label{rich}
\begin{eqnarray}
  Z
  &=&
  \int \cd\cu \cd\c \cd\bc\;
   \exp[-S_g(\cu) -\bc \Dstag(\cu) \c]
\label{richa}
\\
  && \rule{0ex}{4ex} \times
  \sumx{0}\; \int \cd\cv^{(0)} \cd\j^{(0)} \cd\bj^{(0)} \;
  \exp\Big[ - \cb_0\left(\cv^{(0)},\cu\right) -\cn_0(\cu) \Big]\;
\label{richb}
\\
  && \hspace{5ex} \times
  \exp\Big[ -\a_0
  \Big(\bj^{(0)} - \bc\, Q^{(0)\dagger}(\cu)\Big)
     \Big(\j^{(0)} - Q^{(0)}(\cu) \c\Big) \Big]
\NON
  &=& \rule{0ex}{4ex} \int \cd\cv^{(0)} \cd\j^{(0)} \cd\bj^{(0)}\;
  \exp\Big[ -S_0 \Big( \cv^{(0)},\j^{(0)},\bj^{(0)} \Big) \Big] \,.
\label{richc}
\end{eqnarray}
\end{subequations}
The blocking transformation is introduced on line (\ref{richb}).
As promised, it consists of a coherent superposition of
``elementary'' blocking transformations,
each corresponding to a particular set of values of
all the disorder fields.
In going from Eq.~(\ref{richb}) to Eq.~(\ref{richc})
we both integrate out the original fields,
and sum over all values of the disorder fields.
Similar coherent superpositions of blocking transformations
are introduced in all subsequent steps.

With all the disorder fields in place,
the coarse-lattice theory obtained at the $n^{\rm th}$ step
is manifestly invariant under hypercubic rotations. To see this, observe that
all the blocking kernels in Eq.~(\ref{K}) become hypercubic-rotation invariant
thanks to the transformation properties endowed to the disorder fields.
The original action is invariant too, and the sum of the original action
plus the blocking kernels may be regarded as a generalized action,
which is hypercubic-rotation invariant as well.
The effective action $S_n$
obtained after integrating out any number of fields retains
the same invariance.

Because the values of the disorder \textit{fields} can vary locally,
the resulting action $S_n$ can be written as a sum over coarse-lattice sites
of a local hypercubic scalar.  The same would not be true had we restricted
the disorder fields to take globally constant values only.
Such a global sum would amount to averaging correlation functions
of different (blocked) theories, and the result would in general violate
clustering. If (manifest) hypercube symmetry was enforced by global averaging,
violation of clustering would occur at every blocking level $k$,
where it is expected to scale like a power of $a_k$
(and, ultimately, like a power of $a_c$), which is unacceptable.
This unpleasant situation is avoided, however,
because the disorder fields are local fields.

Turning to the representation (\ref{gauss}),
in order to maintain the gaussian nature of the remaining fermion integral
one must refrain from integrating out any fields other than fermions.
This means that summations over the disorder fields must not be carried out
explicitly as well.  The representation then takes the form
\begin{eqnarray}
  Z
  &=&
  \sumx{0} \sumx{1} \cdots \sumx{n}
  \int \cd\cu
  \prod_{k=0}^{n} \Big[ \cd\cv^{(k)} \Big]\;
  \bltz_n\Big(1;\cu,\{\cv^{(k)}\}\Big)
\NON
  && \hspace{13ex} \times
  \int \cd\j^{(n)} \cd\bj^{(n)}
  \exp\Big(-\bj^{(n)} D_n\, \j^{(n)}\Big)\,,
\label{gaussdis}
\end{eqnarray}
Related subsequent equations (\eg in Sec.~\ref{plan})
are modified accordingly.  We should also modify the process of
generating ensembles of blocked-lattice configurations.
Integrating out the remaining fermion fields as well as all the blocked
gauge fields in Eq.~(\ref{gaussdis}) we obtain
\begin{equation}
  Z =
  \sumx{0} \sumx{1} \cdots \sumx{n} \int \cd\cu \exp(-S_g)\; \det(\Dstag)\,,
\label{weightdis}
\end{equation}
which is to be compared with Eq.~(\ref{weightb}).\footnote{
  The generalization to $n_r \ne 1$, cf.\ Eq.~(\ref{weight}), is straightforward.
}
This equation states the (obvious)
result that, with no more blocked gauge fields around,
the disorder fields decouple from the original theory.
Therefore, the original fine-lattice gauge field is to be generated
as always with its usual Boltzmann weight $Z^{-1} \exp(-S_g)\, \det(\Dstag)$,
while all the disorder fields are to be generated with a flat measure.
Any ``tensor-product'' configuration made of a fine-lattice
gauge-field configuration
and a configuration of all the disorder fields then serves as a mother
configuration for the production of the chain of
daughter configurations of the blocked gauge fields.
In practice, this entails the simple instruction that a new set of values
of the disorder fields is to be picked up at random for any new evaluation
of a blocking kernel.

Last let me address the following question.
The fermion blocking kernels are not invariant under ordinary
chiral transformations, and this leads directly to the GW relation,
and to the replacement of any ordinary chiral symmetry by its GWL cousin,
as discussed for the case at hand in Refs.~\cite{rg,BGS}.
In comparison to hypercubic rotations,
the invariance (of the massless limit)
under ordinary chiral symmetries
is in fact lost already in the free theory after one or more blocking steps.
One may wonder whether a similar trick with some new
disorder fields would help us retain the invariance under ordinary chiral
transformations in the blocked theory.  The answer is yes,
but it carries with it very little gain as I will now explain.

In the free theory, the generator of (ordinary) chiral transformations
in the taste basis is $[\g_5\otimes \x_5]$.
Within the blocking process, we may enforce the invariance
under the same (global) chiral symmetry
by augmenting each of the fermion blocking kernels
with a new disorder field $M^{(k)}=M^{(k)}(\tx^{(k)})$
transforming like a mass spurion.
Specifically, $M^{(k)}(\tx^{(k)})$ is a sixteen by sixteen matrix, labeled
by a double, Dirac and taste, index.
It takes values in the $U(1)$ group $\exp(i\th [\g_5\otimes \x_5])$.
Again taking the $k=0$ step as an example,
the new blocking kernel would take the form
\begin{equation}
  \ck_{F}^{(0)} = \a_0
  \Big(\bj^{(0)} - \bc\, Q^{(0)\dagger}(\cu)\Big) \, M^{(0)} \,
     \Big(\j^{(0)} - Q^{(0)}(\cu) \c\Big)\,,
\label{spurion}
\end{equation}
where the dependence on all other disorder fields has been suppressed.
Under a global $U(1)$ chiral rotation,
and assuming that the taste-basis field transforms as
$\j \to \exp(i\th [\g_5\otimes \x_5])\, \j$,
the new disorder field transforms as
\begin{equation}
  M^{(k)}(\tx^{(k)})
  \to \exp(-2i\th [\g_5\otimes \x_5])\, M^{(k)}(\tx^{(k)})\,.
\label{transM}
\end{equation}
Taken alone, the new blocking kernel (\ref{spurion}) is in fact
invariant not only under global but also
under local $U(1)$ chiral transformations.  However,
the original action is only invariant under the corresponding
global transformations (the $U(1)_\e$ symmetry) in the massless limit.
Hence, the blocked action obtained after integrating over
the original staggered fields, as well as over all values of
the new disorder field, will be invariant under global
chiral rotations only, as it should.
Of course, that blocked action is not bilinear in the fermion fields,
nor can it be reasonably approximated by any bilinear fermion action
even in the free case.
This has to be so, or else the Nielsen-Ninomiya theorem \cite{NN}
would be violated.

More relevant is the role of the chiral disorder fields
within the representation (\ref{gaussdis}).
Particularly illuminating is to consider their effect
on the fermions pull-back mapping (\ref{pullF}).
It is easily seen that the role of $M^{(k)}(\tx^{(k)})$ is
to multiply the contact term in Eq.~(\ref{mappf}),
obtained while undoing the $k^{\rm th}$ blocking step,
by a chiral phase which depends on the Dirac and the taste indices.
Under a ``complete'' fermion pull-back all the way to the staggered
theory on the original fine lattice,
nothing would depend on the chiral disorder fields,
apart from the contact terms encountered along the way.
The integration at each blocking step
over all values of $M^{(k)}(\tx^{(k)})$ for all $\tx^{(k)}$
would thus wipe out all the contact terms.

For non-coinciding points $\tx^{(n)}\ne \ty^{(n)}$ on
the ``last'' coarse lattice, however, contact terms are absent anyway.
The upshot is that, when evaluating blocked fermion propagators on
blocked ensembles (cf.~App.~\ref{ensemble}),  we have the following choice
for coinciding coarse-lattice points.
We may either evaluate the contact terms generated by
the fermions pull-back mapping, assuming there were no
chiral disorder fields; or else, we are free to drop them,
assuming that these disorder fields were present.
Whichever choice we make,
it is of no consequence for any non-coinciding coarse-lattice points.
The coarse-lattice fermion propagator between non-coinciding points
is independent of the chiral disorder fields.

\section{\label{nnlcl} Non-locality of the interacting theory
at non-zero lattice spacing}
In the free theory, a local {\it square}-root operator may be constructed
at non-zero lattice spacing in the massive case \cite{DA}.
This is not possible in the interacting case: the fourth-root theory,
or the square-root theory for that matter,
are non-local for any non-zero fine-lattice spacing $a_f$.
This paper shows that the magnitude of all the non-local terms,
(but not their range!)
goes to zero with the fine-lattice spacing.

Here I briefly repeat the argument
why, in the interacting fourth-root theory, the range of the non-local terms
must be a physical scale \cite{BGS}.
With Eqs.~(\ref{gauss}) and (\ref{fctrn}) in mind let us assume on the contrary
that, after $n+1$ blocking steps,
a local fourth root exists in the sense that
\begin{equation}
  \det^{1/4}(D_n) = \exp(-\quart\d\Seff)\, \det(\tD\,)\,,
\label{dSeff}
\end{equation}
where $\d\Seff$ is local,
and where $\tD$ is a local lattice Dirac operator
which describes one quark in the continuum limit.\footnote{
  Both $\d\Seff$ and $\tD$ may in general depend on the original
  as well as on all the blocked gauge fields.
  The arguments simplify a bit if no blocking steps are done,
  as was assumed in Ref.~\cite{BGS}.
}

Let us now compare the actual Goldstone-boson spectrum of the ordinary
staggered theory (no roots) to that dictated by Eq.~(\ref{dSeff}).
Substituting the fourth power of Eq.~(\ref{dSeff})
back into Eq.~(\ref{Zn}) and noting that
$\det^4(\tD\,)=\det(\tD \otimes {\bf 1})$, the assumed locality of $\d\Seff$
would imply that the RG-blocked theory is in fact
a local four-taste theory with an exact $U(4)$ taste symmetry.
This would imply, in turn, that the
fifteen pseudo-Goldstone pions must be exactly degenerate.
This conclusion is wrong, however!
As explained above (see in particular Sec.~\ref{pullb} and App.~\ref{sym}),
the RG-blocked theory has the same
low-energy spectrum as the original staggered theory. This spectrum
constitutes of fifteen {\it non}-degenerate pseudo-Goldstone pions
for any non-zero fine-lattice spacing \cite{MG,LS,milc}.
Thus, the different lattice symmetries of
the staggered theory and of the putative theory defined by
the Dirac operator $\tD\otimes {\bf 1}$ rule out a local $\d\Seff$.
As discussed in Sec.~\ref{plan} and on, however, the notion
of a reweighted theory, namely of a taste-symmetric theory
that only {\it approximates} the
staggered theory, can be very useful.

\section{\label{mbound} The propagator bounds}
In this appendix I collect a few observations on the propagator bounds
(\ref{mscale}) and (\ref{m4scale}).  The first thing to notice
is that the precise form of these bounds is not important,
so long as it is known that the norm of the inverse Dirac operator
in question is bounded by some non-zero constant in the limit $a_f\to 0$.
That constant will depend on $m_r(a_c)$, and may depend on
$a_c$ and $\L$ as well.

A configuration for which the bound (\ref{mscale})
is nearly, but, not quite, saturated is an instanton with size $\r \sim a_c$.
The bound is not fully saturated because
there is no index theorem for staggered fermions.
(For related observations, see Refs.~\cite{BGS,BGSS};
for related numerical work, see Ref.~\cite{evs}.)
For zero modes of larger-size instantons,
the bound (\ref{mscale}) may be corrected by factors of $\log(\r/a_c)$
due to both wave-function and mass renormalizations over
the range from $a_c$ to $\r$.
Because $a_c \le \r \le \L^{-1}$, and both $a_c$ and $\L$ are held fixed,
I have neglected such logarithmic corrections.

My remaining comments concern the bound (\ref{m4scale}) in
the reweighted theory. (The bound (\ref{m4scale}) pertains to the one-taste
reweighted theory derived from the fourth-root staggered theory, but
similar comments apply to the four-taste
reweighted theory derived from the ordinary staggered theory.)
The $U(1)_\e$ symmetry of the massless staggered theory is disguised
as a  GWL chiral symmetry by the RG blocking.
This chiral symmetry is broken only
by the (staggered) mass term, which, in turn, is protected
against additive mass renormalization.
The same is not true for the reweighted theories:
for fixed $a_f>0$, the taste-invariant operator $D_{inv,n}$ has
no chiral symmetry in the limit where the staggered mass goes to zero.
Thus, Eq.~(\ref{minv}) reflects the presence of an additive fermion-mass
renormalization in the reweighted theory.

One should, however, be careful with the interpretation of Eq.~(\ref{minv}).
First, as noted above, already in the staggered theory itself
the bound (\ref{mscale}) is not expected to be completely saturated.
Second, in this paper I do not consider the reweighted theories
as ``stand-alone'' coarse-lattice theories.  Their renormalization
is \textit{defined} with references to the underlying fine-lattice
cutoff $a_f$ (see Sec.~\ref{rrew}).
Thus, the rightmost term in Eq.~(\ref{minv})
vanishes in the continuum limit $a_f\to 0$
with, in particular, $a_c$ fixed, which implies that \textit{no} fine-tuning
of the fermion mass in the reweighted theory is needed.

The conclusion that an additive mass renormalization of a certain size
is present cannot, in any case, be drawn based on the magnitude
of $\D_n$ alone, as can be seen from the following example.
In Ref.~\cite{BGS} another family of reweighted theories
was constructed with a taste-invariant Dirac operator
$D_{ov,n}(m)=\tD_{ov,n}(m)\otimes\id$ such that once again
$D_n(m)-D_{ov,n}(m)$ scales in essentially the same way
as does $\D_n(m)=D_n(m)-D_{inv,n}(m)$.  Nevertheless,
$D_{ov,n}(m)$ satisfies a GW relation in
the limit where the \textit{staggered} mass $m$ goes to zero.
This implies that, just like the original staggered mass,
the fermion mass residing in $D_{ov,n}(m)$
renormalizes multiplicatively.\footnote{
  This observation is relevant for Ref.~\cite{HH} where a comparison
  of the staggered ensemble to a reweighted overlap ensemble was attempted.
  See also Ref.~\cite{BGS}.
}

In the physical one-flavor theory the only chiral symmetry is anomalous,
and instantons modify the quark's mass.
Correspondingly, there is a related tiny correction
to the denominator in Eq.~(\ref{m4scale}), coming from the integration over
instantons with size in the range $a_f \le \r \le a_c$.
By choosing $a_c$ small enough, instantons with size $\r\le a_c$ are
strongly suppressed, and this correction can be made arbitrarily small
relative to $m_r(a_c)$.  No similar correction exists in a theory with more
than one degenerate flavor.\footnote{
  I thank Mike Creutz for a discussion of this point.
}
For related observations on the fourth-root
regularization of one-flavor QCD (prompted by the claims made in Ref.~\cite{Mike}),
see Refs.~\cite{BGSS,Steve}.

\section{\label{iroprep} Scaling and the multi-gauge-field representation}
In this appendix I further expand on scaling issues
within the multi-gauge-field diagrammatic expansion,
discussed in Sec.~\ref{rewpt} of the main text.
This appendix is organized around a few examples
that each illuminate some particular aspect.
In App.~\ref{sclF} I discuss the free theory,
and in App.~\ref{sclI} the interacting theory.

\subsection{\label{sclF} Free theory: small-momentum expansion of $D_n$}
Extending the result obtained in Ref.~\cite{rg} to $m\ne 0$ and
adapting to the present conventions and notation,
the blocked free propagator takes the form
\begin{equation}
  D_n^{-1}
  = \cm_n - \sum_\m \Big( i [\g_\m\otimes \id] \ca_\m^n
    + [\g_5 \otimes \x_5\x_\m] \cb_\m^n \Big)\,,
\label{freeDinv}
\end{equation}
where, in the massless limit,
\begin{equation}
  \cm_n\Big|_{m=0} = R_n = \sum_{k=0}^n (16)^{k-n}/\a_k \,.
\label{Rn}
\end{equation}
A straightforward calculation gives\footnote{
  Equation (\ref{estfree}) follows from Eq.~(11) of Ref.~\cite{rg}.
  The free propagator $D_n^{-1}$ is constructed as a sum over terms with
  fine-lattice momentum $p+(2\p/a_c)k^{(n)}$,
  where the coarse-lattice momentum $p$ is fixed,
  and $k_\m^{(n)}$ takes integer values
  such that the $k_\m^{(n)}$-summation samples all of the Brillouin zone
  of the original fine lattice.
  Thanks to the suppression provided by the blocking kernels,
  cf.~Eq.~(11d) therein, any term with $k_\m^{(n)} \ne 0$
  is $O(p^2)$ at most.
}
\begin{subequations}
\label{estfree}
\begin{eqnarray}
  \ca_\m^n &=& \frac{p_\m}{p^2+m^2} \Big(1+ O(p^2)\Big) \,,
\label{estfreea}
\\
  \cb_\m^n &=& 2^{-n-1}\, a_c\; \frac{p_\m^2}{p^2+m^2} \Big(1+ O(p^2)\Big)\,.
\label{estfreeb}
\end{eqnarray}
\end{subequations}
Inverting Eq.~(\ref{freeDinv}) we find\footnote{
  Equation (\ref{pexp}) corrects a mistake in Eq. (3.21) of Ref.~\cite{BGS}.
}
\begin{equation}
  D_n(p)
  =
  m + i[\sl{p}\otimes \id]
   + a_f \sum_\m [\g_5 \otimes \x_5\x_\m]\, p_\m^2
  - R_n \Big(m+i[\sl{p}\otimes \id] \Big)^2 + \cdots \,.
\label{pexp}
\end{equation}
The ellipsis stand for terms of homogeneity degree three or higher
in $p_\m$ and $m$.
Observe that, even though $D_n$ lives on the coarse lattice,
the first three terms on the right-hand side of Eq.~(\ref{pexp})
are exactly the same as in the usual taste-basis Dirac operator
\cite{taste,saclay}.  In particular, the ``skewed Wilson term'' (that
comprises the leading taste-breaking term) has a coefficient
that scales with the fine lattice spacing $a_f = 2^{-n-1}\, a_c$.
Because $-\p/a_c \le p_\m \le \p/a_c$, the leading taste-breaking term
is indeed of order $a_f/a_c^2$. Considering the massless limit we see that
there is also a term $R_n\, p^2$ with the same structure as
an ordinary Wilson term.  The coefficient $R_n$
scales with the coarse-lattice spacing because $\a_k^{-1}=O(a_k)$
by assumption.  The $R_n$-dependent terms reflect
the fact that $D_n$ satisfies a GW relation in the massless limit.

\subsection{\label{sclI} Aspects of scaling in the interacting theory}
The scaling of $\D_n$ was derived in Sec.~\ref{rewpt} for any
theory whose partition function can be cast in the
multi-gauge-field form of Eq.~(\ref{Znr}).
This includes as special cases the ordinary and fourth-root
staggered theories, with or without reweighting.
Here I illustrate some of the ``inner working'' of
the scaling of $\D_n$.
I first discuss two terms that (should) occur
in $\D_n$, and how their functional form depend on the blocking level $n$.
I then discuss how the two terms scale.

The free-theory result (\ref{pexp})
together with gauge invariance requires the presence in $\D_n$ of
a covariant, skewed Wilson term
\begin{equation}
  \co_D^{(n)} =
  a_f z_n^{-2}
  \sum_\m \bj^{(n)} [\g_5\otimes \x_\m\x_5] \nabla_\m^{[n]}\, \j^{(n)} \,,
\label{Odn}
\end{equation}
where, using Eq.~(\ref{Dnb}) and the overall normalization
of the blocking kernels in Eq.~(\ref{covQ}),
the wave function renormalization factor is
\begin{equation}
  z_n = \prod_{k=1}^n z^{(k)}\,.
\label{prodz}
\end{equation}
The covariant laplacian $\nabla_\m^{[n]}$ reduces to $p_\m^2 +O(p^4)$
in the free-theory limit.  The superscript notation $[n]$ is meant
to remind us that, excepting $\cv^{(n)}$, this covariant laplacian
depends on the entire ``tower'' of gauge fields,
$\nabla_\m^{[n]}=\nabla_\m^{[n]}(\cu,\cv^{(0)},\ldots,\cv^{(n-1)})$.
The explicit (complicated!) form can in principle be computed using Eq.~(\ref{Dn}).

Another example is based on the result of Ref.~\cite{Luo},
where it was shown that the taste-basis Dirac operator of the $k=0$ step
(cf.\ Eq.~(\ref{Dna}) and Sec.~\ref{rewpt}) contains a term with the generic form
\begin{equation}
  \co_F^{(0)} =  a_f\, \bj_{\a i}^{(0)} \cf^{[0]}_{\m\n}\, \j_{\b j}^{(0)}\,
  c^{(0)}_{\a\b ij \m\n}\,.
\label{Of}
\end{equation}
The notation $\cf^{[0]}_{\m\n}=\cf^{[0]}_{\m\n}(\cu)$ is a shorthand
for $(1-\cw_{\m\n}(\cu))/(ia_{\! f} g_0)$,
where $\cw_{\m\n}(\cu)$ is a Wilson-loop operator
(without a color trace), and $g_0$ is the bare coupling; $\cf^{[0]}_{\m\n}$
reduces to $F_{\m\n}$ in the classical continuum limit.
In Eq.~(\ref{Of}) all indices except color are explicitly shown,
and pairs of indices are summed over.
The dimensionless tensor $c_{\a\b ij \m\n}$ has $O(g_0)$ entries,
and its explicit form is such that the operator $\co_F^{(0)}$
violates taste symmetry.

The presence of both $\co_D^{(0)}$ and $\co_F^{(0)}$ in the taste-basis
Dirac operator
is dictated by shift symmetry, which mixes the leading, dimension-four,
taste-invariant part of this Dirac operator
with taste non-invariant terms of dimension five and higher.\footnote{
  The taste-symmetric propagator usually used in staggered perturbation
  theory is related to the taste-basis propagator
  by a non-local unitary transformation.
}
The precise form of $\co_D^{(0)}$ and $\co_F^{(0)}$ depends on the
covariant ``blocking'' kernel (\ref{tstQ})
that has been chosen for the $k=0$ step.
While Ref.~\cite{Luo} discusses the taste-basis Dirac operator
in the case $\a_0\to\infty$, the result (\ref{Of})
generalizes to $\a_0<\infty$.

How the operator $\co_F^{(0)}$ ``evolves'' with blocking
is less certain than in the case of $\co_D^{(n)}$, where we could appeal
to gauge invariance (and the free theory) to determine the overall
normalization.  Still, based on the fact that the initial taste-basis
operator $D_0$ is known to contain $\co_F^{(0)}$,
one would expect that $D_n$ contains a similar-looking taste-breaking term,
\begin{equation}
  \co_F^{(n)}
  = a_f \, \bj_{\a i}^{(n)} \cf^{[n]}_{\m\n}\, \j_{\b j}^{(n)}\,
  c^{(n)}_{\a\b ij \m\n}\,,
\label{Ofn}
\end{equation}
where the coefficients $c^{(n)}_{\a\b ij \m\n}$ evolve logarithmically.
The operator $\cf^{[n]}_{\m\n}$ has similar properties
to $\cf^{[0]}_{\m\n}$, except that now it depends on the ``tower''
of gauge fields
$\cf^{[n]}_{\m\n}=\cf^{[n]}_{\m\n}(\cu,\cv^{(0)},\ldots,\cv^{(n-1)})$.

Let me now consider the contribution of $\co_F^{(n)}$ to the taste-violating
part of blocked observables.  The point to make is that
any operator of the general form (\ref{Ofn}) will be suppressed
by the (effective) gauge-field action
in the Boltzmann weight (cf.\ Eqs.~(\ref{Znr}) and (\ref{contB})).
The underlying reason that this
works is that the discussion below is nothing but a reconstruction,
using the terminology of non-perturbative ensemble averages,
of the familiar diagrammatic argument why taste-violating processes
mediated by a hard-gluon exchange are suppressed
by powers of the lattice cutoff \cite{SinP,asq,Steve}.

Suppose that a ``big part'' of $\cf^{[n]}_{\m\n}$,
which accounts for the gauge-field dependence of $\co_F^{(n)}$,
comes from the fine-lattice gauge
field, or from a blocked gauge field $\cv^{(k)}$ with $k\ll n$.
Then one can devise gauge-field configurations
for which $\co_F^{(n)}$ is too large.
In fact, there exist configurations for which $\co_F^{(n)}$
will be $O(1/a_f)$.  However, all such configurations are rare.
A particular example consists of a
fine-lattice vector potential $A_{\m,x}$ with the shape of
a wave packet whose average momentum is $p \sim 1/a_f$ and
whose width is $\D p \sim 1/a_c$.  Such a vector potential
is coherent over the coarse-lattice scale;
its amplitude, $\ahat$, can in principle get as large as $O(1/a_f)$.
Were it to happen,
this would give rise to an $O(1/a_f)$ value of $\co_F^{(n)}$.
However, large values of $\ahat$ are suppressed.
By expanding the fine-lattice gauge-field action to quadratic order
one finds that the action of the ``wave packet'' is
$\sim \ahat^2 a_c^4 / a_f^2$.
Therefore the average value of $\ahat$ is $O(a_f/a_c^2)$.
(I have neglected the coupling-constant dependence, together with any other
corrections that scale logarithmically with the lattice cutoff.)
This, in turn, implies that $\co_F^{(n)} = O(a_f/a_c^2)$ as well.
Obviously, this particular ``wave packet'' configuration is further suppressed
in the ensemble because it has a limited phase space.
But a similar conclusion is reached for generic fluctuations
of the fine-lattice gauge field, when taking into account
that these fluctuations are uncorrelated over the coarse-lattice scale.
Individual (local) fluctuation will typically be $O(1/a_f)$.
The random-walk sum of $O((a_c/a_f)^4)$ such fluctuations is $O(a_c^2/a_f^3)$,
if the sum is over a fine-lattice region with roughly the size
of a coarse-lattice hypercube.
The average value, which is the only thing that a coarse-lattice
field will be sensitive to, is $O(a_f/a_c^2)$.
Thus, again, one finds that $\co_F^{(n)} = O(a_f/a_c^2)$.
Last, in the case of the local four-taste staggered theory
it is clear that, upon integrating the tower of gauge fields,
and the fine-lattice gauge field in particular, $\co_F^{(n)}$ will
induce four-fermi (or higher dimensional) terms in the coarse-lattice
action $S_n$, which are suppressed by $a_f^2$ (at least).\footnote{
  A parallel statement in the context
  of the fourth-root theory would only be meaningful within
  a diagrammatic expansion augmented by the replica trick.
  I stress that the discussion of Sec.~\ref{rooted}
  is non-perturbative, and free of this limitation.
}

I now turn to the role of $\co_D^{(n)}$.
Recall that, according to Sec.~\ref{rewpt},
$\D_n$ scales like $a_f/a_c^2$ (up to logarithms)
on the ensemble of any theory defined by Eq.~(\ref{Znr}).
However, one cannot deduce that $\co_D^{(n)}$ scales like $a_f/a_c^2$
all by itself.  We only know that the sum of all contributions to $\D_n$,
coming not only from $\co_D^{(n)}$ but from many other (higher-dimensional)
terms, must scale like $a_f/a_c^2$.  As I will now explain,
this scaling depends not only on the suppression provided by
the gauge-field action (as was the case for $\co_F^{(n)}$),
but also on the underlying staggered symmetries and,
in particular, on shift symmetry.  The example below
also shows that, in other circumstances, there exist operators
that would fit the description of $\co_D^{(n)}$ as given around Eq.~(\ref{Odn}),
and yet they would scale as badly as $1/a_f$.

In order to illustrate in what way things could be different,
let us consider the proposal \cite{BJ}
to couple the taste-basis fermions directly to a gauge field
on the lattice with spacing $a_0=2a_f$.
The resulting Dirac operator $D_0^{DK}$ has no shift symmetry
(the superscript ``$DK$'' stands for Dirac-K\"ahler formulation,
which was the main thrust of Ref.~\cite{BJ}).
The one-loop calculation of Ref.~\cite{MW} proves that
a taste-violating $O(1/a_f)$ mass term is induced in that theory.
Thus, $\D^{DK}_n$ (defined in analogy with Eq.~(\ref{Dinv}))
will scale like $1/a_f$ on the corresponding ensemble,
and will diverge in the limit $a_f\to 0$.  In more detail,
before doing any blocking steps, the $1/a_f$ divergence
originates directly from the skewed Wilson term \cite{MW}.
By itself, RG blocking obviously cannot ``eliminate''
any of the divergences of the underlying theory.
(Assuming on the contrary that a divergent mass term is present
in the fine-lattice propagator, but absent from the blocked propagator,
one reaches a contradiction by invoking the pull-back mapping,
cf.\ App.~\ref{sym}.)
Thus, after $n$ blocking steps, the $O(1/a_f)$ scaling is expected to come
from an operator $\co_D^{(n)DK}$ with the same general form as in Eq.~(\ref{Odn}).
Note that, because the kinetic term is only marginal, the range of
the blocked propagator $1/D_n^{DK}$ rapidly tends to zero with $n$.
This is, of course, nothing but the decoupling of a fermion
with a cutoff-scale mass.

Let us add to the Dirac operator of the DK theory
a (taste non-symmetric) mass counterterm:
\begin{equation}
  D^{DK}_{sub}= D^{DK}_0+\co_M^{DK}\,, \qquad
  \co^{DK}_M = m^{DK} \sum_\m [\g_5\otimes \x_\m\x_5]\,,
\label{mut}
\end{equation}
where $m^{DK}=O(1/a_f)$ too;
we moreover fine-tune $m^{DK}$ so that taste symmetry is restored,
and the correct physical quark masses are obtained in the continuum limit.
(Any additional,
taste-symmetric mass term present in $D^{DK}_0$ will renormalize
multiplicatively \cite{mgjs,MW}.)
With the subtracted operator $D^{DK}_{sub}$ at the starting point,
the taste-violating effects of the DK theory have become \textit{irrelevant}.
The desired $O(a_f/a_c^2)$ scaling
of the taste-breaking $\D^{DK}_{sub,n}$
will now be recovered on the corresponding ensemble.

The example of the DK theory illustrates that there are
two distinct issues here: separation of
relevant and irrelevant operators; and the scaling of irrelevant
operators (in particular, within the current RG framework)
once we have actually determined what they are.
The separation of relevant from
irrelevant terms will in general require subtractions (i.e.\
additive renormalizations).  This is indeed the case in the DK theory.
No such subtractions are needed in the staggered theory, thanks
to its extended symmetry.

As soon as the appropriate counterterms have been added to
the underlying theory, all terms in the RG-blocked lattice action
that break any of the symmetries of the continuum theory
must have become irrelevant, and will scale accordingly
on the corresponding ensemble.\footnote{
  In a strict technical sense, this statement directly applies to
  internal symmetries.  The role of rotation symmetry in the RG-blocked
  action is more involved, see App.~\ref{sym} and App.~\ref{disorder}.
}
This amounts to a standard lore
in the case of local theories.  In this paper I have extended
this conclusion to the fourth-root theory (under plausible assumptions).

In summary, what the example of the DK theory shows is that the only
conceivable way for the scaling of $\D_n$ to go wrong,
is when we overlook some of the necessary counterterms of the underlying
theory. Once all the counterterms needed for the desired (universal)
continuum limit are introduced, the anticipated scaling of
the taste-breaking part of the blocked Dirac operator, as discussed
in Sec.~\ref{rewpt}, is a generic property of the RG transformation.
In the staggered case, however,
it turns out that no counterterms are necessary \cite{mgjs},
thanks in particular to shift symmetry.



\end{document}